\documentclass[twocolumn]{aastex6}
\usepackage{natbib}
\usepackage{longtable}
\usepackage{multirow}
\newcommand{\kms}{km\,s$^{-1}$}

\shorttitle{LOSS Classifications and Stripped-Envelope Rates}
\shortauthors{Shivvers et al.}

\begin{document}

\title{Revisiting the Lick Observatory Supernova Search Volume-Limited Sample:
         Updated Classifications and Revised Stripped-envelope Supernova Fractions}

\author{Isaac Shivvers,\altaffilmark{1} Maryam Modjaz,\altaffilmark{2} WeiKang Zheng,\altaffilmark{1}  Yuqian Liu,\altaffilmark{2}   \\
    Alexei V. Filippenko,\altaffilmark{1} Jeffrey M. Silverman,\altaffilmark{3} Thomas Matheson,\altaffilmark{4} Andrea Pastorello,\altaffilmark{5} \\
    Or Graur,\altaffilmark{6,2,7,8} Ryan J. Foley,\altaffilmark{9} Ryan Chornock,\altaffilmark{10} Nathan Smith,\altaffilmark{11} \\
    Jesse Leaman,\altaffilmark{12} Stefano Benetti\altaffilmark{5} \\
}
\altaffiltext{1}{Department of Astronomy, University of California, Berkeley, CA 94720-3411, USA}
\altaffiltext{2}{Center for Cosmology and Particle Physics, New York University, New York, NY 10003, USA}
\altaffiltext{3}{Department of Astronomy, University of Texas at Austin, Austin, TX 78712, USA}
\altaffiltext{4}{National Optical Astronomy Observatory, Tucson, AZ 85719, USA}
\altaffiltext{5}{INAF-Osservatorio Astronomico di Padova, Vicolo dell'Osservatorio 5, I-35122 Padova, Italy}
\altaffiltext{6}{Harvard-Smithsonian Center for Astrophysics, 60 Garden St., Cambridge, MA 02138, USA}
\altaffiltext{7}{Department of Astrophysics, American Museum of Natural History, New York, NY 10024-5192, USA}
\altaffiltext{8}{NSF Astronomy and Astrophysics Postdoctoral Fellow}
\altaffiltext{9}{Department of Astronomy and Astrophysics, University of California, Santa Cruz, CA 95064, USA}
\altaffiltext{10}{Department of Physics and Astronomy, Ohio University, Athens, OH 45701, USA}
\altaffiltext{11}{Steward Observatory, University of Arizona, 933 N. Cherry Ave., Tucson, AZ 85721, USA}
\altaffiltext{12}{Advanced Robotics and Automation Lab, Department of Computer Science
and Engineering, University of Nevada, Reno, NV 89557, USA}

\email{ishivvers@berkeley.edu}

\begin{abstract}
We re-examine the classifications of supernovae (SNe) presented in the Lick Observatory
Supernova Search (LOSS) volume-limited sample with a focus on the stripped-envelope SNe.
The LOSS volume-limited sample, presented by \citet{2011MNRAS.412.1419L} and \citet{2011MNRAS.412.1441L},
was calibrated to provide meaningful measurements of SN rates in the local universe;
the results presented therein continue to be used for comparisons to theoretical and modeling efforts.
Many of the objects from the LOSS sample were originally classified based upon only a small subset
of the data now available, however, and recent studies have both updated some
subtype distinctions and improved our ability to perform robust classifications, especially for stripped-envelope SNe.
We re-examine the spectroscopic classifications of all events in the LOSS volume-limited sample
(180 SNe and SN impostors) and update them if necessary. We discuss
the populations of rare objects in our sample including broad-lined Type Ic SNe,
Ca-rich SNe, SN~1987A-like events (we identify SN~2005io as SN~1987A-like
here for the first time), and peculiar subtypes.  The relative fractions of Type Ia SNe, Type II SNe, and stripped-envelope SNe 
in the local universe are not affected, but those of some subtypes are. 
Most significantly, after discussing the often unclear boundary between SNe~Ib and Ic when only noisy
spectra are available, we find a higher SN Ib fraction and a lower SN Ic fraction than calculated by \citet{2011MNRAS.412.1441L}:
spectroscopically normal SNe~Ib occur in the local universe $1.7 \pm 0.9$ times more often than do normal SNe~Ic.
\end{abstract}


\section{Introduction}
\label{sec:intro}

The Lick Observatory Supernova Search (LOSS) has been a long-running
project at the University of California, Berkeley, using
the Katzman Automatic Imaging Telescope at Lick Observatory 
\citep[KAIT; e.g.,][]{2000AIPC..522..103L,2001ASPC..246..121F,2003fthp.conf..171F,2005ASPC..332...33F},
with many spectroscopic follow-up observations obtained with the 3\,m Shane telescope at Lick and the 10\,m telescopes at Keck Observatory.
LOSS/KAIT has been discovering and observing SNe since first light in 1996;
these data have contributed to several PhD theses and formed the foundation of many research projects on SNe.
A detailed examination of the relative rates of nearby SNe was one of those projects,
and was published as a series of papers in 2011
\citep{2011MNRAS.412.1419L,2011MNRAS.412.1473L,2011MNRAS.412.1441L,2011MNRAS.412.1508M,2011MNRAS.412.1522S}.
The second of these, \citet[][L11 hereafter]{2011MNRAS.412.1441L}, presents a sample
of 180 events that occurred within 80\,Mpc (for Type Ia SNe) or 60\,Mpc (for
core-collapse SNe), all of which were spectroscopically classified
\citep[the classes of SNe are differentiated primarily via spectroscopy; e.g.,][]{1997ARAA..35..309F}.
Most SN classifications from this time period were performed via visual inspection and comparisons
with spectra of a few SNe of well-understood types and subtypes.

Over time we have found that a small fraction of the objects in L11
deserve reclassification; in some cases this is because the original 
classifications were made using only a subset of the now-available data on the objects,
while in other cases our more modern classification methods are less prone to errors than the methods
used at the time of classification.
Independent of data quality or cadence, there is a 
history of debate in the literature over the exact distinction (if any) between 
SNe~Ib and SNe~Ic and whether transitional events  showing weak helium lines exist
\citep[e.g.,][]{1990AJ....100.1575F,1990RPPh...53.1467W,1994ApJ...436L.135W,1996ApJ...462..462C,2001AJ....121.1648M,2006PASP..118..791B}.

The results of recent efforts by \citet{2014arXiv1405.1437L}, \citet{2014AJ....147...99M}, and \citet{2016ApJ...827...90L}
argue that the distinction between SNe~Ib and SNe~Ic is useful, and they
offer a clearly defined scheme for discriminating between them alongside updated software tools to perform
those classifications in a repeatable manner.
\citet{2014AJ....147...99M} identify as SNe~Ib all events with detections of 
both the \ion{He}{1}\,$\lambda$6678 and \ion{He}{1}\,$\lambda$7065 lines at phases between
maximum light and $\sim$50\,days post-maximum, regardless of line strengths
(the stronger \ion{He}{1} $\lambda$5876 line is also present, but overlaps with \ion{Na}{1}).  
They find that at least one good spectrum observed at these phases is necessary and sufficient to detect the helium lines, which
are often absent at pre-maximum and nebular phases even for helium-rich events. Using
this classification scheme, they find
evidence for a transitional population of ``weak helium'' SNe~Ib \citep{2011MNRAS.416.3138V,2014AJ....147...99M,2016ApJ...827...90L}.

Clarifying the distinction between SNe~Ib and SNe~Ic is important given the surprising ratio of population fractions for these subtypes
found by LOSS \citep[SNe~Ic/SNe~Ib = $14.9^{+4.2}_{-3.8}\% / 7.1^{+3.1}_{-2.6}\%$;][]{2011MNRAS.412.1522S},
which has proven difficult to reproduce with stellar modeling efforts
\citep[e.g.,][]{2009AA...502..611G,2010ApJ...725..940Y,2015PASA...32...15Y},
though see also \citet{2013AA...558L...1G}.
\citet{2014AJ....147...99M} show that a subset of the objects originally labeled SNe~Ic in their
sample in fact do qualify for the SN Ib label according to the definition above, and so they relabel these events as SNe~Ib
(see their discussion of all such cases, in their \S 4.2).

For some of those SNe, the spectra that were used to classify them and
thus announce their types were obtained before
the helium lines became prominent; for some, applying proper telluric corrections made the
\ion{He}{1}\,$\lambda$6678 or $\lambda$7065 lines more apparent; for others
the spectra show clear helium but the exact division between SNe~Ib and SNe~Ic was under debate in the 
literature at the time of classification \citep[e.g., SN 1990U;][]{1990IAUC.5069....1F,1990IAUC.5111....2F,2001AJ....121.1648M}.
We explore these issues within the LOSS sample and also find that some events with helium lines were systematically labeled as SNe~Ic ---
we update the classifications for these events and recalculate the relative fractions of core-collapse events.

In this article we re-examine the classifications of the 180 events in the volume-limited sample of L11
and we make public all spectra of them we have been able to locate.
This work was performed in conjunction with \citet{2016arXiv160902921G,2016arXiv160902923G}, who re-examine correlations between SN rates
and galaxy properties.
Note that much of the spectroscopy discussed herein has already been described in the literature and
made publicly available by, for example, \citet[SNe~Ia]{2012MNRAS.425.1789S}, 
\citet[SNe~II]{2014MNRAS.445..554F,2014MNRAS.442..844F}, \citet[SNe~IIb/Ib/Ic]{2001AJ....121.1648M}. We collect
these spectra, light curves obtained by LOSS, as-yet unpublished spectra
from our archives, and as-yet unpublished spectra contributed from other SN research groups' archives, and analyze the complete set.

We present 151 newly published spectra of 71 SNe and 20 rereduced KAIT light curves.
In \S\ref{sec:data} we describe these data, in \S\ref{sec:methods} we detail our methods
for classification, in \S\ref{sec:updated} we present all updated classifications and discuss notable events within the sample,
in \S\ref{sec:rates} we calculate updated core-collapse SN rates in the local universe,
in \S\ref{sec:progenitor} we discuss the implications these updates have for our understanding of the progenitors of stripped-envelope SNe,
and in \S\ref{sec:conclusion} we conclude.

\section{Data}
\label{sec:data}

Spectra were collected from our own UC Berkeley Supernova Database
 \citep[UCB SNDB;][]{2012MNRAS.425.1789S},\footnote{The SNDB was updated in 2015 and is available online at \url{http://heracles.astro.berkeley.edu/sndb/}.}
from the literature, and from WISeREP \citep[the Weizmann Interactive Supernova
Data REPository;][]{2012PASP..124..668Y}.\footnote{\url{http://wiserep.weizmann.ac.il/}}
We do not include any results from spectropolarimetric or nonoptical observations;
we know of no such observations that would help for the few events
we cannot robustly classify using optical data.
We made an effort to track down as-yet unpublished spectra for all objects
with sparse or no spectral data in our database or in the public domain.
All objects in this sample were classified in the 
Central Bureau of Electronic Telegrams (CBETs), and we
contacted original authors to request data whenever possible.
Contributions were made by the Center for Astrophysics 
(CfA) SN group \citep{2008AJ....135.1598M},\footnote{\url{https://www.cfa.harvard.edu/supernova/}}
the Padova-Asiago SN group \citep{2014AN....335..841T},\footnote{\url{http://sngroup.oapd.inaf.it/}}
the Carnegie Supernova Project \citep[CSP;][]{2006PASP..118....2H},\footnote{\url{http://csp.obs.carnegiescience.edu/}}
and the National Astronomical Observatories, Chinese Academy of Sciences (NAOC)
SN group \citep{1999ScChA..42..220Q,1999ScChA..42.1075L}.

We publish spectra from the following observatories and instruments:
\begin{itemize}
\item the Kast double spectrograph \citep{kast} mounted on the Shane 3\,m telescope at Lick Observatory;
\item the Low Resolution Imaging Spectrometer \citep[LRIS;][]{1995PASP..107..375O} 
and the Echellette Spectrograph and Imager \citep[ESI;][]{2002PASP..114..851S} on the 10\,m Keck I \& II telescopes at Keck Observatory;
\item the FAST spectrograph \citep{1998PASP..110...79F} on the Tillinghast 60\,inch telescope 
and the Blue Channel spectrograph \citep{1989PASP..101..713S} on the 6.5\,m Multiple Mirror Telescope (MMT) at the Fred Lawrence Whipple Observatory (FLWO);
\item the European Southern Observatory (ESO) Faint Object Spectrograph and Camera \citep[EFOSC;][]{1984Msngr..38....9B} on the ESO 3.6\,m telescope,
the Danish Faint Object Spectrograph and Camera \citep[DFOSC, modeled after EFOSC;][]{1995Msngr..79...12A} on the Danish 1.54\,m telescope,
and the ESO Multi-Mode Instrument in medium resolution spectroscopy mode \citep[EMMI;][]{1986SPIE..627..339D} on the ESO 3.58\,m New Technology Telescope,
all at La Silla Observatory;
\item the Asiago Faint Object Spectrograph and Camera (AFOSC, modeled after EFOSC) on the 1.82\,m Copernico telescope 
and the Boller and Chivens spectrograph (B\&C$_{1.2}$) on the 1.2\,m Galileo telescope
at Asiago Observatory;
\item the Boller and Chivens spectrograph (B\&C$_{2.5}$)
and the Wide Field Reimaging CCD Camera in long-slit spectroscopy mode \citep[WFCCD, described by][]{2006PASP..118....2H} on the 2.5\,m du Pont telescope 
and the Low Dispersion Survey Spectrograph \citep[LDSS-2;][]{1994PASP..106..983A} on the 6.5\,m Magellan Clay telescope at Las Campanas Observatory;
\item and the Optomechanics Research, Inc.\footnote{\url{http://www.echellespectrographs.com/about.htm}}
spectrograph (OMR) mounted on the NAOC 2.16\,m telescope at Xinglong Observatory near Beijing, China.
\end{itemize}

Details of the spectral reduction pipeline used by the UCB team are described by
\citet{2012MNRAS.425.1789S}. \citet{2008AJ....135.1598M}, \citet{2012AJ....143..126B}, and \citet{2014AJ....147...99M}
discuss the reduction process performed on the
CfA spectra, and \citet{2006PASP..118....2H} outline the reduction process performed on the CSP spectra.
Standard IRAF\footnote{\url{http://iraf.noao.edu/}} reduction packages were used by the Padova-Asiago and NAOC groups.
Most spectra presented here have resolutions
of $\sim$10\,\AA, were observed at or near the parallactic angle \citep{1982PASP...94..715F}, 
and were flux calibrated with bright standard stars observed at similar airmasses.
Most spectra have also been corrected for wavelength-dependent telluric absorption.
Details of the observations and data-reduction methods vary from group to group,
and we discuss any possible data-quality issues for the spectra most vital to our classification effort in the 
text. 

All photometry used by L11 and in this effort was observed at Lick Observatory with KAIT
or the Nickel 1\,m telescope, and all SNe discussed here were discovered by
LOSS/KAIT \citep[e.g.,][]{2000AIPC..522..103L,2001ASPC..246..121F}.
KAIT photometry is generally performed on unfiltered images (the {\it clear} band), 
though filtered {\it BVRI} KAIT images of some events are available. Nickel data are observed
through a {\it BVRI} filter set. Details for both instruments and for our photometry reduction pipeline are given by \citet{2010ApJS..190..418G},
and we present these light curves as observed, without correcting for Milky Way (MW) or host-galaxy dust absorption, unless otherwise stated.
All of the spectra and photometry used in this project will be made public through the
UCB SNDB, WiseREP, and the Open Supernova Catalog \citep{2016arXiv160501054G}.\footnote{\url{https://sne.space/}}
See Appendix~\ref{appendix} for logs of the data released publicly here for the first time.

\section{Classification Methods}
\label{sec:methods}

Following \citet{2012MNRAS.425.1789S} and \citet{2014AJ....147...99M},
we use the SN IDentification code\footnote{\url{http://people.lam.fr/blondin.stephane/software/snid/index.html}}
\citep[SNID;][]{2007ApJ...666.1024B} as our primary classification tool.
SNID classifies SNe by cross-correlating an input (optical) spectrum against a library of template spectra \citep{1979AJ.....84.1511T}.
Updated sets of template spectra have been released since the original
release of SNID --- for this study we use the BSNIP v7.0 templates \citep{2012MNRAS.425.1789S}
augmented by the \citet{2014arXiv1405.1437L} stripped-envelope templates (and following
all suggestions from their Table 4).  When running SNID, we set the SN redshift with the {\it forcez} keyword
using observed host-galaxy redshifts from the NASA/IPAC Extragalactic Database (NED).\footnote{\url{https://ned.ipac.caltech.edu/}}
For those SNe that SNID alone cannot identify, we incorporate results 
from two other spectral identification codes, Superfit\footnote{\url{http://www.dahowell.com/superfit.html}}
\citep{2005ApJ...634.1190H}
and GELATO\footnote{\url{https://gelato.tng.iac.es/}} \citep{2008AA...488..383H},
and for some stripped-envelope SNe we also compare to the average spectra of \citet{2016ApJ...827...90L}.

As shown by (for example) L11, the light curves
of SNe~Ic, Ib, and IIb are similar to each other, but are generally distinguishable from those of SNe~Ia and hydrogen-rich core-collapse SNe.
We incorporate light-curve information in our classifications when it proves useful, comparing the light curves of individual objects to the {\it clear}-band templates from L11
and providing constraints on the phases of spectra.
Recent studies have advanced our understanding of stripped-envelope SN light-curve evolution
\citep[e.g.,][]{2011ApJ...741...97D,2013MNRAS.434.1098C,2014ApJS..213...19B,2015AA...574A..60T,2016MNRAS.457..328L,2016MNRAS.458.2973P}.
\citet{2011ApJ...741...97D} present template SN~Ib/c light curves in the $R$ and $V$ bands assembled from 25 events
and \citet{2016MNRAS.457..328L} give template bolometric light curves assembled from 38 events,
while L11 produce four template light curves for stripped-envelope SNe: ``Ibc.fast,'' ``Ibc.ave,'' ``Ibc.slow,'' and ``IIb.''
The Ibc.ave template is very similar to the $R$-band SN~Ibc template of \citet{2011ApJ...741...97D} and the
SN~Ib and SN~Ic templates of \citet{2016MNRAS.457..328L}.
The SN~IIb templates from L11 and \citet{2016MNRAS.457..328L} are also in good agreement, and both show 
cooling envelope emission followed by a dip and a rise to a second radioactively-powered peak, 
with a post-peak evolution basically indistinguishable from that of SNe~Ib/c.  
L11 do not produce a template for broad-lined Type Ic SNe (labeled SNe~Ic-BL here), but other authors show that SN~Ic-BL light curves are quite similar to those
of other SNe~Ib/c though trending toward higher absolute luminosities \citep[e.g.,][]{2011ApJ...741...97D,2015AA...574A..60T,2016MNRAS.458.2973P}.

Several recent large-scale SN data releases have relied on SNID classifications 
using relatively stringent requirements for a robust identification, requiring a high {\em rlap} value for the top match 
({\em rlap} is a quality parameter used by SNID --- a higher value corresponds to a more trustworthy classification)
and that the first few matches be of the same subtype
\citep[e.g.,][]{2012MNRAS.425.1789S,2013MNRAS.430.1746G,2014AJ....147...99M,2015MNRAS.450..905G}.
We follow these methods whenever possible, and for most of the SNe in our sample
they clearly indicate a single type and subtype.

All of the SNe~Ia in this sample have been examined in detail by other authors
\citep[e.g.,][]{2012AJ....143..126B,2012MNRAS.425.1789S,2013ApJ...773...53F}.
We follow the methods of \citet{2012MNRAS.425.1789S} to determine SN Ia
subtypes, and we do not attempt to identify subpopulations within the normal SNe~Ia
--- i.e., high-velocity events \citep{2009ApJ...699L.139W} or the subgroups
defined by \citet{2005ApJ...623.1011B}.
We discuss the more peculiar SNe~Ia from this sample in \S\ref{sec:ia}.

We do not attempt to sort the hydrogen-rich SNe~II into the
IIP and IIL subtypes.  Type II SNe have long been sorted into those that exhibit
a clear plateau phase and those that decline linearly in magnitudes
\citep[IIP and IIL, respectively; e.g.,][]{1979AA....72..287B,1997ARAA..35..309F}.
L11 used spectra to identify H-rich SNe, and then labeled
as Type IIL those that decline more than 0.5\,mag in the $R$ band during the first 50\,d after explosion
and the rest as IIP, but recent work has shown that the issue may be more complex.
While \citet{2012ApJ...756L..30A} find there to be distinct SN~IIP and SN~IIL subclasses 
among the $R$-band light curves of 21 H-rich noninteracting Type II SNe, \citet{2014ApJ...786...67A}
show that their sample of $V$-band light curves for 116 SNe~II indicate that there is
a continuous distribution of properties for these events.
\citet{2016ApJ...820...33R} present an analysis of the early light curve rise for 57 events finding
only a weak correlation between rise times and decline rates, \citet{2015ApJ...799..208S} and \citet{2016MNRAS.459.3939V}
argue that there exists a continuous distribution of properties for SNe~II and that there is no evidence for separate SNe~IIP and SNe~IIL subclasses,
while \citet{2016ApJ...828..111R} argue for a type II subclassification system based upon
both the rise and the fall of the light curves and \citet{2014MNRAS.445..554F} argue that
a simple subclass definition based upon the light curve decline alone remains reasonable.
Throughout this article, we group the SN~IIP-like and SN~IIL-like events under the label ``SNe Type II,'' but
when comparing to data from other sources we preserve the SN~IIP/SN~IIL labels if given by the
original authors.

Though we do not differentiate between SNe~IIL and SNe~IIP, we do identify other H-rich subclasses.
We identify the H-rich SNe with narrow spectral emission lines indicative of interaction with dense
circumstellar material \citep[SNe~IIn; e.g.,][]{1990MNRAS.244..269S,1991MNRAS.250..513C,1991ESOC...37..343F} and the SN impostors,
thought to be nonterminal ejections or explosions from the surface of massive stars
\citep[labeled ``IIni'' in L11; e.g.,][]{2000PASP..112.1532V,2006MNRAS.369..390M,2011MNRAS.415..773S}.
We  also identify three slow-rising (SN~1987A-like) Type II SNe in our sample \citep[e.g.,][]{1989ARAA..27..629A,1993ARAA..31..175M}.

Our main focus is on the stripped-envelope SNe. We divide this class
into those with some hydrogen features (SNe~IIb), those without hydrogen features
\citep[or with only very weak hydrogen features;][]{2016ApJ...827...90L} but with 
clear helium features (SNe~Ib), and those exhibiting neither clear hydrogen nor clear helium features (SNe~Ic).
The exact distinction between SNe~Ib and Ic has been an issue of some debate in the literature;
we follow \citet{2014AJ....147...99M} and \citet{2016ApJ...827...90L} to define the differences
between these subclasses.  
We differentiate between Type Ic SNe and Ic-BL SNe \citep[see][]{2016ApJ...832..108M},
and we identify the ``calcium-rich'' SNe separately
\citep[included in the Ibc-pec category by L11, this class of events has been described by, e.g.,][]{2003IAUC.8159....2F,2010Natur.465..322P,2012ApJ...755..161K}.

\begin{figure}[t]
\epsscale{1.0}
\plotone{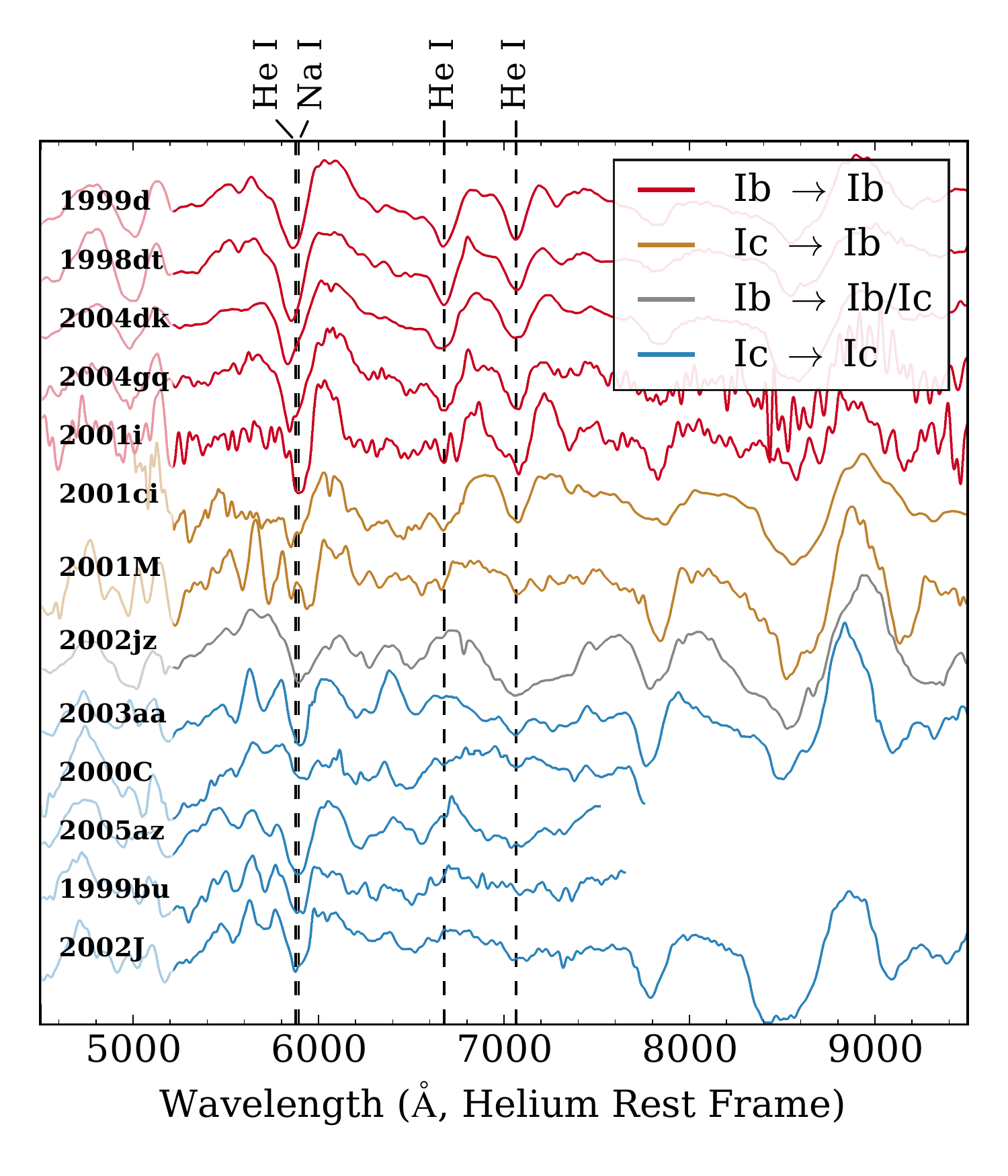}
\caption{Representative sample of spectra of the SNe~Ib and Ic in our sample, observed
     between 5 and 20\,days after peak brightness, as well as spectra of three SNe for which
     we provide updated classifications (SNe 2001M, 2001ci, and 2002jz).
     We have subtracted a spline continuum from these spectra, smoothed them with a 50\,\AA\
    Gaussian kernel, and shifted them in velocity space to align their \ion{He}{1} features
    (or \ion{Na}{1}\,$\lambda$5892 if no helium is detected).
    \label{fig:IbVSIc}}
\end{figure}

For events with spectra that match both the SN~Ib and SN~Ic SNID templates equally well, 
we discuss the available data in detail and assign
the label ``Ib/Ic'' (i.e., unsure) if we remain unable to determine a single best classification.
Figure~\ref{fig:IbVSIc} shows that this classification scheme tends to move
events that were previously labeled SNe~Ic into the Ib or
Ib/Ic categories (there are no SNe~Ib that we reclassify as SNe~Ic in this work).

We follow \citet{2011MNRAS.412.1522S} and group the SNe~IIb with other stripped-envelope SNe in our sample,
although L11 included them with the Type II SNe.
SNe~IIb show a strong H$\alpha$ line at early times, as do normal SNe~II, but then 
the hydrogen fades and the later spectra of SNe~IIb resemble those of SNe~Ib
\citep[e.g.,][]{1988AJ.....96.1941F,1993ApJ...415L.103F,1997ARAA..35..309F,2008MNRAS.389..955P,2011ApJ...739...41C,2013ApJ...767...71M}.
Several authors have claimed the detection of weak high-velocity hydrogen features in
SNe Ib and Ic \citep[e.g.,][]{2006PASP..118..791B,2007PASP..119..135P,2016ApJ...820...75P,2016ApJ...827...90L}.
However, \citet{2016ApJ...827...90L} argue that the putative weak H$\alpha$ absorption line often present in SNe~Ib
is, at all phases, weaker than the H$\alpha$ line in SNe IIb, and that SNID capably distinguishes
between SNe~Ib and SNe~IIb even after the strong H$\alpha$ feature of the SNe~IIb has
faded, so long as spectra were obtained during the photospheric phase.  The nebular spectra of SNe~IIb and SNe~Ib,
on the other hand, are often very similar and are not well separated by SNID.
When discriminating between Types IIb and Ib, we trust the SNID result if obtained 
from spectra of the photospheric phase.

To examine possible biases SNID may introduce when classifying stripped-envelope subtypes,
we ran a series of trials introducing wavelength restrictions, noise, and artificial dust reddening to spectra of Type IIb, Ib, and Ic 
SNe at two different phases in their evolution (near maximum brightness and 2--4\,weeks post-maximum).
We classified the degraded spectra with the methods described above and compared the results to those
obtained from the original data.
We chose events that are not included in the SNID template set
and for which we have relatively high signal-to-noise ratio (S/N) spectra (S/N $> 30$) at these phases covering $\sim3500$--10,000\,\AA:
SN~IIb~2003ed, SN~Ib~1998dt, and SN~Ic~2003aa. The spectra used in this study either cover a similar wavelength
range or $\sim3500$--7500\,\AA; we test the efficacy of SNID using both the full spectra and
spectra trimmed to match the smaller wavelength range.

We find that, regardless of subtype or which of the two wavelength ranges is used, SNID capably classifies events in
the presence of strong reddening ($E(B-V) \sim 2.0$\,mag and $R_V = 3.1$), so long as the spectra exhibit
a S/N $\gtrsim 1$--3.  This is to be expected, as SNID divides input spectra by a pseudo-continuum fit and discards
the spectral color information before performing cross-correlation \citep{2007ApJ...666.1024B}.
In contrast, classifications performed via visual comparison may be prone to error when strong reddening is present.

At moderate and high noise levels (S/N $\lesssim$ 3), the post-maximum spectrum of SN~Ic~2003aa 
could be confused with a SN~Ib spectrum while the near-maximum-light spectrum is still identified as that of a SN~Ic
\citep[the \ion{He}{1} lines in SNe~Ib fade as the event nears the nebular phase; e.g.,][]{2014AJ....147...99M}.
At extremely high noise levels (S/N $\lesssim$ 1) SNID prefers a SN~Ic-BL classification for SN~2003aa,
especially when examining the post-maximum spectrum or using spectra covering only the smaller wavelength range,
which does not capture the strong \ion{Ca}{2} near-infrared (IR) triplet feature.

We also find that,
if the SN redshift is uncertain and the S/N is low, SN~Ic spectra can be confused with SN~Ia spectra
\citep[as shown by][]{2007ApJ...666.1024B},
and so incorporating an independently measured redshift is helpful.  Though SN~2004aw is not in
the sample discussed here, it offers a nice illustration of the sometimes-confounding similarities between
spectra of Type Ia and Type Ic SNe \citep[e.g.,][]{2004IAUC.8311....3M,2004IAUC.8312....3B,2004IAUC.8331....2F}.

Examining SN~1998dt, we find that SNID correctly identifies it as a SN~Ib at both phases using spectra with S/N $\gtrsim$ 1,
and though the classification becomes very uncertain using only spectra with S/N $\lesssim$ 1 and a restricted wavelength
range, SNID never prefers an incorrect label.

SN~2003ed was correctly identified as a Type IIb SN so long as we examined spectra with S/N $\gtrsim$ 1.
Unlike SN~1998dt, however, including spectra covering an extended wavelength range did not significantly affect the results
even at low S/N, since the \ion{He}{1}\,$\lambda$5876 and H$\alpha$ features proved the most useful and they are 
captured by all of our spectra.

Based on the above discussion, we adopt the following guidelines to avoid systematic biases 
when using SNID for our stripped-envelope classifications. 
First, it is difficult to discriminate between SNe Ib and Ic or between SNe IIb and Ib
if only noisy (S/N $\lesssim 3$) spectra observed at more than a few weeks post-maximum are available. 
Second, SNe Ic can be mislabeled as SNe Ic-BL when only low-S/N spectra are available.

\section{Updated Classifications}
\label{sec:updated}

For the bulk of the SNe in the sample, especially the SNe Ia,
our methods robustly confirm the L11
classifications.\footnote{Note that, in a small number of cases, L11 reclassified some 
events from the original type announced in the CBETs; in this article we 
only discuss differences relative to the L11 labels.}
Table~\ref{tab:all} lists all SNe in this sample, the
type and subtype labels used by L11, and our updated labels.
In the ensuing subsections, we
discuss each changed classification individually as well
as the rare subtypes and the uncertain and peculiar events
we find.  Many of the comparison spectra used in this section were drawn from the
updated SNID template set; citations to the 
original publications are given.

\begin{deluxetable*}{ l | c c | c || l | c c | c }
\tablewidth{0pt}
\tabletypesize{\scriptsize}
\tablecaption{Updated Classifications of SNe in the LOSS Volume-Limited Sample \label{tab:all}}
\tablehead{
  \colhead{Name} & \colhead{Previous (L11)} &
  \colhead{This Work} & 
  \colhead{Ref.} &
\colhead{Name} & \colhead{Previous (L11)} &
  \colhead{This Work} & 
  \colhead{Ref.}
}
\startdata
SN 1998de  &  Ia-91bg  &  Ia-91bg  &  1,2,2  & SN 1998dh  &  Ia-norm  &  Ia-norm  &  2,3 \\ 
SN 1998dk  &  Ia-norm  &  Ia-norm  &  2,3  & SN 1998dm  &  Ia-norm  &  Ia-norm  &  2,3 \\ 
SN 1998dt  &  Ib  &  Ib  &  4  & SN 1998ef  &  Ia-norm  &  Ia-norm  &  3 \\ 
SN 1998es\tablenotemark{1}  &  Ia-91T & Ia-99aa &  2,3  & SN 1999D  &  IIP  &  II  &  5 \\ 
SN 1999aa\tablenotemark{1}  &  Ia-91T & Ia-99aa &  2,3,6  & SN 1999ac\tablenotemark{1}  &  Ia-91T & Ia-99aa/Ia-norm &  2,3,7 \\ 
SN 1999an  &  IIP  &  II  &  -  & SN 1999bg  &  IIP  &  II  &  5 \\ 
SN 1999bh\tablenotemark{1}  &  Ia-02cx & Ia-02es &  8,9  & SN 1999br  &  IIP  &  II  &  - \\ 
SN 1999bu  &  Ic  &  Ic  &  -  & SN 1999bw  &  {\it impostor}  &  {\it impostor}  &  10 \\ 
SN 1999by  &  Ia-91bg  &  Ia-91bg  &  3  & SN 1999cd  &  IIb  &  IIb  &  - \\ 
SN 1999cl  &  Ia-norm  &  Ia-norm  &  2,3  & SN 1999cp  &  Ia-norm  &  Ia-norm  &  3 \\ 
SN 1999da  &  Ia-91bg  &  Ia-91bg  &  3  & SN 1999dk  &  Ia-norm  &  Ia-norm  &  3 \\ 
SN 1999dn  &  Ib  &  Ib  &  4  & SN 1999dq\tablenotemark{1}  &  Ia-91T & Ia-99aa &  2,3 \\ 
SN 1999ej  &  Ia-norm  &  Ia-norm  &  2  & SN 1999ek  &  Ia-norm  &  Ia-norm  &  3 \\ 
SN 1999el  &  IIn  &  IIn  &  -  & SN 1999em  &  IIP  &  II  &  5,11 \\ 
SN 1999gd  &  Ia-norm  &  Ia-norm  &  2,3  & SN 1999gi  &  IIP  &  II  &  12 \\ 
SN 1999go  &  IIL  &  II  &  13  & SN 2000C  &  Ic  &  Ic  &  - \\ 
SN 2000H  &  IIb  &  IIb  &  14  & SN 2000L  &  IIP  &  II  &  - \\ 
{\bf SN 2000N}  &  {\bf IIb/IIL}  &  {\bf II}  &  -  & SN 2000cb\tablenotemark{2}  &  IIP & II-87A &  15 \\ 
SN 2000ch  &  {\it impostor}  &  {\it impostor}  &  10,16  & SN 2000dc  &  IIL  &  II  &  17 \\ 
SN 2000dm  &  Ia-norm  &  Ia-norm  &  3  & SN 2000dr  &  Ia-norm  &  Ia-norm  &  3 \\ 
SN 2000el  &  IIP  &  II  &  -  & SN 2000eo  &  IIn  &  IIn  &  - \\ 
SN 2000ex  &  IIP  &  II  &  -  & SN 2001E  &  Ia-norm  &  Ia-norm  &  3 \\ 
{\bf SN 2001J}  &  {\bf IIP}  &  {\bf II/IIb}  &  -  & SN 2001K  &  IIP  &  II  &  - \\ 
SN 2001L  &  Ia-norm  &  Ia-norm  &  3  & {\bf SN 2001M}  &  {\bf Ic}  &  {\bf Ib}  &  - \\ 
SN 2001Q  &  IIb  &  IIb  &  -  & SN 2001V\tablenotemark{1}  &  Ia-91T & Ia-99aa &  2,3 \\ 
SN 2001ac  &  {\it impostor}  &  {\it impostor}  &  10  & SN 2001bq  &  IIP/IIL  &  II  &  5 \\ 
{\bf SN 2001ci}  &  {\bf Ic}  &  {\bf Ib}  &  -  & SN 2001cm  &  IIP  &  II  &  5,18 \\ 
SN 2001dc  &  IIP  &  II  &  19  & SN 2001dn  &  Ia-norm  &  Ia-norm  &  3 \\ 
SN 2001do  &  IIL  &  II  &  17  & SN 2001en  &  Ia-norm  &  Ia-norm  &  3 \\ 
SN 2001ep  &  Ia-norm  &  Ia-norm  &  3,20  & SN 2001fh  &  Ia-norm  &  Ia-norm  &  3 \\ 
SN 2001fz  &  IIP  &  II  &  -  & SN 2001hf  &  IIL  &  II  &  - \\ 
SN 2001is  &  Ib  &  Ib  &  -  & SN 2002J  &  Ic  &  Ic  &  - \\ 
SN 2002an  &  IIL  &  II  &  5  & SN 2002ap  &  Ic-pec  &  Ic-BL  &  21 \\ 
SN 2002bo  &  Ia-norm  &  Ia-norm  &  3  & SN 2002bu\tablenotemark{3}  &  IIn & {\it impostor} &  10 \\ 
SN 2002bx  &  IIP  &  II  &  5  & SN 2002ca  &  IIP  &  II  &  5 \\ 
SN 2002ce  &  IIP  &  II  &  -  & SN 2002cf  &  Ia-91bg  &  Ia-91bg  &  3 \\ 
SN 2002cr  &  Ia-norm  &  Ia-norm  &  3  & SN 2002dj  &  Ia-norm  &  Ia-norm  &  3 \\ 
SN 2002dk  &  Ia-91bg  &  Ia-91bg  &  3  & SN 2002do  &  Ia-norm  &  Ia-norm  &  3 \\ 
SN 2002dq  &  IIP  &  II  &  -  & SN 2002ds\tablenotemark{4}  &  IIP & II &  - \\ 
SN 2002er  &  Ia-norm  &  Ia-norm  &  3  & SN 2002es\tablenotemark{1}  &  Ia-02cx & Ia-02es &  22 \\ 
SN 2002fb  &  Ia-91bg  &  Ia-91bg  &  3  & SN 2002fk  &  Ia-norm  &  Ia-norm  &  3 \\ 
SN 2002gd  &  IIP  &  II  &  5  & SN 2002gw  &  IIP  &  II  &  - \\ 
SN 2002ha  &  Ia-norm  &  Ia-norm  &  3  & SN 2002hh  &  IIP  &  II  &  5 \\ 
SN 2002hw  &  Ia-norm  &  Ia-norm  &  3  & SN 2002jg  &  Ia-norm  &  Ia-norm  &  3 \\ 
{\bf SN 2002jj}  &  {\bf Ic}  &  {\bf Ic/Ic-BL}  &  -  & SN 2002jm  &  Ia-91bg  &  Ia-91bg  &  3 \\ 
{\bf SN 2002jz}  &  {\bf Ic}  &  {\bf Ib/Ic}  &  -  & SN 2002kg  &  {\it impostor}  &  {\it impostor}  &  23,24 \\ 
SN 2003E  &  IIP  &  II  &  -  & SN 2003F  &  Ia-norm  &  Ia-norm  &  3 \\ 
SN 2003G  &  IIn  &  IIn  &  -  & SN 2003H\tablenotemark{5}  &  Ibc-pec & Ca-rich &  25 \\ 
SN 2003Y  &  Ia-91bg  &  Ia-91bg  &  3  & SN 2003Z  &  IIP  &  II  &  5 \\ 
SN 2003aa  &  Ic  &  Ic  &  -  & SN 2003ao  &  IIP  &  II  &  - \\ 
SN 2003bk  &  IIP  &  II  &  -  & {\bf SN 2003br}  &  {\bf IIP}  &  {\bf II/IIb}  &  - \\ 
{\bf SN 2003bw}\tablenotemark{4}  &  {\bf IIP} & {\bf II/IIb} &  -  & SN 2003cg  &  Ia-norm  &  Ia-norm  &  26 \\ 
SN 2003dr\tablenotemark{5}  &  Ibc-pec & Ca-rich &  25  & SN 2003du  &  Ia-norm  &  Ia-norm  &  3,27 \\ 
SN 2003dv  &  IIn  &  IIn  &  28  & SN 2003ed  &  IIb  &  IIb  &  - \\ 
SN 2003ef  &  IIP  &  II  &  -  & SN 2003gm  &  {\it impostor}  &  {\it impostor}  &  23 \\ 
SN 2003gt  &  Ia-norm  &  Ia-norm  &  3  & SN 2003hg  &  IIP  &  II  &  13 \\ 
SN 2003hl  &  IIP  &  II  &  5  & {\bf SN 2003id}  &  {\bf Ic-pec}  &  {\bf Ic-pec}  &  - \\ 
SN 2003iq  &  IIP  &  II  &  5  & SN 2003kf  &  Ia-norm  &  Ia-norm  &  3 \\ 
SN 2003ld  &  IIP  &  II  &  -  & {\bf SN 2004C}  &  {\bf Ic}  &  {\bf IIb}  &  - \\ 
SN 2004ab  &  Ia-norm  &  Ia-norm  &  9  & {\bf SN 2004al}  &  {\bf IIb/IIL}  &  {\bf II}  &  - \\ 
SN 2004aq  &  IIP  &  II  &  -  & SN 2004bd  &  Ia-norm  &  Ia-norm  &  3 \\ 
SN 2004be  &  IIb  &  IIb  &  -  & SN 2004bl  &  Ia-norm  &  Ia-norm  &  3 \\ 
{\bf SN 2004bm}  &  {\bf Ibc-pec/IIb}  &  {\bf IIb/IIb-pec}  &  -  & SN 2004bv  &  Ia-91T  &  Ia-91T  &  3 \\ 
SN 2004ca  &  Ia-norm  &  Ia-norm  &  3  & {\bf SN 2004cc}  &  {\bf Ic}  &  {\bf Ib/Ic}  &  13 \\ 
SN 2004ci  &  IIP  &  II  &  -  & SN 2004dd  &  IIP  &  II  &  - \\ 
SN 2004dk  &  Ib  &  Ib  &  -  & SN 2004er  &  IIP  &  II  &  - \\ 
SN 2004et  &  IIP  &  II  &  5  & SN 2004fc  &  IIP  &  II  &  - \\ 
SN 2004fx  &  IIP  &  II  &  29  & SN 2004gq  &  Ib  &  Ib  &  14 \\ 
SN 2005A  &  Ia-norm  &  Ia-norm  &  3,30  & SN 2005E\tablenotemark{5}  &  Ibc-pec & Ca-rich &  25,31 \\ 
{\bf SN 2005H}  &  {\bf IIb}  &  {\bf II/IIb}  &  13  & SN 2005J  &  IIL  &  II  &  - \\ 
SN 2005U  &  IIb  &  IIb  &  14  & SN 2005W  &  Ia-norm  &  Ia-norm  &  3,30 \\ 
SN 2005ad  &  IIP  &  II  &  -  & SN 2005am  &  Ia-norm  &  Ia-norm  &  3,30 \\ 
SN 2005an  &  IIL  &  II  &  -  & SN 2005aq  &  IIn  &  IIn  &  - \\ 
SN 2005as  &  Ia-norm  &  Ia-norm  &  3  & SN 2005ay  &  IIP  &  II  &  5 \\ 
SN 2005az  &  Ic  &  Ic  &  14  & SN 2005bb  &  IIP  &  II  &  - \\ 
SN 2005bc  &  Ia-norm  &  Ia-norm  &  3  & SN 2005bo  &  Ia-norm  &  Ia-norm  &  3,30 \\ 
SN 2005cc  &  Ia-02cx  &  Ia-02cx  &  9  & SN 2005cf  &  Ia-norm  &  Ia-norm  &  32 \\ 
SN 2005ci\tablenotemark{2}  &  IIP & II-87A &  15  & SN 2005de  &  Ia-norm  &  Ia-norm  &  3 \\ 
SN 2005el  &  Ia-norm  &  Ia-norm  &  3,30  & SN 2005hk  &  Ia-02cx  &  Ia-02cx  &  3,33,34 \\ 
{\bf SN 2005io}\tablenotemark{2}  &  {\bf IIP} & {\bf II-87A} &  -  & SN 2005kc  &  Ia-norm  &  Ia-norm  &  3,30 \\ 
SN 2005ke  &  Ia-91bg  &  Ia-91bg  &  3,30  & {\bf SN 2005lr}  &  {\bf Ic}  &  {\bf IIb}  &  - \\ 
{\bf SN 2005mg}  &  {\bf IIP}  &  {\bf II/IIb}  &  -  & SN 2005mz  &  Ia-91bg  &  Ia-91bg  &  9 \\ 
SN 2006F  &  Ib  &  Ib  &  -  & SN 2006T  &  IIb  &  IIb  &  - \\ 
SN 2006X  &  Ia-norm  &  Ia-norm  &  3,30  & SN 2006ax  &  Ia-norm  &  Ia-norm  &  3,30 \\ 
SN 2006be  &  IIP  &  II  &  -  & SN 2006bp  &  IIP  &  II  &  - \\ 
SN 2006bv\tablenotemark{3,4}  &  IIn & {\it impostor} &  10  & SN 2006ca  &  IIP  &  II  &  - \\ 
SN 2006cm\tablenotemark{1}  &  Ia-91T & Ia-99aa/Ia-norm &  3  & SN 2006dy  &  Ia-norm  &  Ia-norm  &  3 \\ 
SN 2006ef  &  Ia-norm  &  Ia-norm  &  3,30  & {\bf SN 2006eg}  &  {\bf Ic}  &  {\bf IIb/Ib/Ic/Ic-BL}  &  - \\ 
SN 2006ke  &  Ia-91bg  &  Ia-91bg  &  3  & SN 2006le  &  Ia-norm  &  Ia-norm  &  3 \\ 
SN 2006lf  &  Ia-norm  &  Ia-norm  &  3  & SN 2006qr  &  IIP  &  II  &  - \\ 
\enddata

    \vspace{1mm}
    This table lists the previous classifications for all objects in L11
    and our confirmed or updated classifications. Notable objects, discussed 
    individually within the text, are printed in boldface.
    We also list references to the original publishers for all data already in the literature
    that were used in this effort.
    When we cannot confim a single clear classification, we list more than one
    possible type or subtype.
    See \S\ref{sec:methods} for a detailed
    description of each type and subtype label. \\
    {\bf References:}
  [1] \citet{2001PASP..113..308M}; 
  [2] \citet{2008AJ....135.1598M}; 
  [3] \citet{2012MNRAS.425.1789S}; 
  [4] \citet{2001AJ....121.1648M}; 
  [5] \citet{2014MNRAS.442..844F}; 
  [6] \citet{2004AJ....128..387G}; 
  [7] \citet{2005AJ....130.2278G}; 
  [8] \citet{2001ApJ...546..734L}; 
  [9] \citet{2012AJ....143..126B}; 
  [10] \citet{2011MNRAS.415..773S}; 
  [11] \citet{2002PASP..114...35L}; 
  [12] \citet{2002AJ....124.2490L}; 
  [13] \citet{2008AA...488..383H}; 
  [14] \citet{2014AJ....147...99M}; 
  [15] \citet{2011MNRAS.415..372K}; 
  [16] \citet{2004PASP..116..326W}; 
  [17] \citet{2014MNRAS.445..554F}; 
  [18] \citet{2009ApJ...694.1067P}; 
  [19] \citet{2004MNRAS.347...74P}; 
  [20] \citet{2008MNRAS.391.1605S}; 
  [21] \citet{2003PASP..115.1220F}; 
  [22] \citet{2012ApJ...751..142G}; 
  [23] \citet{2006MNRAS.369..390M}; 
  [24] \citet{2006astro.ph..3025V}; 
  [25] \citet{2010Natur.465..322P}; 
  [26] \citet{2006MNRAS.369.1880E}; 
  [27] \citet{2005ApJ...632..450L}; 
  [28] \citet{2015MNRAS.450..246B}; 
  [29] \citet{2006PASP..118....2H}; 
  [30] \citet{2013ApJ...773...53F}; 
  [31] \citet{2009AJ....138..376F}; 
  [32] \citet{2009ApJ...697..380W}; 
  [33] \citet{2006PASP..118..722C}; 
  [34] \citet{2007PASP..119..360P} 
  \tablenotetext{1}{See \S\ref{sec:ia}}
  \tablenotetext{2}{See \S\ref{sec:87A}}
  \tablenotetext{3}{See \S\ref{sec:2ni}}
  \tablenotetext{4}{See \S\ref{sec:nospec}}
  \tablenotetext{5}{See \S\ref{sec:carich}}
\end{deluxetable*}

\subsection{Reclassified Objects}
\label{sec:reclassified}

\subsubsection{SN 2000N (IIb/IIL\,$\rightarrow$\,II)}

SN~2000N was discovered in MCG-02-34-054 \citep{2000IAUC.7374....1S} and
classified as a Type II SN from a spectrum with a low S/N
\citep{2000IAUC.7377....2J}. However, because the data they had on this object were quite sparse,
L11 could not determine if SN~2000N was  a Type IIb or a Type II SN.
We were able to obtain the spectrum originally used to classify the SN 
as well as additional spectra, including one near peak brightness
when the spectra of SNe IIb and II are more clearly differentiable; see Figure~\ref{fig:2000n}.
SNID identifies SN~2000N as a Type II SN, and our rereduced light curve indicates
that the SN~IIL template from L11 is a better match than the SN~IIb one.

\begin{figure}[ht]
\epsscale{1.0}
\plotone{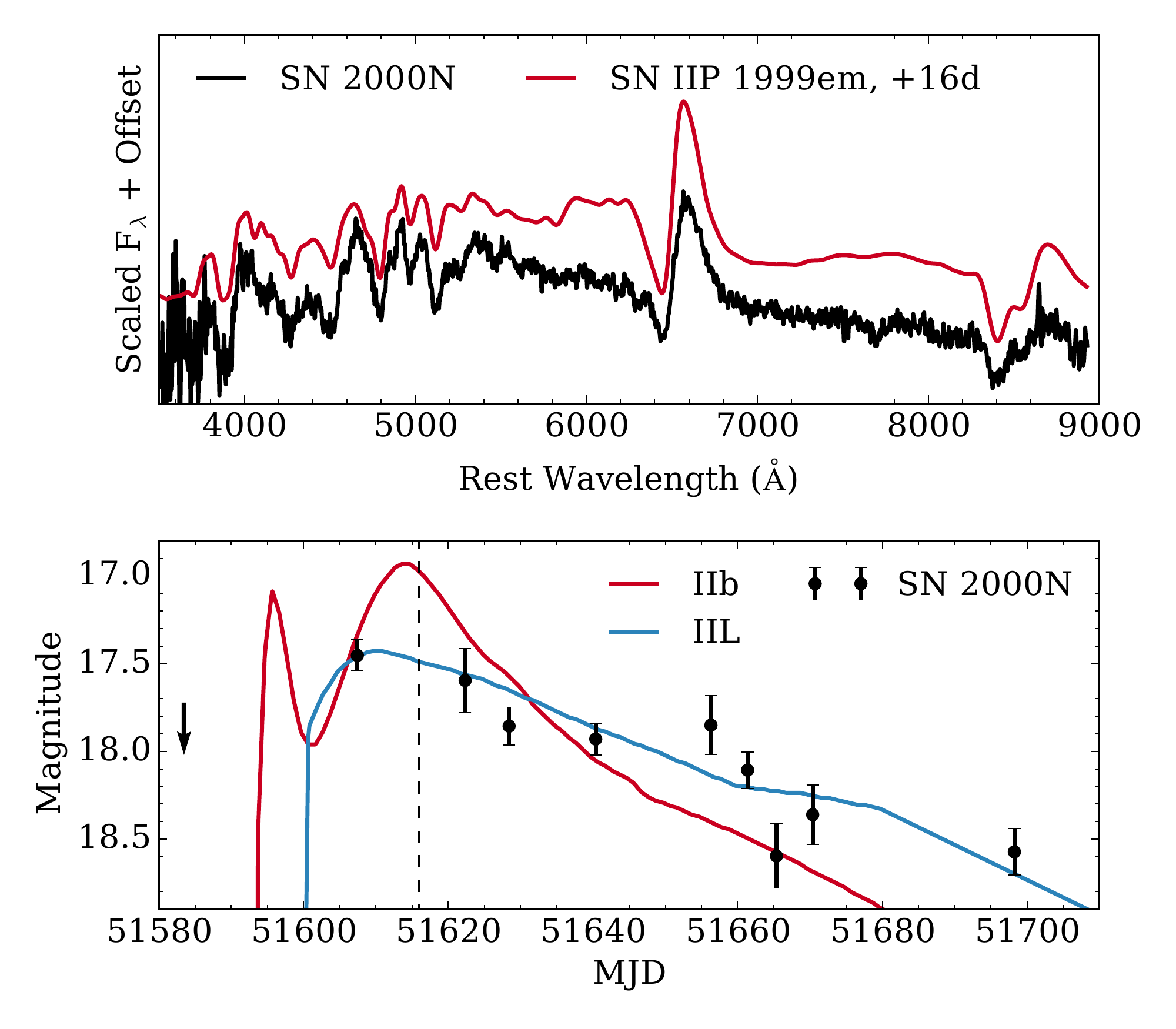}
\caption{Top: the near-maximum-light
    spectrum of SN~2000N alongside that of the Type IIP SN~1999em \citep{2002PASP..114...35L}.
    Bottom: the light curve of SN~2000N, with a vertical line
    showing the date the spectrum shown was observed, and colored lines showing the
    template light curves from L11.
    \label{fig:sn2000n}}
\end{figure}

\subsubsection{SN 2001M (Ic\,$\rightarrow$\,Ib) }

\begin{figure}[ht]
\epsscale{1.4} 
\hspace*{-8mm}
\plotone{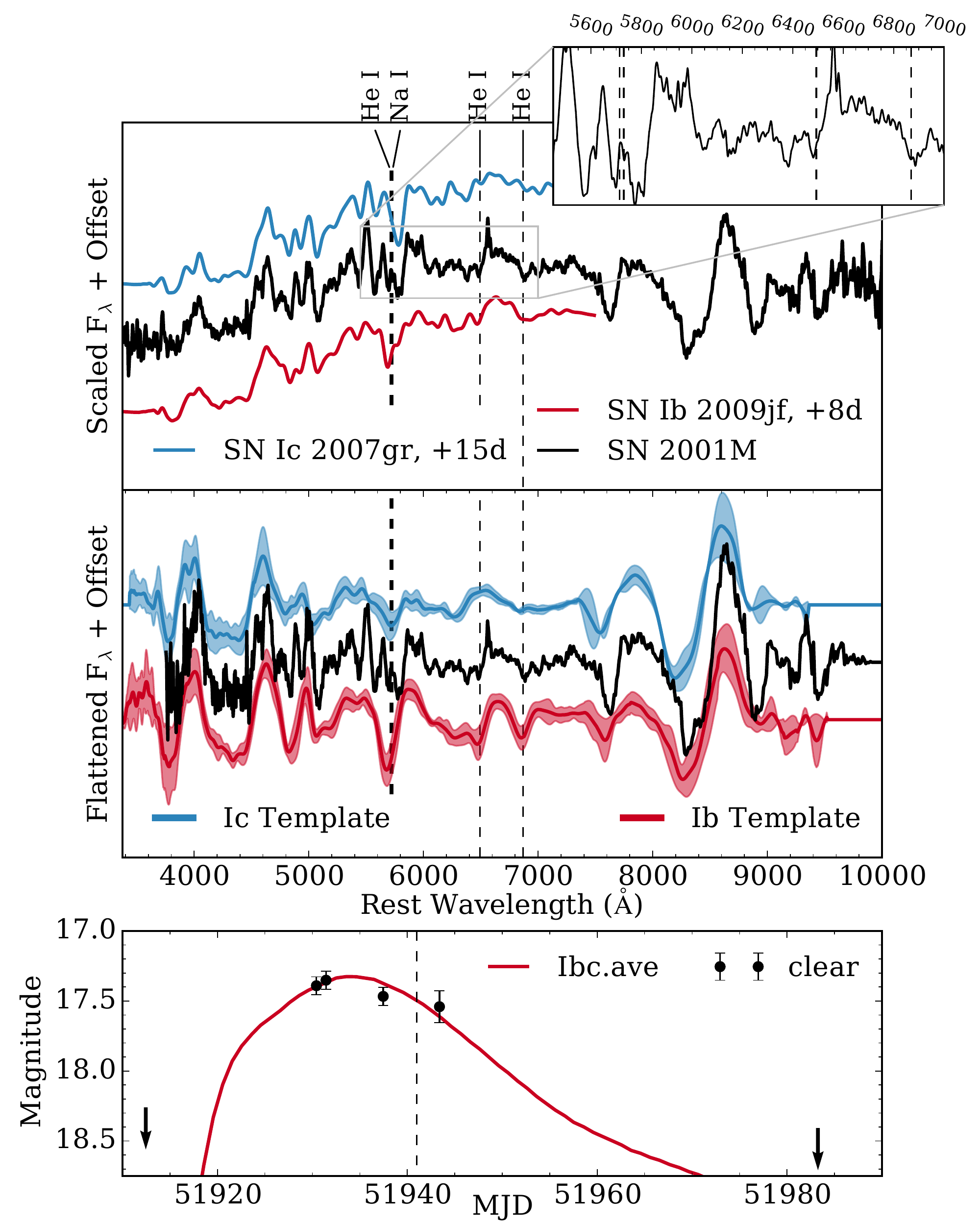}
\caption{ Top: spectrum of SN~2001M smoothed with a 20\,\AA~Gaussian kernel, and
    with weak helium features identified.
    We also show spectra of SN Ic 2007gr and SN Ib 2009jf \citep{2014AJ....147...99M},
    and the wavelengths of \ion{He}{1} and \ion{Na}{1} features are marked 
    at $v = 8300$\,\kms.
    The inset shows the \ion{He}{1} lines we identify in the SN~2001M.
    Middle: the continuum-normalized spectrum of SN~2001M compared to
    the $+$10\,d average spectra of Ib and Ic SNe from \citet{2016ApJ...827...90L},
    with the standard deviation shown as shaded regions.
    Bottom: light curve of SN~2001M compared to the average Ib/c template
    light curve from L11, with the date of spectrum shown with a dashed line. 
    \label{fig:sn2001m}}
\end{figure}

SN~2001M was discovered in NGC 3240 and classified as a SN~Ic \citep{2001IAUC.7568....1A,2001IAUC.7576....5S,2001IAUC.7579....2F}.
SNID identifies reasonably good cross-correlations with spectra of both SNe~Ib and
SNe~Ic. We have been able to locate only a single spectrum of this object and our light curve
is sparse, so the phase of our spectrum is somewhat uncertain; see Figure~\ref{fig:2001m}.
Narrow, unresolved H$\alpha$ emission from the star-forming host galaxy partially obscures
the \ion{He}{1}\,$\lambda$6678 absorption line, but we do identify probable weak 
\ion{He}{1}\,$\lambda$6678 and \ion{He}{1}\,$\lambda$7065
lines at $v \approx 8500$\,km\,s$^{-1}$.  The region around \ion{He}{1}\,$\lambda$5876
is complex, with what must be several overlapping absorption lines; we believe it to
be consistent with \ion{He}{1}\,$\lambda$5876 absorption but not dominated by it.

We compare our spectrum to the $+$10\,day average SN~Ib and SN~Ic spectra of \citet{2016ApJ...827...90L},
after estimating a date of peak of 24 January 2001 from the light curve
(implying a phase of $+$7.5\,d for our spectrum).
This comparison shows many similarities between SN~2001M and both classes, but 
reinforces our identifications of the \ion{He}{1} lines.
Given our detections of both \ion{He}{1}\,$\lambda$6678 and \ion{He}{1}\,$\lambda$7065,
we relabel SN~2001M as a Type Ib SN, though we note that the 
helium lines are weak compared to those in most SNe~Ib.

\subsubsection{SN 2001ci (Ic\,$\rightarrow$\,Ib) }
\label{sec:sn2001ci}

SN~2001ci was discovered in NGC~3079 \citep{2001IAUC.7618....1S} and announced
as a SN Ic heavily obscured by host-galaxy dust absorption \citep{2001IAUC.7638....1F}.
A re-examination of the spectrum cited therein
(observed UT 2001-05-30; we were not able to locate any other spectra of this object)
confirms that it is heavily reddened by host-galaxy dust.
The MW contribution to the reddening is only $E(B-V) = 0.0097$\,mag
\citep[][used for all subsequent MW reddening measures]{2011ApJ...737..103S},
but the spectrum clearly shows narrow \ion{Na}{1} absorption features from the host galaxy
indicating significant reddening produced by host-galaxy dust.
The spectrum is of sufficiently high resolution to resolve both components of 
the \ion{Na}{1}~D doublet, but their equivalent widths (EWs) are well outside the range of the
empirical relations of \citet{2012MNRAS.426.1465P}: $EW_{D1} \approx 2.4$\,\AA\ and 
$EW_{D2} \approx 2.2$\,\AA.  This implies that $E(B-V) \gtrsim 3.0$\,mag.

\begin{figure}[ht]
\epsscale{1.0}
\plotone{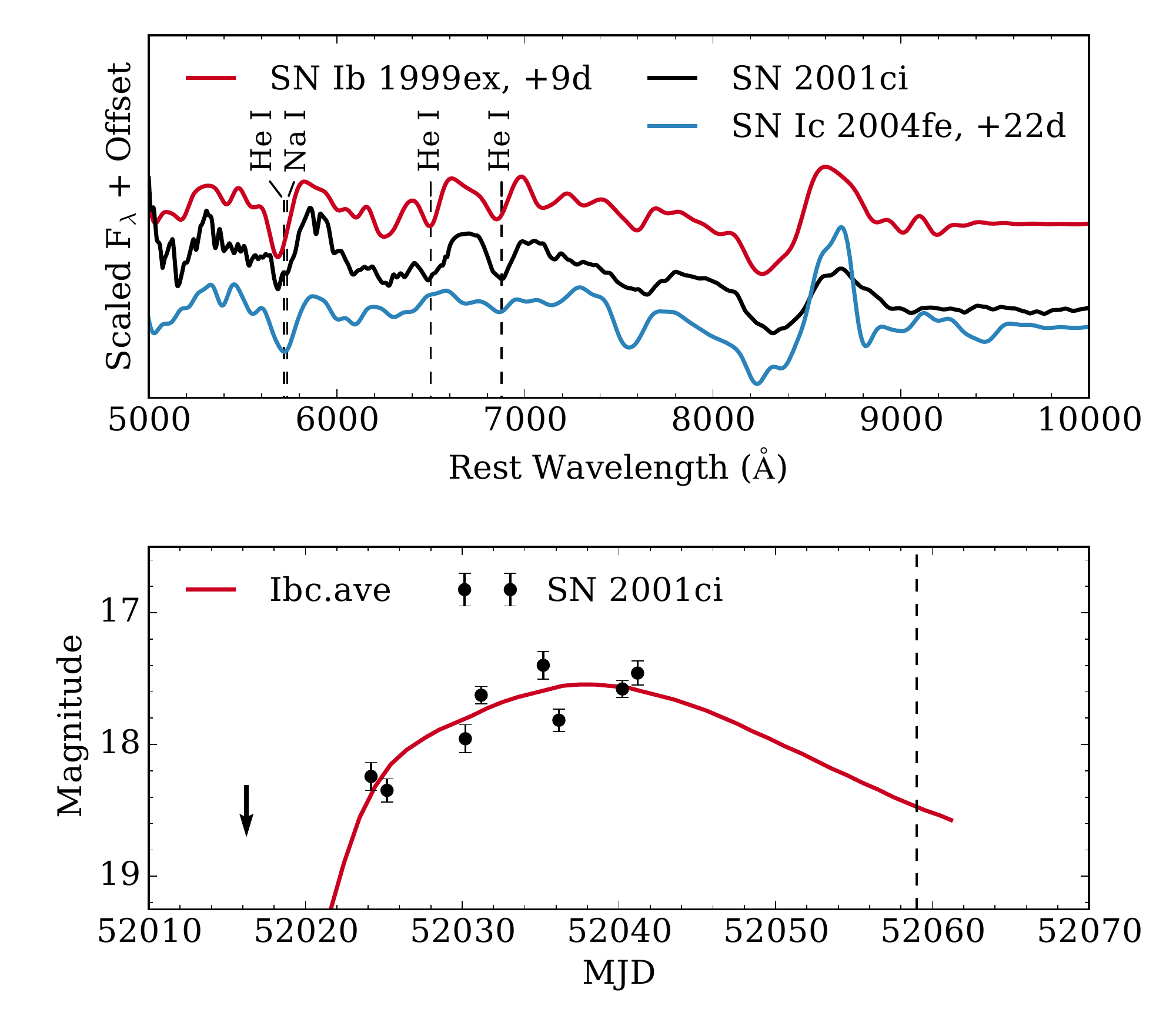}
\caption{Top: spectrum of SN~2001ci (corrected for a reddening of $E(B-V) = 3.0$\,mag 
    and smoothed with a Gaussian kernel 50\AA~wide), compared with that of the Type Ib SN~1999ex \citep{2002AJ....124..417H}
    and the Type Ic SN~2004fe \citep{2014AJ....147...99M}.
    Bottom: the light curve compared to a L11 template with the date of the spectrum marked by a vertical line.
    \label{fig:sn2001ci}}
\end{figure}

Note that SNID, by construction, is insensitive to color information and to uncertainties in
the reddening corrections or flux calibrations \citep{2007ApJ...666.1024B}. Throughout
this paper we apply (often uncertain) reddening corrections to spectra to facilitate visual comparisons,
but they do not appreciably affect the SNID classifications.

Adopting $E(B-V) = 2.0$\,mag and a MW-like dust law is good enough to achieve a robust identification:
SN~2001ci is a Type Ib SN, with a spectrum that is most similar to those of SNe Ib with weak helium 
lines.  See Figure~\ref{fig:sn2001ci} for a comparison to the He-weak SN Ib 1999ex \citep{2002AJ....124..417H}.
\citet{2001IAUC.7638....1F} drew attention to a similarity with SN~1990U, which was (at the time) classified
as a SN~Ic, but \citet{2014AJ....147...99M} have subsequently reclassified SN~1990U as a SN~Ib.

\subsubsection{SN 2004C (Ic\,$\rightarrow$\,IIb)}

\begin{figure}[ht]
\epsscale{1.0}
\plotone{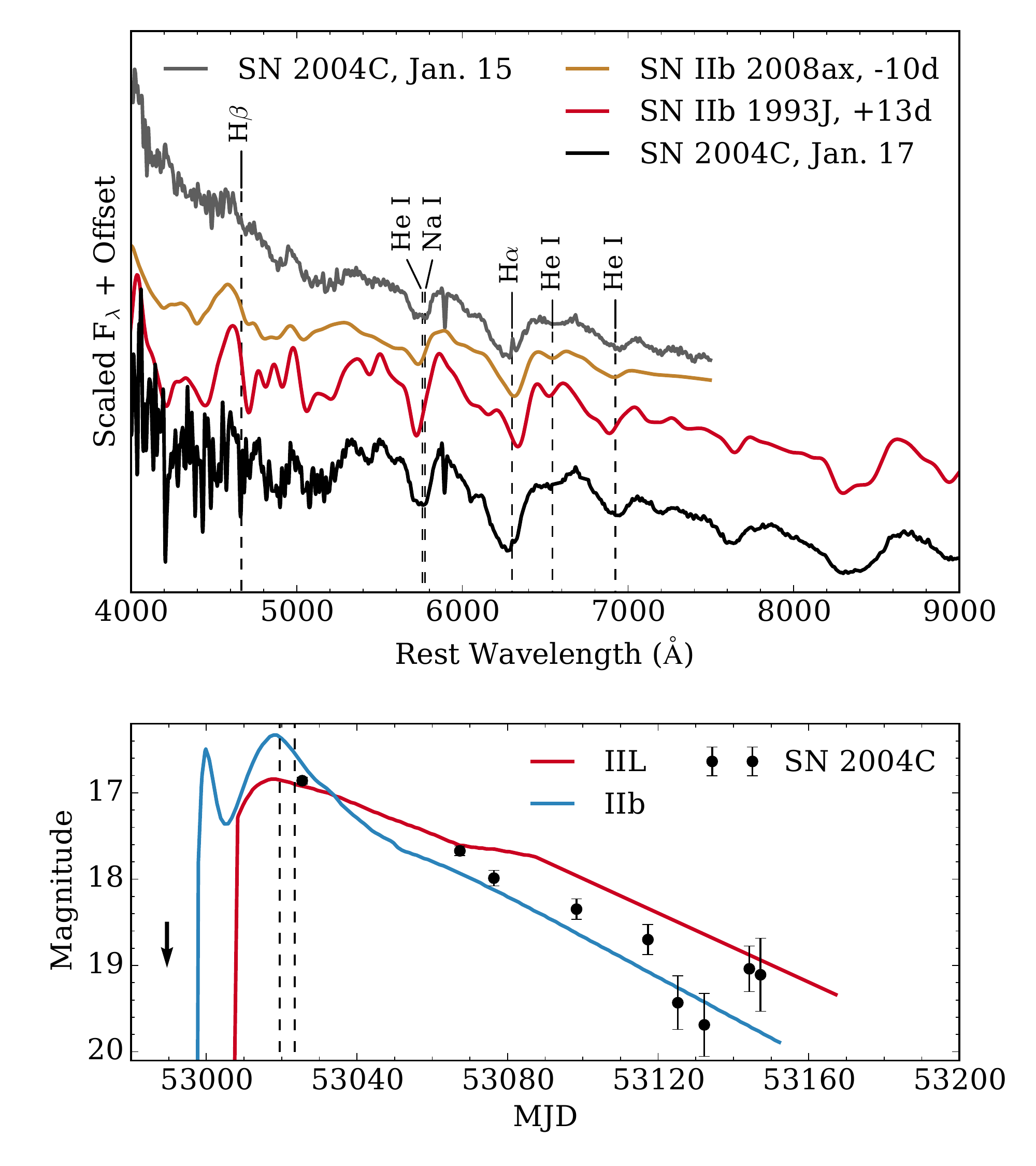}
\caption{Top: spectra of SN~2004C observed on UT~2004-01-15 and on UT~2004-01-17,
    alongside a premaximum spectrum of the Type IIb SN~2008ax \citep{2014AJ....147...99M} 
    and a post-maximum spectrum of the Type IIb SN~1993J \citep{2000AJ....120.1487M}.
    Both spectra of SN~2004C have been dereddened
    by $E(B-V) = 1.0$\,mag and smoothed by a 20\,\AA~Gaussian kernel, 
    and galaxy emission lines have been subtracted by hand.
    Bottom: light curve of SN~2004C compared to templates from L11 with the date of the spectra marked by vertical lines.
    \label{fig:sn2004c}}
\end{figure}

SN~2004C was discovered in NGC~3683 \citep{2004CBET...57....1D} 
and classified as a heavily reddened SN~Ic \citep{2004CBET...57....2M}, spectrally
similar to SN~1990U (see \S\ref{sec:sn2001ci}: SN~1990U was a SN~Ib).
The MW reddening along the line of sight is only  $E(B-V) = 0.0133$\,mag,
but (unresolved) \ion{Na}{1}~D absorption in our spectra indicates strong host-galaxy obscuration.
As with SN~2001ci, the \ion{Na}{1}~D EW is well beyond the empirical relations of
\citet{2012MNRAS.426.1465P}, but the spectra do not appear to be as reddened as
those of SN~2001ci.  Correcting for a total reddening of $E(B-V) = 1.0$\,mag produces
a reasonable result.

Here we publish several spectra of SN~2004C which indicate that SN~2004C
was a Type IIb SN. Figure~\ref{fig:sn2004c} shows a comparison between the spectrum announced
by \citet{2004CBET...57....2M} and the Type IIb SN~2008ax, as well as a later spectrum with broader
wavelength coverage compared to a spectrum of the Type IIb SN~1993J.

\subsubsection{SN 2004al (IIb/IIL\,$\rightarrow$\,II)}

\begin{figure}[ht]
\epsscale{1.0}
\plotone{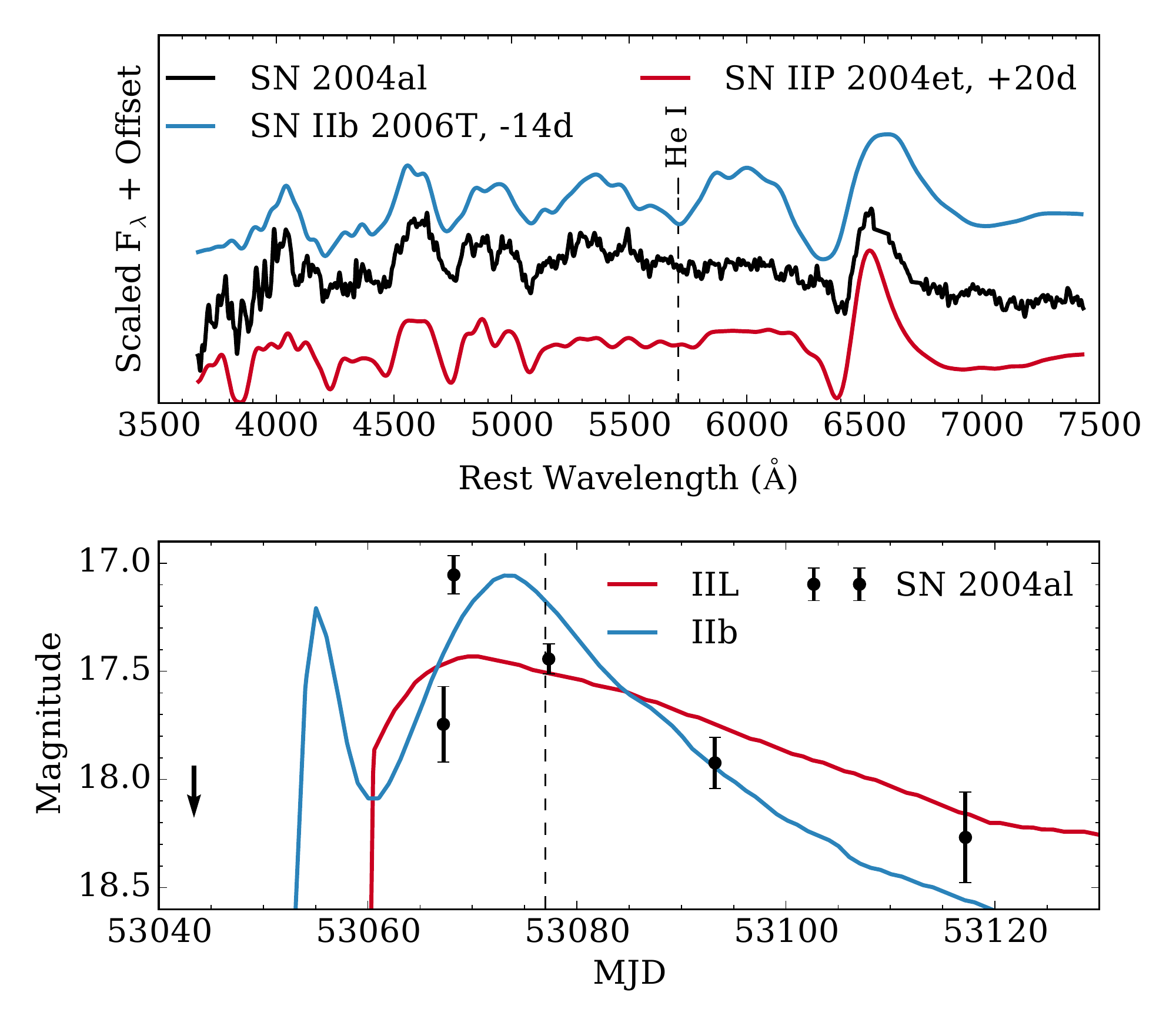}
\caption{Top: spectrum of SN~2004al smoothed with
    a 20\,\AA~Gaussian kernel and compared to spectra of the Type IIP SN~2004et a few
    weeks after maximum \citep{2006MNRAS.372.1315S} and the premaximum SN IIb 2006T
    \citep[the top non-IIP SNID template,][]{2014AJ....147...99M}. 
    Bottom: the light curve compared to templates from L11 with the date of the spectrum marked by a vertical line.
    \label{fig:sn2004al}}
\end{figure}

SN~2004al was discovered in ESO 565-G25 \citep{2004IAUC.8297....2S}
and classified as a Type II SN \citep{2004IAUC.8303....1M}.
As L11 show, SN~2004al had a light curve consistent with either a SN~IIb or a SN~IIL classification, and so
they assign it a 50\%~weight in each class.
Although we have no additional spectra to examine, we were able to obtain the classification spectrum and 
SNID clearly identifies it as a Type II SN, not a SN~IIb, mostly owing to the absence of \ion{He}{1} $\lambda$5876.
Figure~\ref{fig:sn2004al} shows that our spectrum was observed $\sim$1 week post-maximum; 
though \ion{He}{1} lines are often weak in young SNe IIb, they become pronounced by maximum light.
In good agreement with L11, our rereduced light curve exhibits a relatively rapid decline
and is better fit by the SN IIb template; see Figure~\ref{fig:sn2004al}.
Despite this tension, we consider the spectroscopic classification robust and we label SN~2004al a Type II event.

\subsubsection{SN 2005io (IIP\,$\rightarrow$\,II-87A)}
\label{sec:sn2005io}

\begin{figure}[ht]
\epsscale{1.0}
\plotone{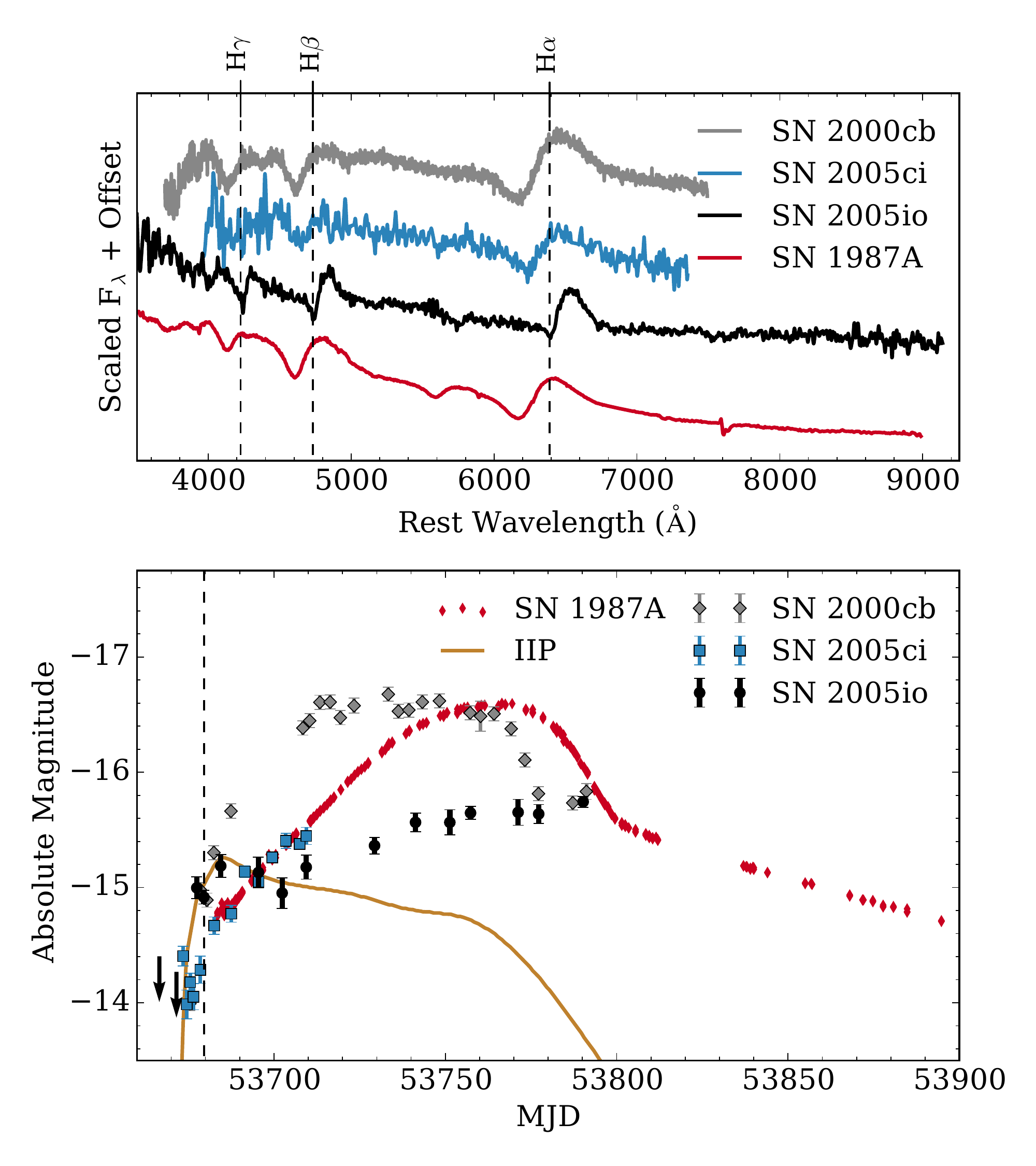}
\caption{Top: spectrum of SN~2005io compared to spectra of the
    young SNe~2000cb, 2005ci, and 1987A
    \citep[SNe 2000cb and 2005ci are also included in the LOSS volume-limited sample;][]{1987ApJ...320..589B,2011MNRAS.415..372K}.
    The spectra of SNe~2005io and 2005ci have been smoothed by a Gaussian kernel 20\,\AA~wide.
    We mark prominent hydrogen features at 8000\,\kms.
    Bottom: light curve of SN~2005io (with a vertical line showing the date the spectrum was observed)
    compared to the light curves of the other SN 1987A-like events after correcting for  
    distance and MW dust absorption along the line of sight, and then offset in time to align their initial rises 
    (\citealt{1990AJ.....99.1146H}; L11). 
    Also shown is the template Type IIP light curve from L11, offset to match the early evolution 
    of SN~2005io.
    \label{fig:sn2005io}}
\end{figure}

SN~2005io was discovered in UGC~3361 \citep{2005IAUC.8628....1L} and
reported as a young Type II SN \citep{2005CBET..274....1F} based upon a Keck
LRIS spectrum. The photometry of SN~2005io shows a peculiar
evolution, however --- see Figure~\ref{fig:sn2005io}.
The light curve initially follows the Type IIP template almost exactly, but after peaking at
an absolute magnitude of $-15.2$\,mag 
SN~2005io goes into a slight decline and then a very slow rise (over $\sim$100 days)
to a second maximum ($<-15.75$\,mag; the peak itself went unobserved).
This is similar to the photometric behavior of 
SN~1987A and other related events, though in SN~1987A the early peak was only
visible in bluer passbands and the rise to its second, more luminous, peak occurred more rapidly.
The spectrum is similar to that of a normal young Type II SN with hydrogen P-Cygni profiles dominated
by the emission component and with an absorption Doppler velocity of $\sim$8,000\,\kms,  slower
than the velocity found for hydrogen in SN~1987A.

Based primarily upon the light-curve evolution, and noting that the subclass is 
heterogeneous \citep{2005MNRAS.360..950P,2011MNRAS.415..372K,2012AA...537A.140T,2012AA...537A.141P,2016AA...588A...5T},
we classify SN~2005io as a SN~1987A-like event
alongside the two other SN 1987A-like events already identified within the L11 sample \citep[SNe 2000cb and 2005ci;][]{2011MNRAS.415..372K}.
It unfortunately appears that the peculiarity of SN~2005io was not recognized while it was bright:
we believe that Figure~\ref{fig:sn2005io}, which shows the classification
spectrum and the KAIT photometry, includes all extant observations of the event.
Though the central wavelength of unfiltered KAIT photometry is quite similar to that of
the $R$ band, the effective passband is significantly wider \citep{2010ApJS..190..418G}.
Our early-time spectrum of SN~2005io shows that it was quite blue while young, and it's likely
that the rapid rise to the first photometric peak was powered by cooling envelope 
emission, as was observed in SN~1987A via photometry in bluer passbands.

The second peak of SN~2005io lasts longer than that of SN~1987A, and
the fade from peak was not observed.  We have upper limits showing that the object
had faded to $\gtrsim 19.7$\,mag (absolute mag $-14.2$) by Sep.~18, but we have
no data obtained between the last detection on Feb.~24 (shown in Figure~\ref{fig:sn2005io}) and then.
SN~2005io shares some similarities with SN~2009E \citep{2012AA...537A.141P},
including a slow rise and faint peak compared to the prototypical SN~1987A.
The early peak and slow rise to second maximum is also reminiscent of
SN~2004ek \citep{2016AA...588A...5T}, although that SN was a great deal 
more luminous ($R \approx -18.5$\,mag) than SN~2005io ($clear \lesssim -15.75$\,mag),
further illustrating the heterogenous nature of these slow-rising SNe~II.

\subsubsection{SN 2005lr (Ic\,$\rightarrow$\,IIb)}

SN~2005lr was discovered in ESO~492-G2 \citep{2005IAUC.8641....1B} and was spectroscopically classified
as a SN~Ic, with spectra similar to those of SN~1990B \citep{2005CBET..321....1H}.
We obtained a copy of the classification spectrum for SN~2005lr from the CSP archives 
(observed 2005 Dec.\ 18) as well as a second (higher S/N) spectrum taken two days later.

\begin{figure}[ht]
\epsscale{1.0}
\plotone{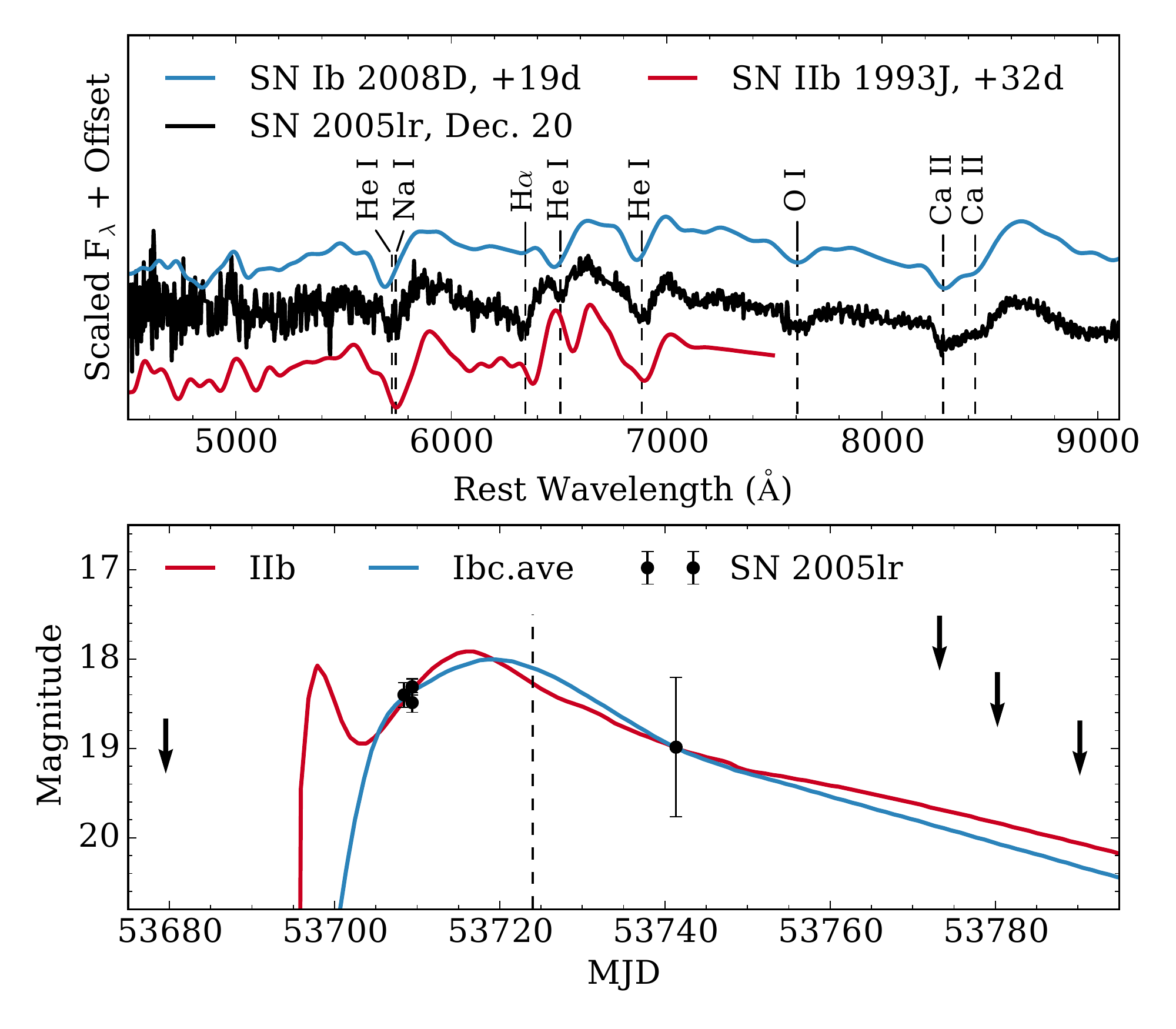}
\caption{Top: the higher S/N spectrum of SN~2005lr smoothed with a Gaussian
    kernel 10\,\AA\ wide and compared to spectra
    of the SN~Ib 2008D and SN~IIb 1993J \citep{2000AJ....120.1487M,2009ApJ...702..226M}.
    Bottom: light curve of the SN compared
    to templates from L11 with date of spectrum shown by a vertical line.
    \label{fig:sn2005lr}}
\end{figure}

The spectra of SN~2005lr show that 
this SN was strongly reddened by host-galaxy dust --- the \ion{Na}{1}\,D line in these low-resolution
spectra exhibits a total EW of $\sim$2.4\,\AA, just beyond the limits of the relations from
\citet{2012MNRAS.426.1465P}.  We again deredden the spectrum by an
arbitrary value of $E(B-V) = 1.0$\,mag to facilitate visual inspection.

SNID identifies SN~2005lr as either a SN~Ib or IIb, with matches to examples of either class.
Distinct \ion{He}{1}, \ion{O}{1}, and \ion{Ca}{2} lines are detected
alongside an H$\alpha$ absorption feature; 
see Figure~\ref{fig:sn2005lr}.  Our spectra appear to be taken well after peak brightness
(although the light curve is extremely sparse) and 
the detection of a relatively strong H$\alpha$ feature at this phase identifies SN~2005lr as a Type IIb SN.

\subsection{Low-Certainty Classifications and Peculiar Events}
\label{sec:lowcert}

We have made an effort to track down spectra of every object in the
volume-limited sample, especially those
spectra originally used for classification in the CBETs, and re-examine the 
classifications of each.  Unfortunately, however, there are several objects in our
sample for which robust classifications are simply
not possible given the peculiarity of the object or the quality of the data.

\subsubsection{SN 2001J (IIP\,$\rightarrow$\,II/IIb)}

\begin{figure}[ht]
\epsscale{1.0}
\plotone{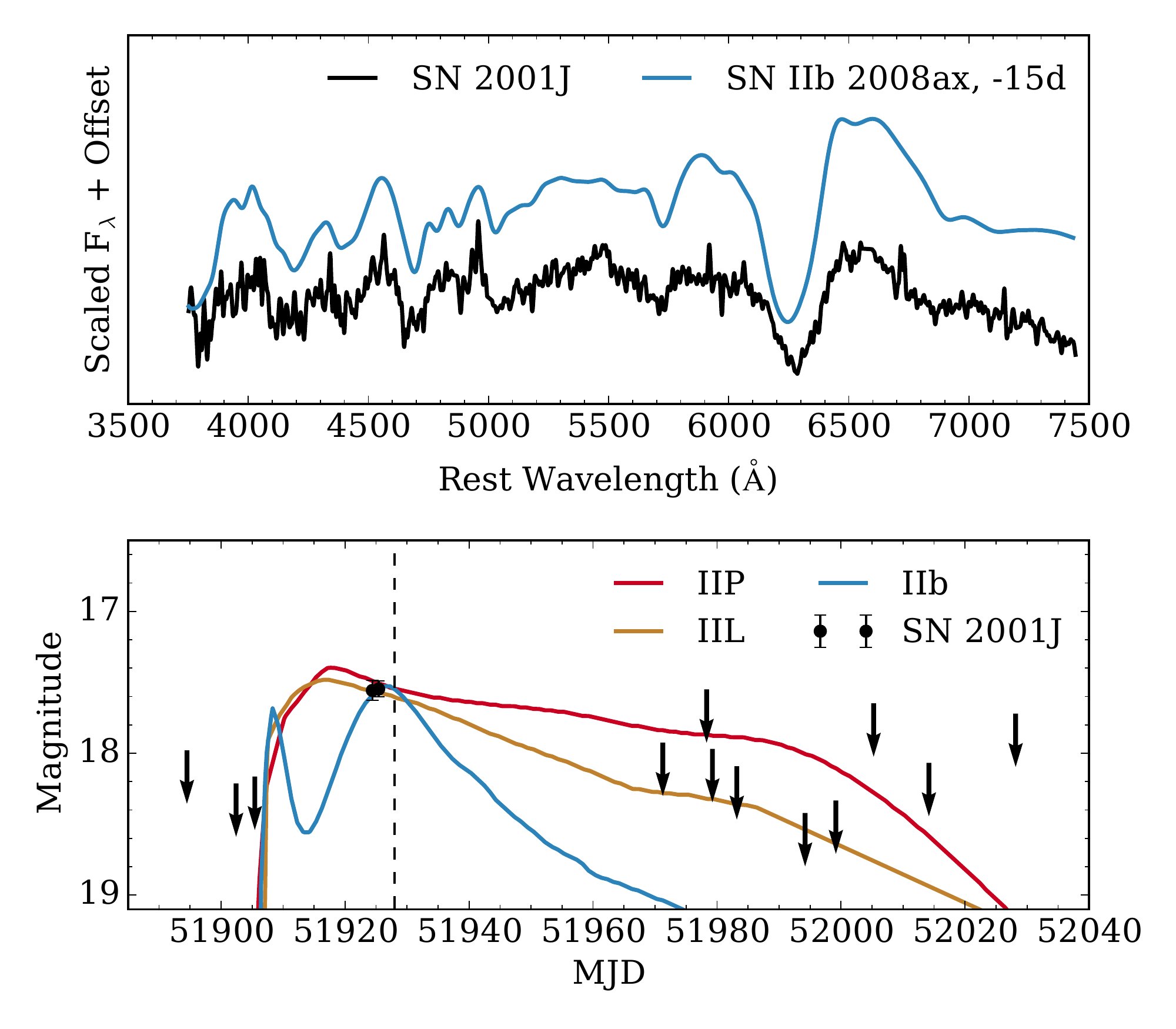}
\caption{Top: the spectrum of SN~2001J (smoothed by a 20\,\AA~Gaussian
    kernel) compared to that of the Type IIb SN~2008ax \citep{2014AJ....147...99M}.
    Bottom: the light curve compared to the Type IIL, IIP, and IIb templates
    from L11.  Upper limits from nondetections are shown as arrows, and the date of the spectrum
    is marked with a dashed vertical line.
    \label{fig:sn2001j}}
\end{figure}

SN~2001J was discovered in UGC~4729 \citep{2001IAUC.7564....2B} and classified
as a Type II SN \citep{2001IAUC.7566....2J}.  SNID identifies the SN~2001J spectrum as 
that of a young SN~IIb, similar to the spectrum of SN~2008ax.  L11
list SN~2001J as a SN~IIP with poor light-curve coverage, but Figure~\ref{fig:sn2001j}
shows our rereduced KAIT light curve (including nondetection upper limits)
indicating there was no bright hydrogen recombination plateau phase.
Taking into account both the SNID result and the rapid light-curve decay,
we prefer the SN Type IIb classification, but cannot rule out the possibility that 
SN~2001J was a hydrogen-rich Type II SN with a relatively rapid decline rate.

\subsubsection{SN 2002jj (Ic\,$\rightarrow$\,Ic/Ic-BL) }

SN~2002jj was discovered in IC~340 \citep{2002IAUC.8026....1H} and classified
as a SN~Ic \citep{2002IAUC.8031....4F}, with a spectrum similar to that of SN~1994I.
We have three spectra of SN~2002jj, all of moderate quality.  SNID prefers
a SN~Ic-BL classification, as do comparisons with the average spectra of \citet{2016ApJ...827...90L},
but the data are not good enough to be sure. 
The light curve (though it is sparse) indicates that all of our spectra were taken well past peak brightness.
SN~2002jj showed a peak absolute magnitude of $-17.66 \pm 0.23$ (L11),
a value in the normal range for both SNe Ic and SNe Ic-BL
\citep[SNe~Ic-BL are, on average, more luminous than other stripped-envelope SNe; e.g.,][]{2011ApJ...741...97D,2016MNRAS.457..328L},
and so it is not clear whether this SN was a bona-fide SN~Ic-BL.

\begin{figure}[ht]
\epsscale{1.0}
\plotone{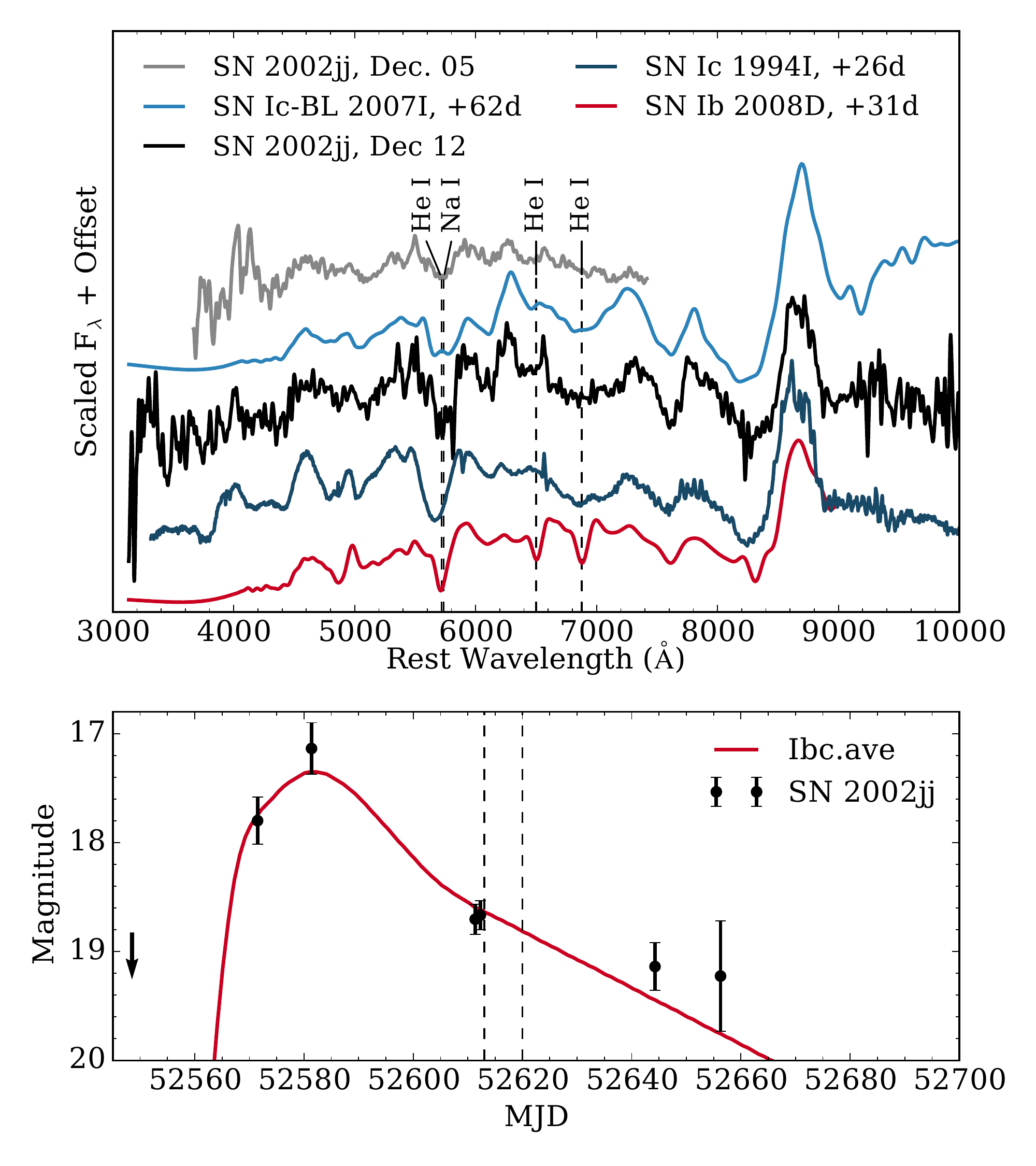}
\caption{Top: spectra of SN~2002jj, smoothed
   by a 40\,\AA~Gaussian kernel and compared to spectra of the
   Type Ic-BL SN~2007I \citep{2014AJ....147...99M}, the Type Ic SN~1994I \citep{1995ApJ...450L..11F},
    and the Type Ib SN~2008D \citep{2009ApJ...702..226M}.
   Bottom: light curve of SN~2002jj compared to the
    average SN~Ib/c template from L11, with the dates of the spectra marked by vertical lines.
    \label{fig:sn2002jj}}
\end{figure}

In addition, the \ion{He}{1} lines used to distinguish SNe Ic from Ib are most apparent 
around the time of peak brightness and then fade, in most cases disappearing entirely by $\sim50$--70\,days
\citep{2014AJ....147...99M,2016ApJ...827...90L}.  The spectroscopic coverage of SN~2002jj
did not begin until $\sim$1\,month post-peak and we cannot rule out the possibility
of weak helium features near peak, but 
we prefer the SN~Ic or SN~Ic-BL classification;
see Figure~\ref{fig:sn2002jj}.

\subsubsection{SN 2002jz (Ic\,$\rightarrow$\,Ib/Ic)}
\label{sec:sn2002jz}

SN~2002jz was discovered in UGC~2984 and classified as a SN~Ic
with a resemblance to SN~1994I \citep{2002IAUC.8037....1P}.
We present three spectra of this object, the most useful
of which was observed on UT 2003-01-07.  There is significant MW dust reddening along
the line of sight ($E(B-V) = 0.4846$\,mag), and the unresolved \ion{Na}{1}~D absorption in our spectra indicates
a similar amount of host-galaxy dust obscuration.

\begin{figure}[ht]
\epsscale{1.0}
\plotone{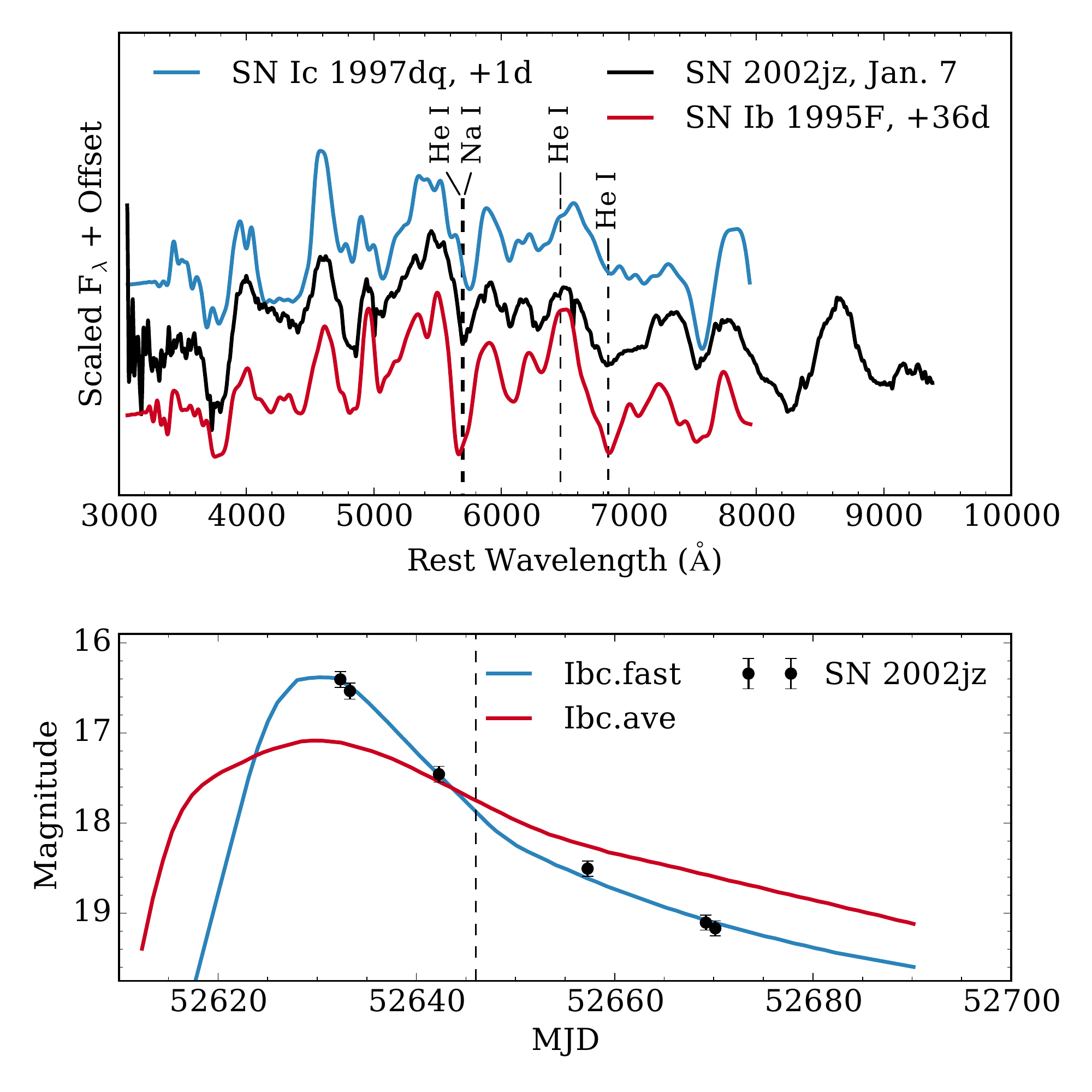}
\caption{Top: spectrum of SN~2002jz from 2003-01-07 (corrected for a MW reddening of $E(B-V) = 0.4846$\,mag)
    compared with that of SN Ic 1997dq and SN Ib 1995F \citep{2001AJ....121.1648M,2014AJ....147...99M}.
    Bottom: light curve of SN~2002jz compared to the stripped-envelope templates of L11, with the date of the spectrum shown with a vertical line.
    \label{fig:sn2002jz}}
\end{figure}

SNID weakly prefers a Type Ib label over the Ic label; the best SNID matches are to the spectra of the
Type Ib SNe 1995F \citep[reclassified from Ic to Ib by][]{2014AJ....147...99M}
and 1999ex \citep[studied in detail by, e.g.,][]{2007PASP..119..135P}.
However, the \ion{He}{1} $\lambda$6678 line in SN~2002jz is 
so weak as to be nearly indiscernible (it is only detected as a notch out of an adjacent P-Cygni emission profile),
and the \ion{He}{1} $\lambda$7065 line is notably weaker than that of SN~1995F --- see Figure~\ref{fig:sn2002jz}.
We give SN~2002jz equal weights in the Ib and Ic categories.

\subsubsection{SN 2003br (II\,$\rightarrow$\,II/IIb) }

SN~2003br was discovered in MCG-05-34-18 \citep{2003IAUC.8090....1S}
and was classified as a Type II SN \citep{2003IAUC.8091....2B}.
We present the classification spectrum of SN~2003br, but this
alone is not enough to distinguish between a Type II 
or a Type IIb classification.  The MW reddening toward the object is small
($E(B-V) = 0.0822$\,mag), but the observed spectral energy distribution (SED) implies that there must
be a moderate-to-large degree of host-galaxy reddening.  
Unfortunately, the spectrum
has neither the signal nor the resolution to measure any
possible narrow
\ion{Na}{1}~D absorption features.  Adopting a correction
of $E(B-V) = 1.0$\,mag appears to roughly correct the SED, and so
we continue with that.

\begin{figure}[ht]
\epsscale{1.0}
\plotone{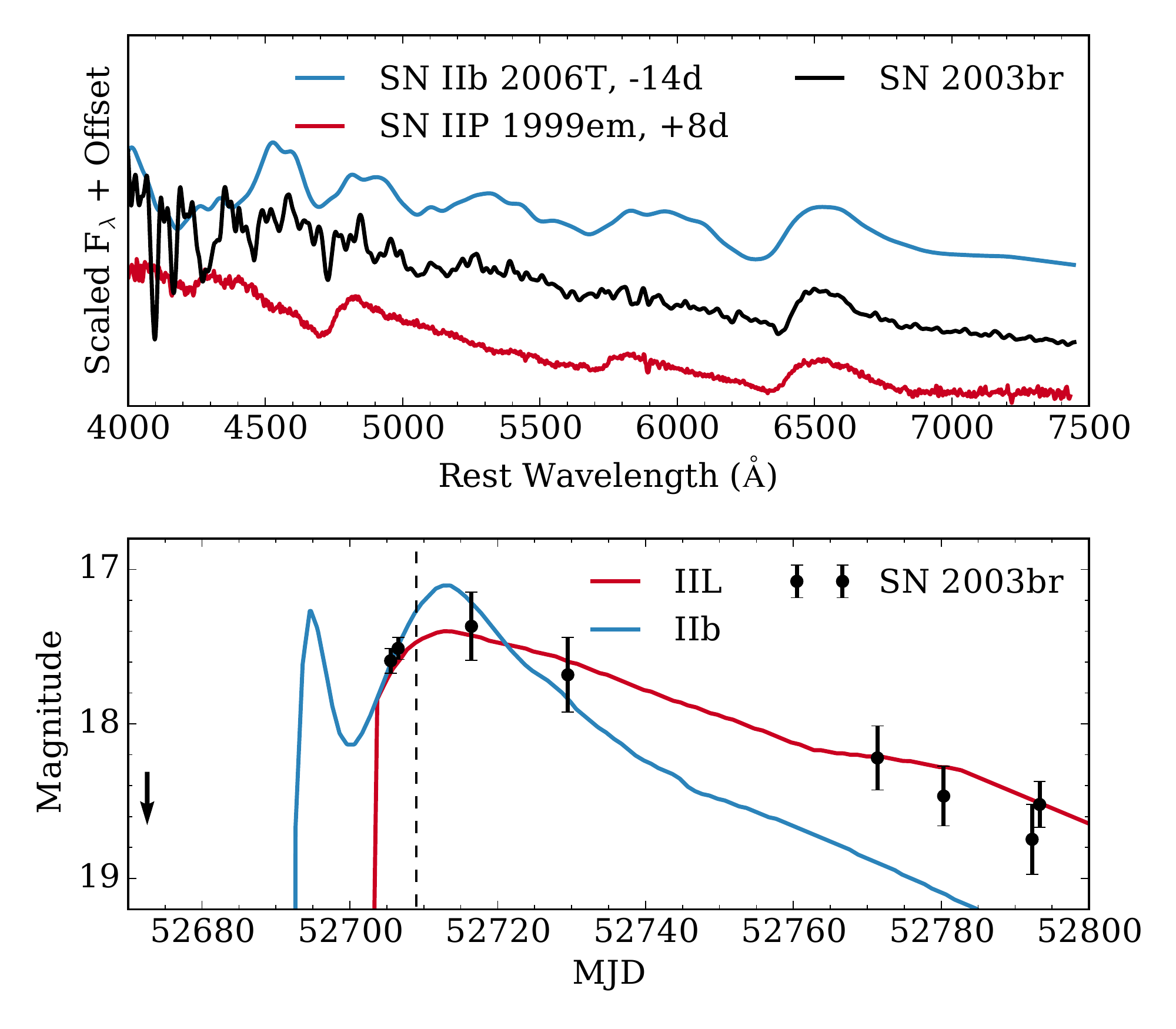}
\caption{Top: spectrum of SN~2003br (corrected for $E(B-V)=1.0$\,mag and smoothed
    with a 40\,\AA~Gaussian kernel),
    compared to that of the Type IIb SN~2006T \citep{2014AJ....147...99M} and
    the Type IIP SN~1999em \citep{2002PASP..114...35L} --- both match well.
    Bottom: the light curve compared to the templates
    from L11, with the date of the spectrum marked.
    \label{fig:sn2003br}}
\end{figure}

SNID indicates that the best match is with the Type IIb SN~2006T,
but the Type IIP SN~1999em also shows a good match.  The light curve
is more similar to the Type IIL template, and so we prefer a Type II classification
but cannot be sure. See Figure~\ref{fig:sn2003br}.

\subsubsection{SN 2003id (Ic-pec) }
\label{sec:sn2003id}

\begin{figure}[ht]
\epsscale{1.35}
\hspace*{-6mm}
\plotone{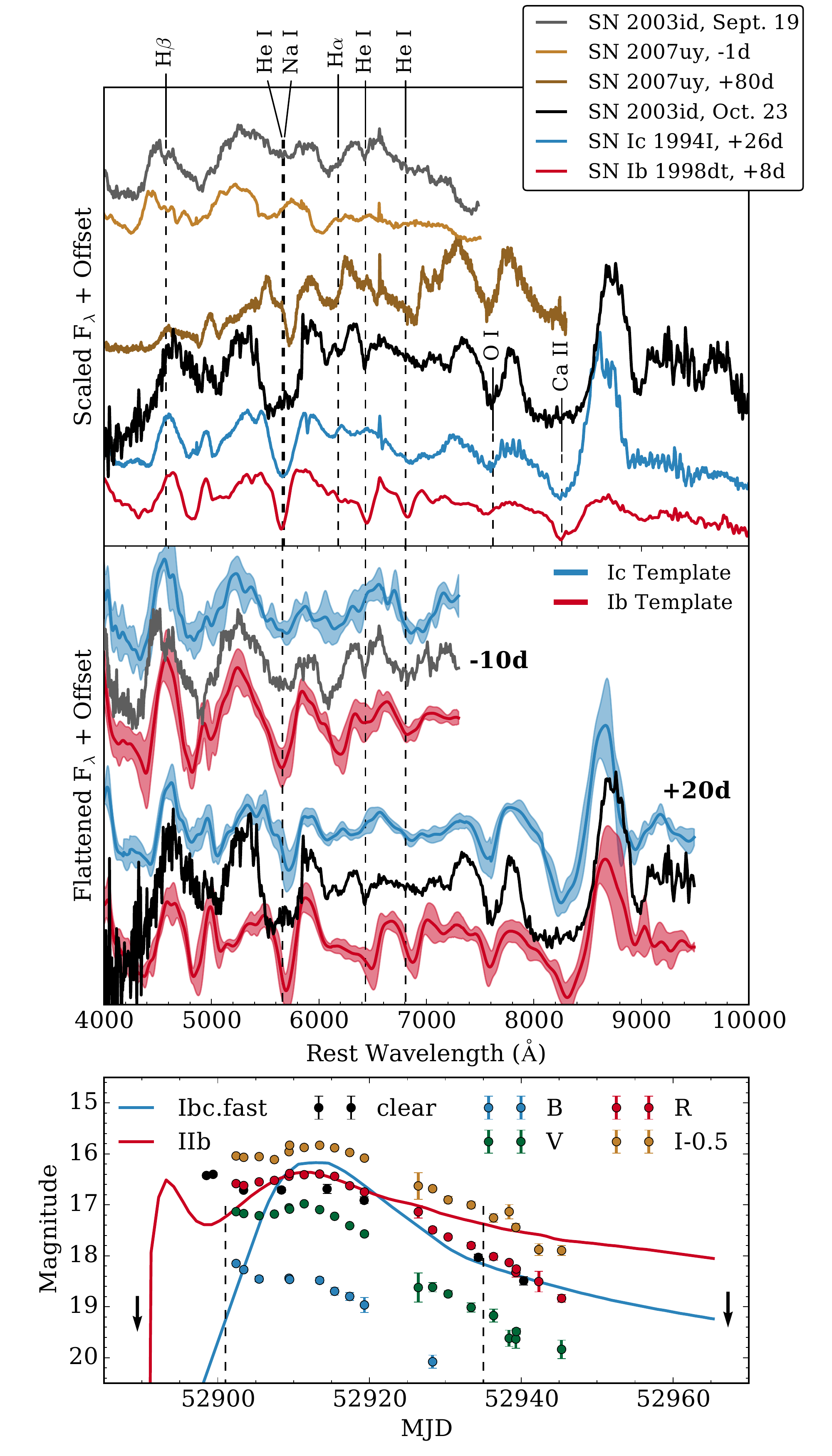}
\caption{ Top: spectra of SN~2003id smoothed with a 20\,\AA~Gaussian kernel
    and compared to spectra of the peculiar Type Ib SN~2007uy \citep{2014AJ....147...99M},
    the normal Type Ic SN~1994I \citep{1995ApJ...450L..11F},
    and the normal Type Ib SN~1998dt \citep{2001AJ....121.1648M}.
    Putative hydrogen and helium lines are marked at $v = 17,500$\,km\,s$^{-1}$
    and $11,000$\,km\,s$^{-1}$, respectively.
    Middle: the continuum-normalized spectra of SN~2003id compared to
    the $-$10\,d and $+$20\,d average spectra of SNe~Ib and SNe~Ic from \citet{2016ApJ...827...90L},
    with the standard deviation shown as shaded regions and
    \ion{He}{1} lines marked as above.
    Bottom: filtered light curves of SN~2003id, with phases
    of spectra marked by the dashed lines.  Template SN~IIb and ``fast'' SN~Ib/c 
    light curves from L11 are overplotted.
    \label{fig:sn2003id}}
\end{figure}

SN~2003id was discovered in NGC 895 and classified as a SN~Ibc-pec by \citet{2003IAUC.8228....2H},
who identify several normal SN~Ic lines as well as a strong blended feature around 5700\,\AA.
Here we analyze spectra observed on Sep.\ 19 and Oct.\ 23, 2003 --- see Figure~\ref{fig:sn2003id}.
Our spectra confirm this to be an odd object with no good matches
in the SNID template set, though with many similarities to
the peculiar SN~Ib~2007uy \citep{2013MNRAS.434.2032R,2014AJ....147...99M}.

The strong blended feature around 5700\,\AA~persists and 
appears to grow stronger over time.  We tentatively identify \ion{He}{1}\,$\lambda$6678
at $v \approx 11,000$\,km\,s$^{-1}$, which implies that the feature at 5700\,\AA\ may include
some absorption from \ion{He}{1}\,$\lambda$5876 in addition to \ion{Na}{1}\,$\lambda\lambda$5890, 5896;
perhaps this feature arises from multiple velocity components of these two ions.
However, we find no clear sign of \ion{He}{1}\,$\lambda$7065 absorption in either spectrum.
(Note that these spectra have undergone a correction for telluric absorption and none of the features 
in them is telluric.)

We also identify a feature near 6140\,\AA~that evolves from a strongly blended state
into two clearly defined components, the redder of which is plausibly high-velocity
H$\alpha$ at $v \approx 17,500$\,km\,s$^{-1}$, and we tentatively identify H$\beta$ at
$v \approx 17,500$--$18,500$\,km\,s$^{-1}$ in the spectrum from Sep.~19.
The bluer feature may be a second, even higher-velocity H$\alpha$ line
at $v \approx 23,500$\,\kms; the bluemost edge of the \ion{Ca}{2} IR triplet absorption implies
similar velocities for the calcium in SN~2003id.
Several authors have previously identified high-velocity H$\alpha$ lines in SN~Ib/c spectra
\citep[e.g.,][]{2006PASP..118..791B,2006AA...450..305E,2007PASP..119..135P}.

We compare the spectra of SN~2003id to the average spectra of \citet{2016ApJ...827...90L} in 
the middle panel of Figure~\ref{fig:sn2003id}. We estimate that the date of $R$-band maximum
was 30 Sep.\ 2003, which means our two spectra were obtained at phases of $-$11 and $+$22 days;
hence, we show the $-$10\,d and $+$20\,d average spectra.
The peculiarity of SN~2003id is apparent here as well: our spectra deviate significantly
from both SNe~Ib and SNe~Ic spectra at several wavelengths.  As above, 
we find that the putative \ion{He}{1}\,$\lambda$6678 line is SN~Ib-like,
while the lack of a \ion{He}{1}\,$\lambda$7065 line is SN~Ic-like, and the extreme
widths of the \ion{Ca}{2} and 5700\,\AA\ features are unlike both.

The light curves provide an additional wrinkle: SN~2003id distinctly
shows a double-peaked evolution, with a rapid decline from
the blue first peak followed by a rise to the second maximum a few days later.
There is no evidence from the later spectrum that the second peak arises from
interaction with dense circumstellar interaction (i.e., we find no narrow emission lines).
Most SNe~Ib and SNe~Ic do not exhibit double-peaked light curves like those of SN~2003id,
though similar behavior is often observed in SNe~IIb.
A very small number of double-peaked SNe~Ib have been discovered
\citep[e.g., SNe~2005bf and 2008D;][]{2005ApJ...633L..97T,2006ApJ...641.1039F,2009ApJ...702..226M},
as has one double-peaked SN~Ic \citep{2016AA...592A..89T}.
However, these events had early peaks which were notably less bright than their main peaks,
they exhibited a diversity of different peak absolute magnitudes, and none show the
peculiar 5700\,\AA\ feature of SN~2003id.

Given the lack of \ion{He}{1}\,$\lambda$7065
and the uncertain (and certainly peculiar) \ion{He}{1}\,$\lambda$6678\,\AA\ and
\ion{He}{1}\,$\lambda$5876\,\AA\ lines, we classify SN~2003id as a peculiar and double-peaked SN~Ic.
We note that this object appears to be different from the other stripped-envelope
SNe in this sample.

\subsubsection{SN 2004bm (Ibc-pec/IIb\,$\rightarrow$\,IIb/IIb-pec)}

\begin{figure}[ht]
\epsscale{1.0}
\plotone{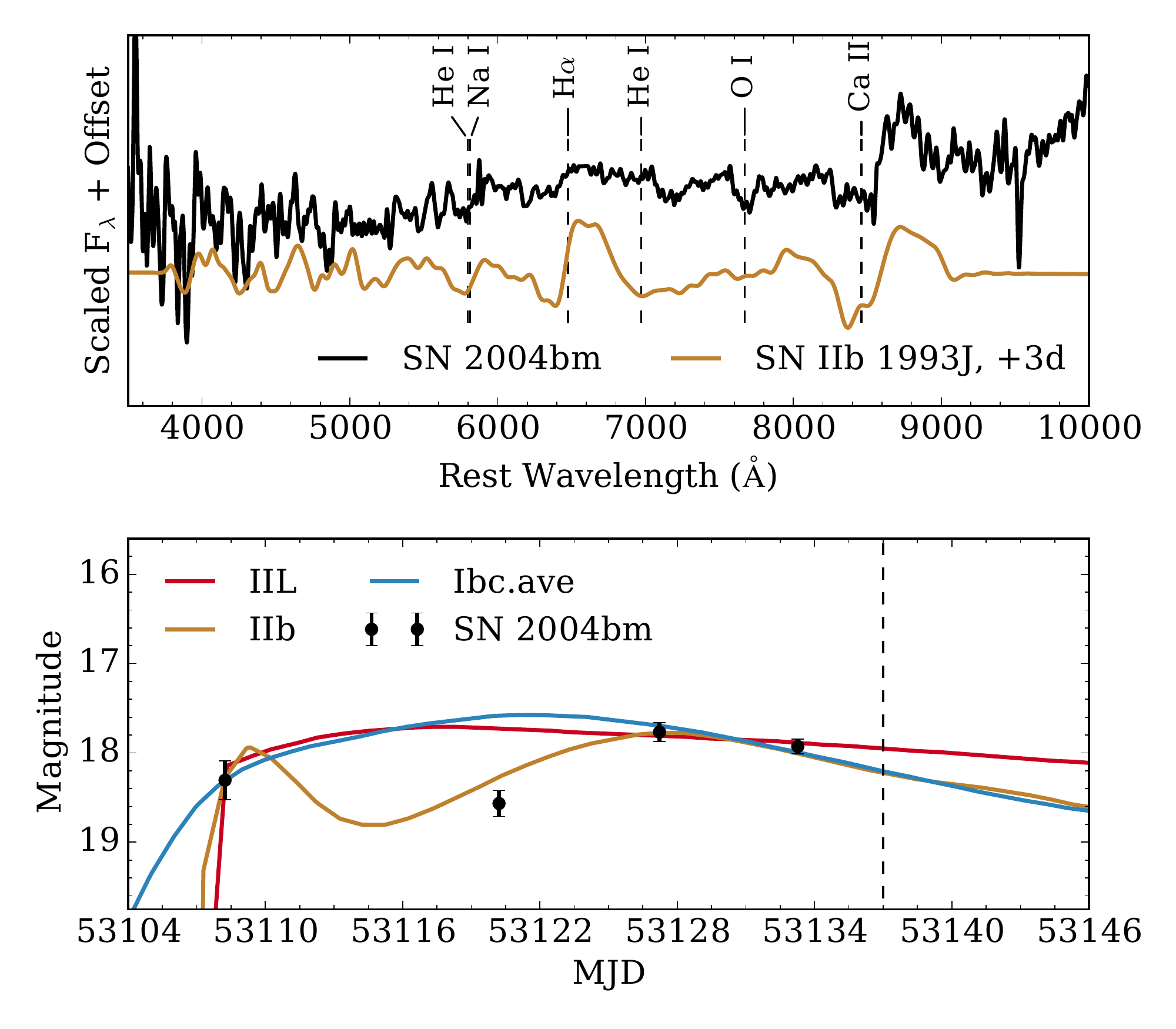}
\caption{Top: spectrum of SN~2004bm smoothed
    with a Gaussian kernel 40\,\AA~wide and compared to the near-maximum
    spectrum of the Type IIb SN~1993J \citep[reddened by $E(B-V) = 0.4$\,mag for comparison;][]{2000AJ....120.1487M}.
    Bottom: the light curve compared to 
    the Type IIL, IIb, and average Ib/c template light curves from L11. A single data point
    indicates a light-curve dip and argues for a Type IIb classification.
    The date of the spectrum is marked by the vertical line.
    \label{fig:sn2004bm}}
\end{figure}

SN~2004bm was discovered in NGC~3437 \citep{2004IAUC.8335....1A} and originally
classified as a SN~Ic \citep[though with some uncertainty;][]{2004IAUC.8339....2F}.
L11 note that the light curve shows a dip. Though the data are sparse, the SN occurred near the core of its host galaxy,
 and this conclusion depends upon only one data point out of four total, our rereduction of the
light curve also shows a dip indicated by the second detection ---
see Figure~\ref{fig:sn2004bm}.  Similar light-curve behavior has been observed in Type IIb SNe
\citep[e.g., L11;][]{2012ApJ...756L..30A}, and L11 used the light curve to argue that SN~2004bm was a SN~IIb.

The only spectrum we have of this SN is the one used for the original classification
and it is of low quality --- SNID does not provide a clear classification, but it does indicate that
the best cross-correlations are with spectra of SNe~IIb and IIP (though the phases are in disagreement).
The spectrum of SN~2004bm does not correlate with spectra of Type Ic SNe well.
Narrow \ion{Na}{1}~D absorption at the host-galaxy
redshift is apparent but unresolved and very noisy, indicating moderate host-galaxy 
reddening atop the MW contribution of $E(B-V) = 0.0159$\,mag.
There are few spectral features in our spectrum, though we identify \ion{Ca}{2} P-Cygni profiles, 
faint \ion{O}{1} absorption, and a very shallow H$\alpha$ P-Cygni line --- see Figure~\ref{fig:sn2004bm}.

Based on the above discussion, we prefer a classification of Type IIb for SN~2004bm, but 
the H$\alpha$ line in the spectrum of SN~2004bm is much too weak for a normal Type IIb SN.
We also note that, if we discard the second photometric data point, the light curve of SN~2004bm is well-fit
by normal Type II events or by stripped-envelope events.

\subsubsection{SN 2004cc (Ic\,$\rightarrow$\,Ib/Ic)}

SN~2004cc was discovered in NGC~4568 \citep{2004IAUC.8350....2M}.
\citet{2004IAUC.8353....2M} note the strong reddening toward SN2004cc
and classify it as a Type I SN, though they prefer no subtype, while 
\citet{2004IAUC.8353....3F} present a SN~Ic classification.

\begin{figure}[ht]
\epsscale{1.0}
\plotone{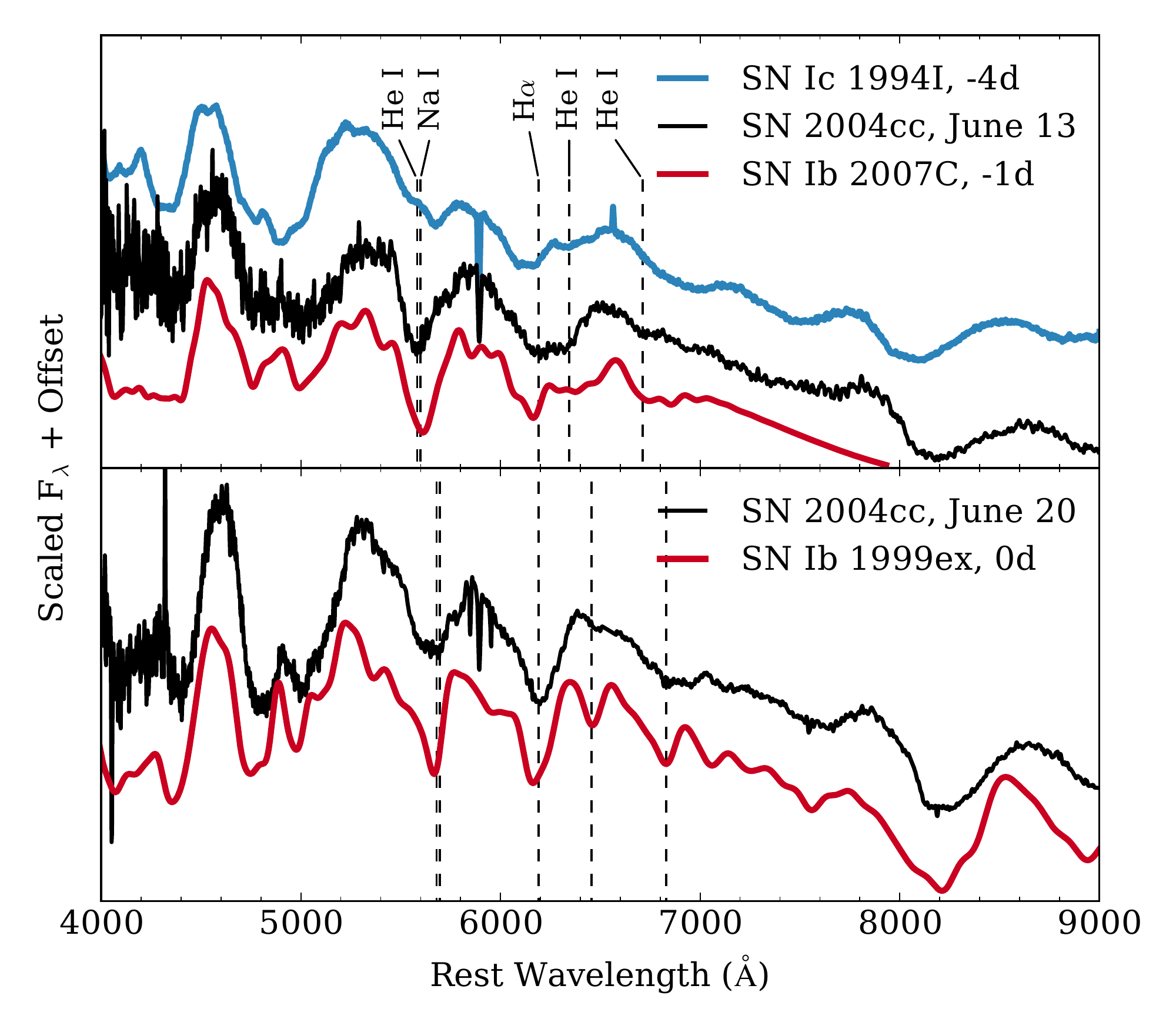}
\caption{Top: spectrum of SN~2004cc observed on UT~2004-06-13 alongside
    premaximum spectra of SN Ic 1994I and SN Ib 2007C \citep{1995ApJ...450L..11F,2014AJ....147...99M},
    and we mark \ion{He}{1} lines at a velocity of 15,000\,\kms.  
   Bottom: spectrum of SN~2004cc observed on UT~2004-06-20
    and a spectrum of the Type Ib SN~1999ex \citep{2002AJ....124..417H}, with the helium line marked at 10,000\,\kms, a deceleration
    consistent with the normal evolution of helium-line velocities in SNe~Ib \citep{2016ApJ...827...90L}.
    Both spectra of SN~2004cc have been dereddened by $E(B-V) = 1.0$\,mag.
    \label{fig:sn2004cc}}
\end{figure}

There is little MW reddening toward SN~2004cc $(E(B-V) = 0.0279$\,mag), but the 
(unresolved) \ion{Na}{1}~D lines in our spectra indicate strong host-galaxy
dust obscuration.  The EW measured from these lines is well outside
the relations of \citet{2012MNRAS.426.1465P}, and so we only roughly
estimate the total reddening, adopting $E(B-V) = 1.0$\,mag for visual comparisons.

SNID identifies reasonable correlations between the spectra
of SN~2004cc and the spectra of both SNe~Ib and SNe~Ic,
strongly disfavoring all other types and slightly preferring the Ib label over Ic.
Unfortunately, we have only a single photometric detection of the SN, so there is little
independent information about the phases of these spectra.

Figure~\ref{fig:sn2004cc} shows that the later spectrum of SN~2004cc matches that of SN~Ib~1999ex quite well,
while the earlier spectrum matches that of SN~Ic~1994I.
Weak H$\alpha$ detections have been claimed for both of these events \citep{2006PASP..118..791B}. 
Though hydrogen absorption may be present, we trust the SNID result (which prefers
a Type Ib or Ic label, rather than a IIb), and we do not consider a Type IIb label for SN~2004cc ---
see the discussion in \S\ref{sec:methods}.
Note also that, if the identification of the H$\alpha$ line in Figure~\ref{fig:sn2004cc} is correct,
it exhibits a much faster (and unchanging)
Doppler velocity ($\sim$17,000\,\kms) than the
\ion{He}{1} lines (which are at $\sim$15,000 and $\sim$10,000\,\kms\ on June 13 and June 20, respectively).
This behavior is peculiar but not unique for this feature 
in stripped-envelope SNe \citep{2016ApJ...827...90L}.

The spectra of SN~2004cc present another puzzle.
The early-time spectrum appears to show a strong \ion{He}{1}\,$\lambda$6678 line
(and a strong \ion{He}{1}/\ion{Na}{1} blend near 5500\,\AA), but very little
\ion{He}{1}\,$\lambda$7065 absorption.
It is difficult to physically explain a strong \ion{He}{1}\,$\lambda$6678 absorption
line without a similarly strong $\lambda$7065 line; other ions may be contributing
to this feature.
Just one week later the \ion{He}{1}\,$\lambda$6678 line has disappeared,
though the width of the line near 5500\,\AA\ implies that \ion{He}{1}\,$\lambda$5876 is
still present.  Given the above uncertainties and the weakness of 
the \ion{He}{1} $\lambda$7065 line,
we assign SN~2004cc the Ib/Ic label.
Interestingly, \citet{2012ApJ...752...17W} present a variable and long-lasting
radio light curve, indicating the presence of a complex circumstellar medium near this object.

\subsubsection{SN 2005H (II\,$\rightarrow$\,II/IIb) }

\begin{figure}[ht]
\epsscale{1.0}
\plotone{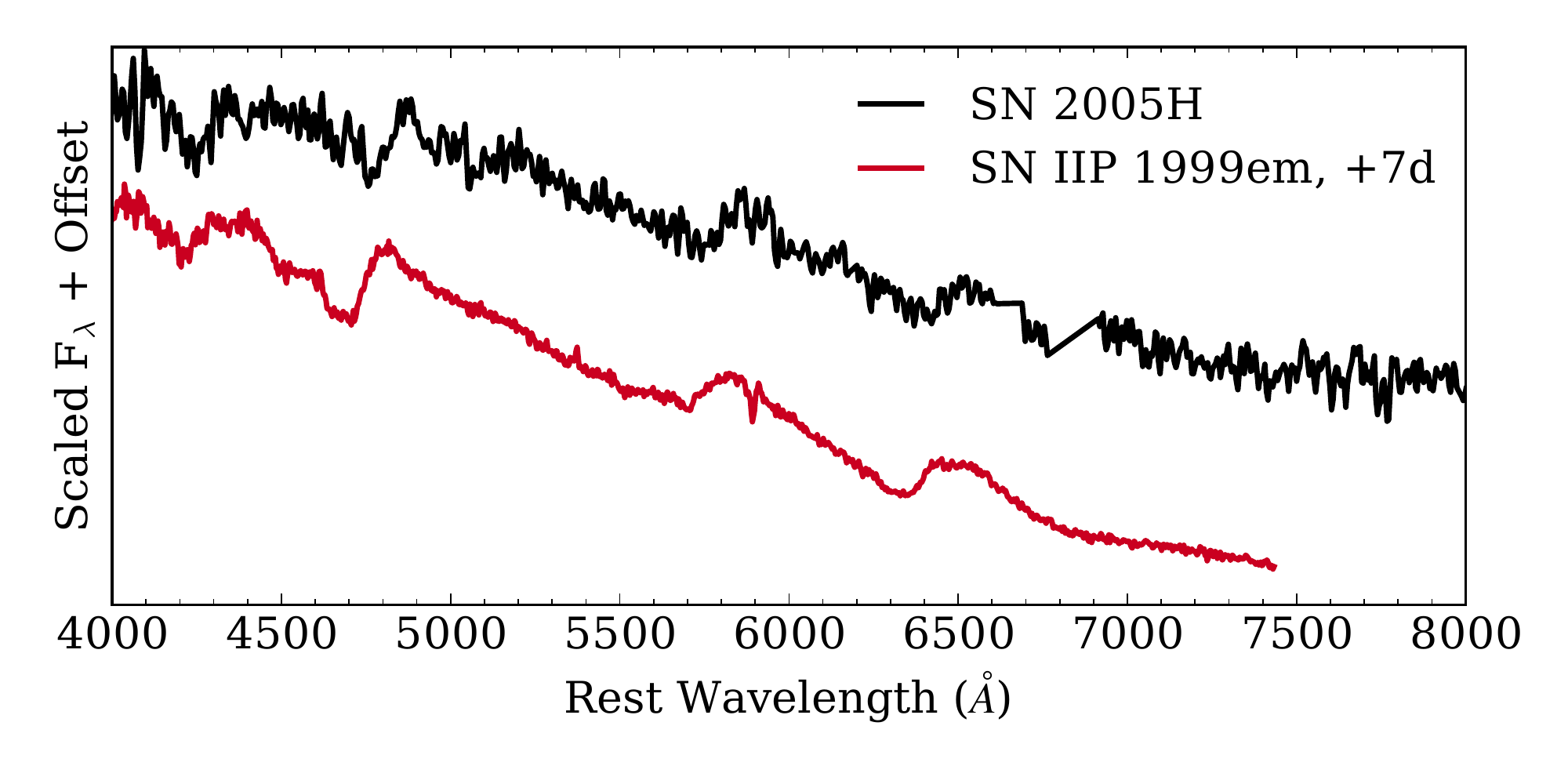}
\caption{Spectrum of SN~2005H alongside a spectrum of
    the Type IIP SN~1999em \citep{2002PASP..114...35L}.
    \label{fig:sn2005h}}
\end{figure}

SN~2005H was discovered in NGC~838 \citep{2005IAUC.8467....1G}
and classified as a Type II SN based upon a noisy spectrum \citep{2005IAUC.8467....2P}.
Very few data exist on SN~2005H, and most of the spectra are entirely
dominated by host-galaxy light.  However, we were able to obtain the original classification
spectrum \citep{2008AA...488..383H} and we find clear detections of
H$\alpha$ absorption and P-Cygni profiles of H$\beta$ and \ion{Na}{1}
on an otherwise smooth blue continuum; see Figure~\ref{fig:sn2005h}.

SNID shows this spectrum to be more similar to spectra of young SNe~II than SNe~IIb.
We have very little information on the light-curve evolution of this object --- the SN
was discovered very near the bright ($R \approx 12-14$\,mag) core of the host and our
data are badly contaminated by galaxy light.  Other than the discovery image we have no
clear detections, only uninformative upper limits.  

\subsubsection{SN 2005mg (II\,$\rightarrow$\,II/IIb)}
\label{sec:05mg}

\begin{figure}[ht]
\epsscale{1.0}
\plotone{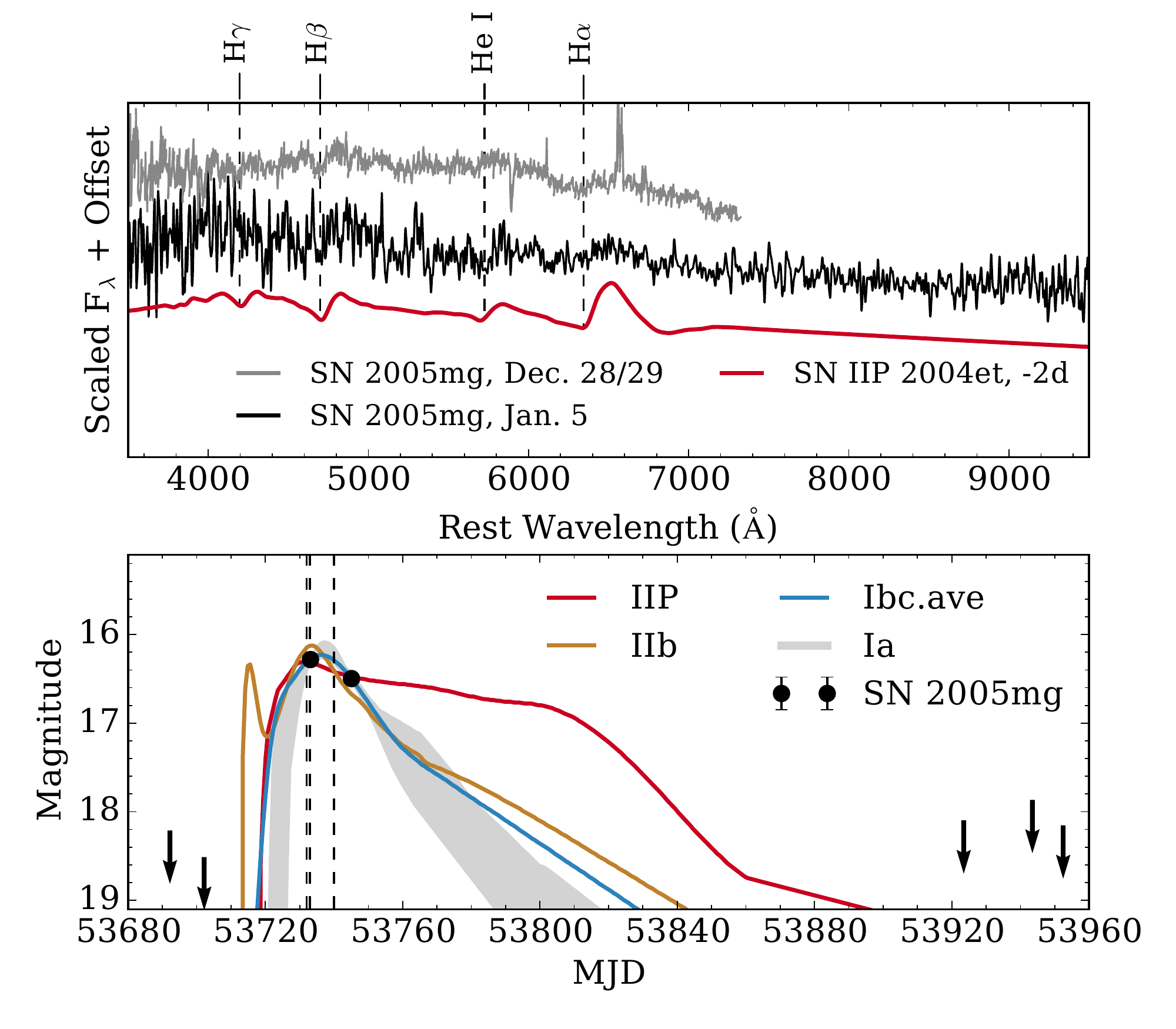}
\caption{Top: spectra of SN~2005mg alongside a spectrum of
    the Type IIP SN~2004et \citep{2006MNRAS.372.1315S}.
    In blue we show the co-addition of the traces obtained from plots
    of the two spectra from Dec.\ 2005, while in black we show the
    full spectrum from Jan.\ 2006.
    Bottom: the light curve compared to 
    template light curves from L11, with the dates of
    spectra shown with dashed lines.
    \label{fig:sn2005mg}}
\end{figure}

SN~2005mg was discovered in UGC~155 \citep{2005CBET..336....1N} and classified as a 
heavily reddened young Type II SN, with detections of H$\alpha$ and H$\beta$ 
in the noisy spectrum \citep{2005CBET..342....1M}.  
We have a spectrum obtained one week later, but
unfortunately we were unable to locate a digital copy of the spectrum
cited above. We did, however, locate plots of both the classification spectrum
and another one obtained the night prior,
both in image form.\footnote{\url{https://www.cfa.harvard.edu/supernova/spectra/}}
Using an online tool,\footnote{\url{http://arohatgi.info/WebPlotDigitizer/app/}}
we traced the spectra from these plots and consider the results in our analysis.

All three spectra exhibit extremely low S/N and SNID identifies no good matches.
The KAIT light curve is similarly uninformative with only two detections
and is consistent with template light curves of all types; see Figure~\ref{fig:sn2005mg}.
However, the spectra do appear to corroborate the 
\citet{2005CBET..342....1M} classification of SN~2005mg as a Type II SN,
with plausible detections of weak H$\alpha$ and H$\beta$
lines showing broad P-Cygni profiles. We therefore consider the original classification robust
though we cannot determine whether SN~2005mg was a Type IIb or a normal Type II SN,
and so we label it as II/IIb (unsure).

\subsubsection{SN 2006eg (Ic\,$\rightarrow$\,IIb/Ib/Ic/Ic-BL) }
\label{sec:sn2006eg}

SN~2006eg was discovered in an anonymous galaxy \citep{2006CBET..600....1M}
and classified as a SN~Ib/c \citep{2006CBET..604....1F}.  SNID, GELATO, and Superfit
identify no good matches, but the best cross-correlations are with spectra of SNe~Ic-norm and SNe~Ic-BL.  
The MW reddening toward SN~2006eg is small ($E(B-V) = 0.0933$\,mag),
and though the spectrum of SN~2006eg is noisy it does not appear that there
could be more than a modest degree of host-galaxy reddening --- we 
detect no narrow \ion{Na}{1}~D lines at the host galaxy's redshift.
Our sparse and noisy light curve is most similar to that of a Type IIL SN, but it is also consistent
with a stripped-envelope classification.

\begin{figure}[ht]
\epsscale{1.0}
\plotone{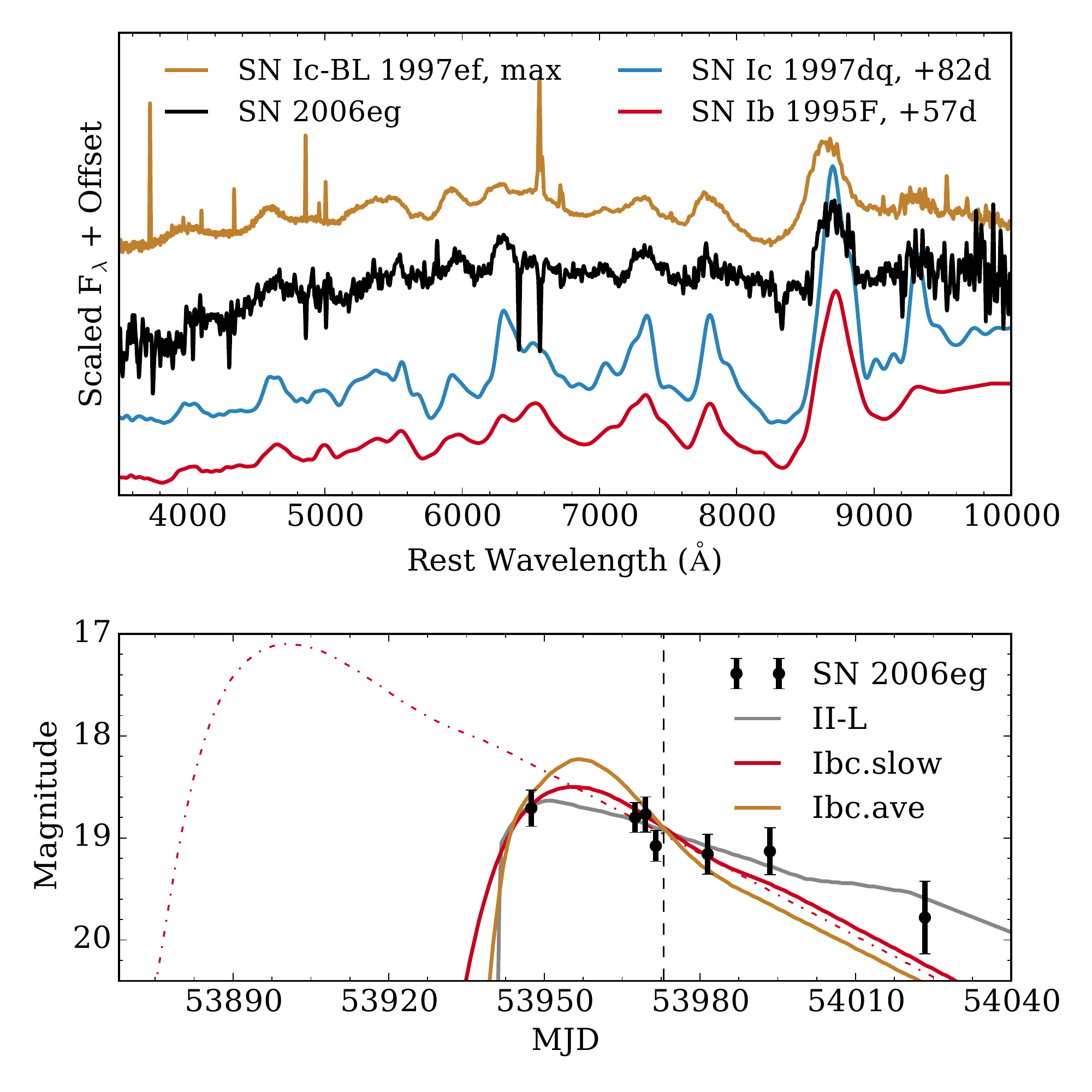}
\caption{Top: spectrum of SN~2006eg (smoothed with a 20\,\AA~Gaussian kernel)
    compared to that of the SN~Ic-BL 1997ef,
    SN~Ic 1997dq, and SN~Ib 1995F \citep{2001AJ....121.1648M,2014AJ....147...99M}.
   Bottom: the light curve, with the date the spectrum was taken marked, alongside template
     light curves from L11.  Our last prediscovery upper limit was 173\,d before the first detection;
    though we show comparison light curves assuming the peak was observed, the true peak could
    have occurred months before discovery.  The red dash-dot line shows the Ibc.slow template
    of L11 offset by 55 days, illustrating how the peak may have occurred prediscovery.
    \label{fig:sn2006eg}}
\end{figure}

SN~2006eg had a peak measured magnitude of $-14.86 \pm 0.23$\,mag (unfiltered; L11),
which makes the event an underluminous outlier from both the SN~Ic-BL
and normal SN~Ib/Ic populations \citep[L11,][]{2011ApJ...741...97D}.
However, as Figure~\ref{fig:sn2006eg} shows, the true peak could have occurred before discovery
and the event may have been significantly more luminous at the true (unobserved) peak.

Our spectrum of SN~2006eg is not only noisy, it also exhibits a low contrast between the continuum and
the SN features --- this event occurred near the center of its spiral host galaxy, and it appears that our
spectrum suffers from significant host-galaxy contamination (the strong, narrow absorption features are likely 
badly subtracted host-galaxy emission lines). It has been our experience that
SNID sometimes mistakenly prefers a SN~Ic-BL classification for low-contrast spectra, and so we 
are hesitant to assign much belief to that result.

The \ion{He}{1} lines that mark the difference between SNe~Ib and SNe~Ic are, in general, time-dependent:
they are most apparent soon after peak and generally fade completely away by $\sim50$--70\,days
\citep{2014AJ....147...99M,2016ApJ...827...90L}.  These types can be 
difficult to differentiate at these ages, and SNe Ic, Ib, and IIb all show
strong features at the same wavelengths as those in SN~2006eg's spectrum.
It is likely that our spectrum of SN~2006eg shows 
a nearly nebular SN IIb, Ib, or Ic ``watered down'' by host-galaxy contamination.
If SN~2006eg was a mostly normal stripped-envelope SN discovered late, then its light curve was 
slowly declining, but it was not an extreme outlier from the diverse
late-time decline rates observed for these SNe \citep[e.g.,][]{2016MNRAS.457..328L}.

Without better data, it is very difficult to firmly label SN~2006eg --- our spectrum
shows it was not a Type Ia or Type II SN, but it could have been a SN IIb, Ib, Ic, or Ic-BL.

\subsection{SNe Ia}
\label{sec:ia}

We follow \citet{2012MNRAS.425.1789S} when identifying the subtypes of SNe Ia, and
in most cases simply duplicate the classifications from their Table~7.
We propose some updated subtype classifications from L11, but no objects appear to be misclassified as SNe~Ia,
nor does it appear that any events labeled otherwise should be reclassified as SNe~Ia.

We update the classifications of SNe~1999bh and 2002es, labeling them with the subtype ``Ia-02es'' 
as identified by \citet{2012ApJ...751..142G}.  These events are subluminous and exhibit low expansion velocities,
sharing properties with both the SN~Ia-2002cx \citep[i.e., SN~Iax;][]{2013ApJ...767...57F} and SN~Ia-1991bg subtypes.
Though L11 note that these two events may form their own subtype, they include them with
the SNe~Iax as the properties of the subtype were only partially understood at the time
\citep[subsequent work has furthered our understanding; e.g.,][]{2015Natur.521..328C,2015ApJ...799...52W,2016ApJ...832...86C}.
\citet{2013ApJ...767...57F} show that SNe~Iax display a wide range of peak luminosities, from the 
extremely subluminous SN~2008ha ($M_V \approx -14.2$\,mag) up into the range of typical SNe~Ia ($M_V \approx -18.5$\,mag),
and the rate calculations of L11 did not account for the low-luminosity members of this class and therefore 
underestimated the true rate of these events.

We also update several events previously labeled SNe~Ia 1991T to the ``SN~Ia-1999aa'' subtype \citep{2004AJ....128..387G},
a subclass that falls in between SNe~Ia-norm and SNe~Ia-1991T and
another distinction intentionally not included in L11 \citep[see][]{2001ApJ...546..734L}.
Three of these events (SNe 1998es, 1999aa, and 1999dq) were previously given a Ia-99aa label by \citet{2012MNRAS.425.1789S}.
The spectral evolution of SN~1999ac was studied in detail by \citet{2005AJ....130.2278G}, who note that early-time spectra are similar
to those of SN~1999aa with relatively weak silicon absorption, but SNID identifies both premaximum and postmaximum spectra as SN~Ia-norm
\citep[][though Ia-99aa templates also provide reasonable fits]{2012MNRAS.425.1789S}
and subtle peculiarities exist throughout this object's evolution. Owing to this peculiarity, we
give this event equal weights in the Ia-norm and Ia-99aa subclasses.

SN~2001V is grouped among the ``shallow silicon'' events by \citet{2009PASP..121..238B}, 
and the premaximum spectra of SN~2001V published by \citet{2012AJ....143..126B} are strongly classified as 99aa-like by SNID.
The data on SN~2006cm are somewhat less conclusive and the early evolution is not well constrained.
The spectra show strong \ion{Na}{1}\,D absorption features from the host UGC 11723
and they are noticeably reddened by host-galaxy dust \citep{2006CBET..526....1B,2011Sci...333..856S}. 
SNID prefers a SN~Ia-normal classification for SN~2006cm \citep{2012MNRAS.425.1789S}, but
the spectra also exhibit good matches to those of Ia-99aa objects and the silicon absorption features are
weaker than those in the SNID-preferred SN~Ia-norm templates.  We give SN~2006cm equal
weights in the SN~Ia-norm and SN~Ia-99aa subclasses.

Finally, SN~2004bv is classified as a SN~Ia-91T event by SNID, but unfortunately the only existing premaximum spectrum
of this SN does not extend to sufficiently blue wavelengths to capture the \ion{Ca}{2}\ H\&K lines, which are the
strongest indicator of a SN~1991T-like event at young epochs \citep[][see their Fig.~5]{2012MNRAS.425.1789S},
and so this classification is somewhat suspect and this event may also have been SN~1999aa-like.
These updates indicate that 91T/99aa-like events exhibit a continuum of spectroscopic properties, with
normal SNe Ia at one end and SN~1991T-like events at the other extreme; most ``shallow silicon'' events
fall somewhere in between.

\subsection{SNe Ic-BL}

L11 grouped the broad-lined SNe~Ic
\citep[SN~Ic-BL, sometimes associated with gamma-ray bursts; e.g.,][]{2006ARAA..44..507W}
into the SN~Ic-pec subclass, though they noted in the text
that SN~2002ap is a member of that group \citep[e.g.,][]{2002ApJ...572L..61M}.
As discussed in \S\ref{sec:lowcert}, SNe 2002jj and 2006eg may plausibly also be of Type Ic-BL.

\subsection{SN~1987A-like SNe}
\label{sec:87A}

As discussed in \S\ref{sec:sn2005io}, SN~2005io was very likely a SN~1987A-like event
\citep[for reviews of SN~1987A and related events, see, e.g.,][]{1989ARAA..27..629A,1993ARAA..31..175M}.
The LOSS volume-limited sample also includes the SN~1987A-like SNe~2000cb and 2005ci \citep{2011MNRAS.415..372K}.
All of these objects were grouped with the Type IIP SNe by L11.

\subsection{Ca-Rich Transients}
\label{sec:carich}

There are three examples of the recently identified class of ``Ca-rich'' SNe in our sample:
SNe 2003H, 2003dr, and 2005E
\citep[e.g.,][]{2003IAUC.8159....2F,2010Natur.465..322P,2012ApJ...755..161K,2015MNRAS.452.2463F}.
All three were identified by \citet{2010Natur.465..322P}.
Though discussed within the text as Ca-rich events, these three SNe were labeled SN~Ibc-pec
by L11 and were grouped with the other stripped-envelope SNe in their analysis.

\begin{figure}[ht]
\epsscale{1.0}
\plotone{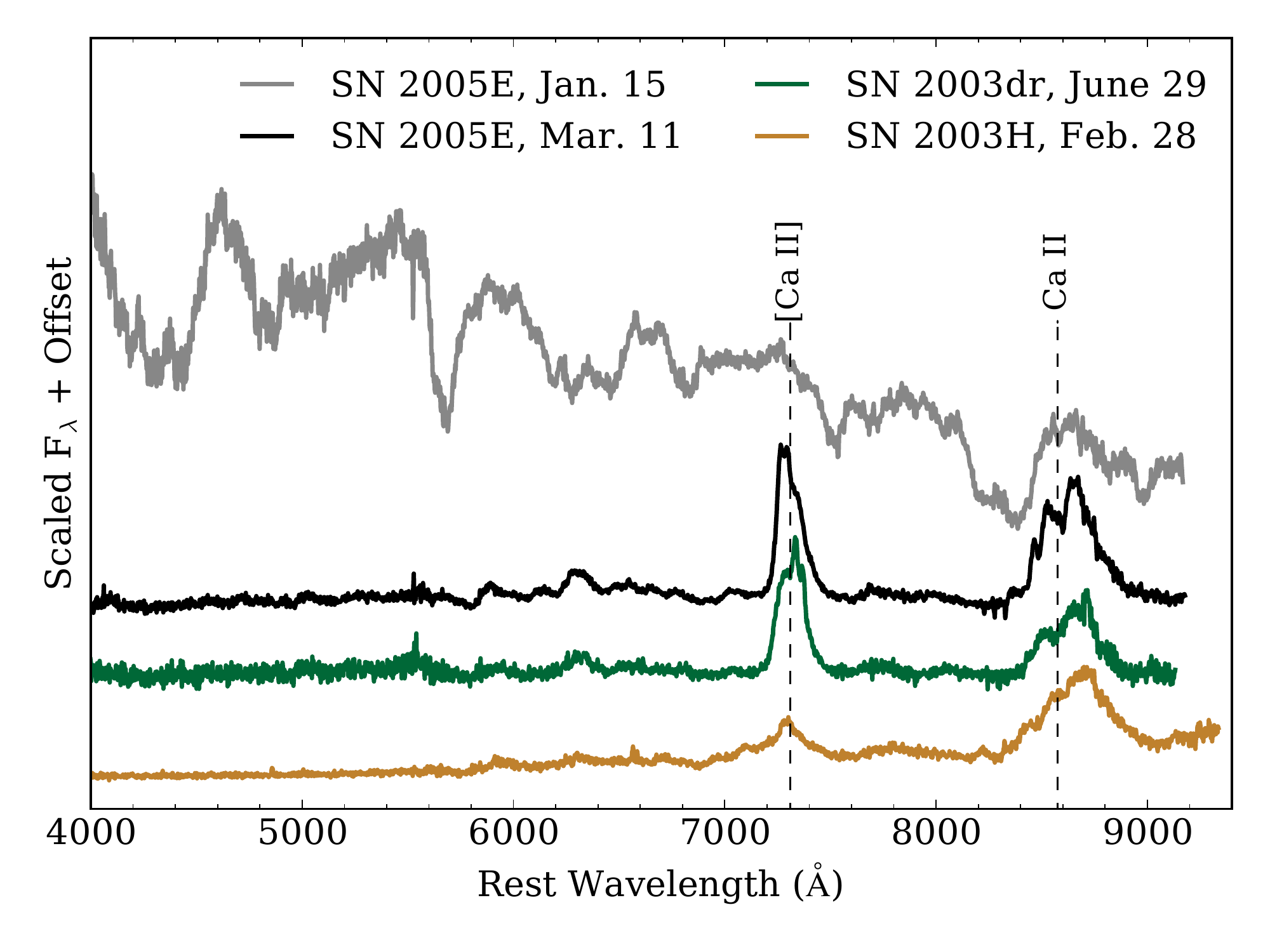}
\caption{Spectra of the Ca-rich objects in our sample.  For SN~2005E 
    we present photospheric and nebular spectra; for the other objects
    we only have nebular spectra.
    \label{fig:ca-rich}}
\end{figure}

Though the provenance of these events is not fully understood,
it now seems likely that they do not arise from the core collapse of massive stars. Removing
these three events from the sample of core-collapse SNe slightly reduces
the ratio of stripped-envelope SNe relative to Type II SNe.
Figure~\ref{fig:ca-rich} shows spectra of all three Ca-rich SNe in the sample.
We note that the photospheric spectra of these events are extremely similar to
those of normal SNe Ib; it is their nebular spectra, their rapid evolution, and their
low peak luminosities that primarily differentiate these events.

\subsection{Type IIn SNe and SN Impostors}
\label{sec:2ni}

Type IIn SNe are hydrogen-rich SNe that exhibit narrow lines in their spectrum ---
indicative of dense circumstellar material surrounding the progenitor at the
time of explosion \citep[see, e.g.,][]{1989AJ.....97..726F,1990MNRAS.244..269S,1991MNRAS.250..513C,2014ARAA..52..487S}.
There were seven Type IIn SNe in the sample identified by L11, 
but two of those (SNe 2002bu and 2006bv) were reclassified as 
SN impostors (luminous but nonterminal outbursts from massive stars)
by \citet{2011MNRAS.415..773S}.  We group SNe 2002bu and 2006bv with
the five other SN impostors from the original sample and do not include them when
calculating the relative fractions of SNe.

\subsection{SNe That Lack Spectra}
\label{sec:nospec}

\begin{figure}[ht]
\epsscale{1.0}
\plotone{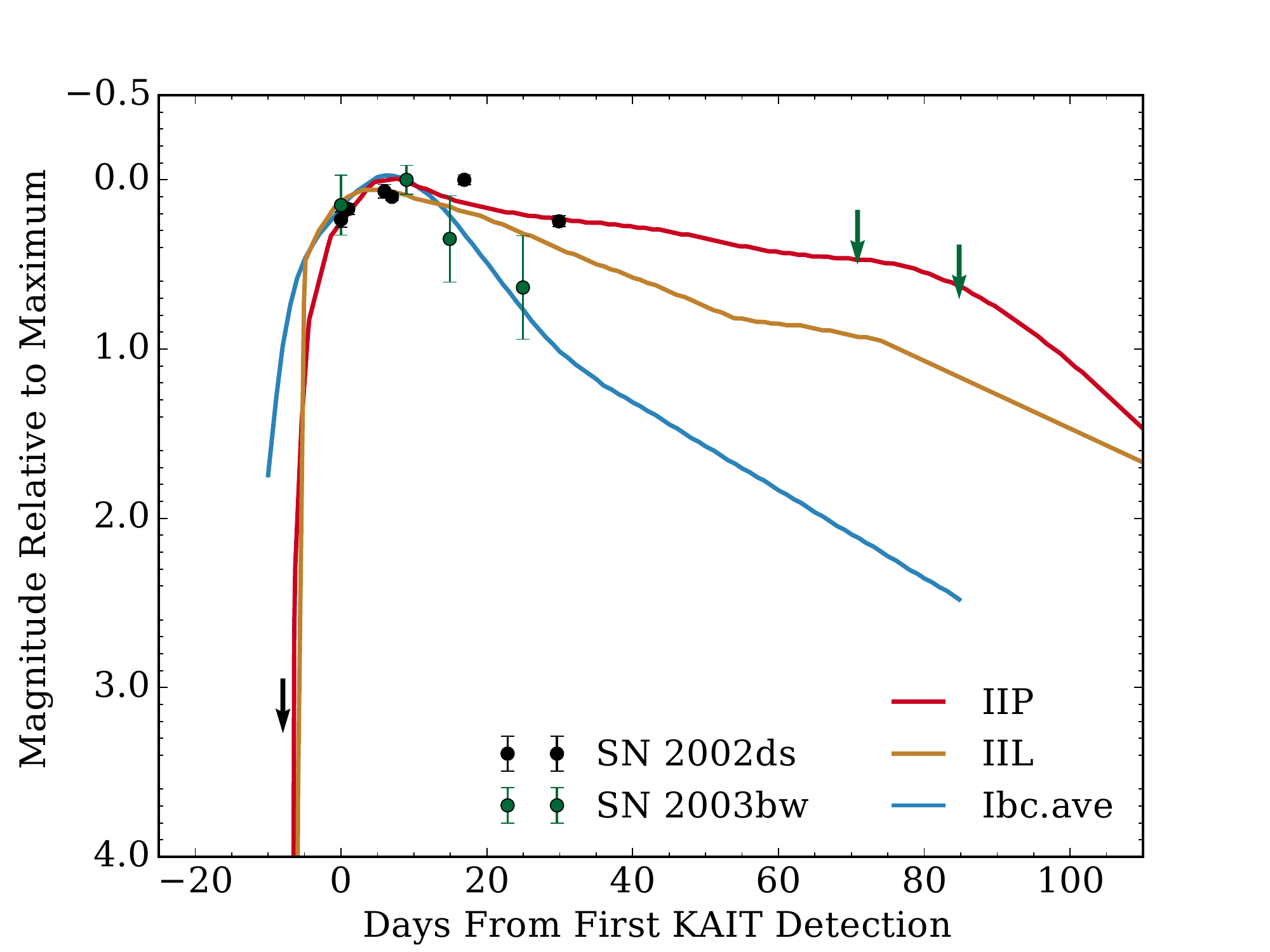}
\caption{Unfiltered light curves for two SNe for which we have not been
    able to collect spectra, along with template light curves from L11 for comparison.
    Upper limits are shown with arrows. For SN~2002ds, the CBET classification of 
    a Type II SN with a hydrogen-recombination plateau is robustly supported.
    The light curve of SN~2003bw appears to match the stripped-envelope template from L11
    better than the Type II templates, but the data are noisy and are consistent with either classification.
    SN~2006bv (not shown) was likely a SN impostor; see \citet{2011MNRAS.415..773S}.
    \label{fig:nospec}}
\end{figure}

Every object in the L11 volume-limited sample was originally classified spectroscopically
and announced through CBETs, but we have not been able to track down spectra for three events:
see Table~\ref{tab:nospec} and Figure~\ref{fig:nospec}.
One of these (SN~2002ds) exhibits a light curve with a pronounced plateau, which corroborates the original
CBET classification of a Type II SN.
The light curve of SN~2003bw, however, does
not rule out the possibility that this event was a low-hydrogen Type IIb SN
\citep[assuming that the hydrogen detection announced in the CBET is robust;][]{2003IAUC.8103....2H}.
For SN~2006bv we adopt the SN impostor reclassification of \citet{2011MNRAS.415..773S}.

\begin{table}[ht] % wrapped in a table so it will float
\begin{deluxetable}{ l | c c c }
\tablecaption{SNe That Lack Spectra \label{tab:nospec}}
\tablehead{
  \colhead{Name} & \colhead{Previous (L11)} & 
  \colhead{This Work} & \colhead{Ref.} 
}
\startdata
SN 2002ds & IIP & II & 1 \\ 
SN 2003bw & IIP & II/IIb & 2  \\  % light curve is inconclusive (4 points)
SN 2006bv & IIn  & {\it impostor}  & 3,4 \\ % discussed, but we have no spectra of this object
\enddata
\vspace{5mm}
{\bf References:}
[1]~\citet{2002IAUC.7929....3L};
[2]~\citet{2003IAUC.8103....2H};
[3]~\citet{2006CBET..493....1S};
[4]~\citet{2011MNRAS.415..773S}
\end{deluxetable}
\end{table}

\section{Updated Fraction Calculations}
\label{sec:rates}

\begin{figure*}
\epsscale{1.0}
\plotone{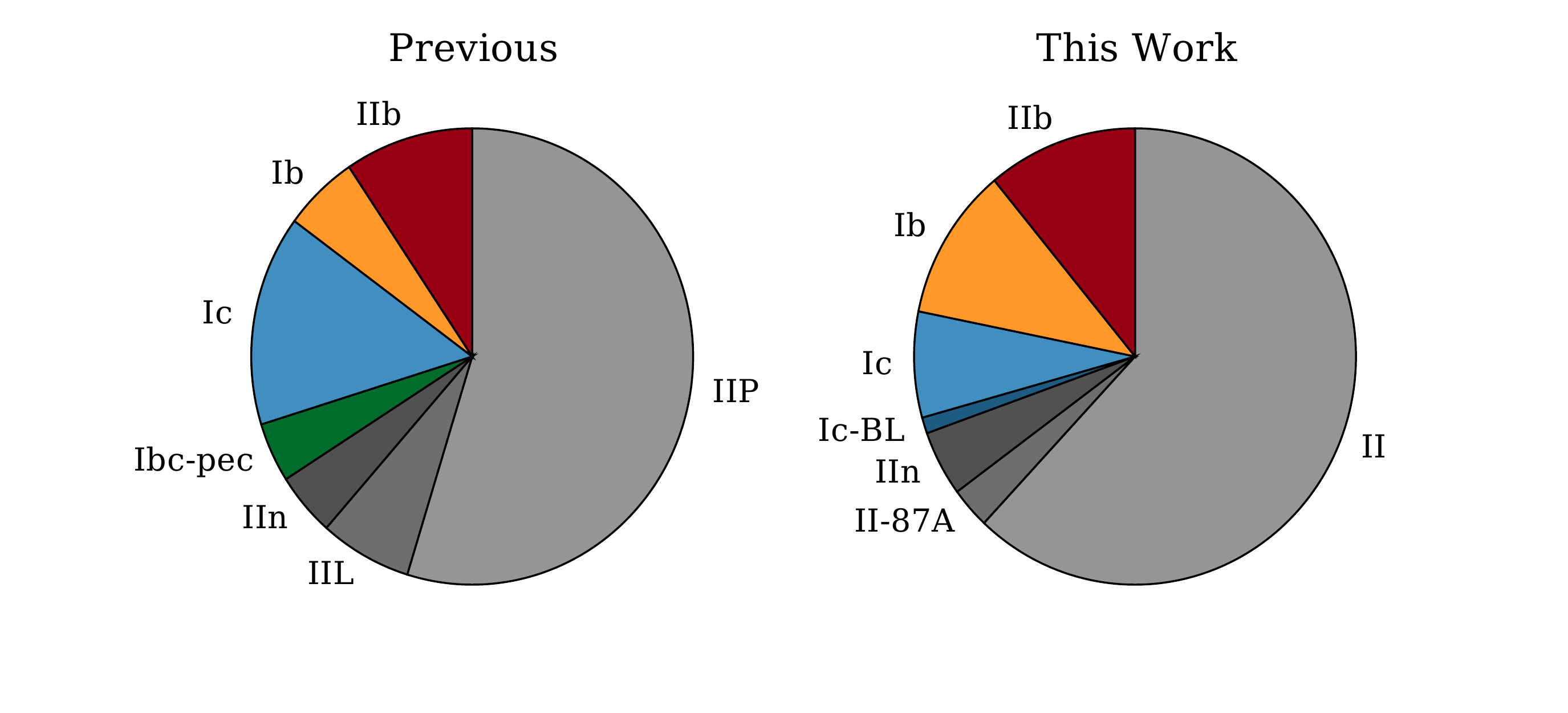}
\caption{Relative fractions of core-collapse SN types within a volume-limited sample
    using the original classifications from L11 (left) compared to the
    updated classifications presented here (right). 
    Subtypes are color-coded along with the other members of 
    their major type, and 
    the ``peculiar'' subtype labels are grouped with the appropriate ``normal'' events
    (except for the SN~Ibc-pec group of L11, which included both SNe~Ic-BL and Ca-Rich transients).
     All fractions are listed in Table~\ref{tab:fractions}
    and any objects listed in Table~\ref{tab:all} with more than one possible
    classification are given a fractional weight in each class, as described
    in \S\ref{sec:rates}.
    \label{fig:pie}}
\end{figure*}

Figure~\ref{fig:pie} and Table~\ref{tab:fractions} summarize our updated fraction calculations.  We follow L11 to estimate our uncertainties, running $10^6$ Monte Carlo
realizations of the sample assuming Poisson statistics and the control-time corrections from L11, \citet{2011MNRAS.412.1419L}, and \citet{2011MNRAS.412.1473L}.
(Note that the stated uncertainties are statistical only.)
We do not recalculate the control times for each event after these updated classifications. Almost all of our updates are swaps
between stripped-envelope subtypes which show very similar light curves, and so the control-time calculations should change very little.
For events with more than one possible classification listed in Table~\ref{tab:all} we assign a fractional weight to each given the 
relative frequencies of the subtypes among the well-classified events.

Many authors, based on both theory and observation, argue for trends in the relative SN rates as a function of metallicity
\citep[e.g.,][]{2008AJ....135.1136M,2010ApJ...721..777A,2011ApJ...731L...4M,2012ApJ...759..107K,2015PASA...32...15Y}.
Adopting the metallicity-luminosity relation of \citet{2002ApJ...581.1019G} and noting that the galaxies hosting core-collapse SNe within the LOSS sample
cover a range of luminosities of $M_K \approx -22$ to $-25$\,mag, \citet{2011MNRAS.412.1522S} estimate metal abundances of
$\sim0.5$--2.0\,Z$_{\odot}$ for these galaxies, and our results are therefore applicable to roughly that range.
\citet{2016arXiv160902921G,2016arXiv160902923G} use the LOSS sample and our updated classifications
to examine correlations between SN rates and host galaxy properties in more detail, 
including the stellar masses, specific star-formation rates, and oxygen abundances (i.e., metallicities) of the host galaxies.
Several authors have shown that local measures of host-galaxy metallicities are more informative than global ones
\citep[e.g.,][]{2008AJ....135.1136M,2010MNRAS.407.2660A,2011ApJ...731L...4M,2016AA...589A.110A};
a study of the explosion-site metallicities of the LOSS sample is a worthy endeavor we leave to future work.

% 1000000 samples

\begin{table}[ht] % wrapped in a table environment so it will float (deluxetable does not float)
\begin{deluxetable}{ l | c c c }
\tablecaption{Updated Relative SN Fractions in a Volume-Limited Survey \label{tab:fractions}}
\tablehead{ 
\colhead{Type} & \colhead{Previous} & \colhead{This Work} & \colhead{Difference}
}
\startdata
\hline 
 & \multicolumn{2}{c}{{\bf Core Collapse}} & \\
\hline
II & $68.9^{+6.0}_{-6.0}$ & $69.6^{+6.7}_{-6.7}$ & - \\ 
IIb+Ib+Ic & $31.1^{+4.6}_{-4.6}$ & $30.4^{+5.0}_{-4.9}$ & - \\ 
\hline
 & \multicolumn{2}{c}{{\bf Stripped Envelope}} & \\
\hline
IIb & $27.6^{+9.1}_{-9.1}$ & $34.0^{+11.1}_{-11.1}$ & $+6.3$ \\ 
IIb-pec &  -  & $2.0^{+1.5}_{-2.0}$ & - \\ 
Ib & $16.1^{+6.8}_{-6.6}$ & $35.6^{+11.4}_{-11.4}$ & $+19.5$ \\ 
Ib-pec &  -  &  -  & - \\ 
Ibc-pec$^{\alpha}$ & $12.4^{+5.9}_{-5.6}$ &  -  & - \\ 
Ic & $41.1^{+11.5}_{-11.4}$ & $21.5^{+8.6}_{-8.6}$ & $-19.6$ \\ 
Ic-pec & $2.8^{+2.6}_{-2.8}$ & $3.2^{+3.1}_{-3.2}$ & - \\ 
Ic-BL &  -  & $3.7^{+2.9}_{-3.7}$ & - \\ 
\hline
 & \multicolumn{2}{c}{{\bf Hydrogen Rich}} & \\
\hline
II$^{\beta}$ & $93.2^{+11.5}_{-11.3}$ & $89.1^{+10.9}_{-10.9}$ & - \\ 
II-87A &  -  & $4.2^{+2.4}_{-2.7}$ & - \\ 
IIn & $6.8^{+3.0}_{-2.9}$ & $6.7^{+3.0}_{-2.9}$ & - 
\enddata
\vspace{5mm}
Relative fractions of core-collapse SNe in the LOSS volume-limited sample, within several
    different subsets, expressed in percentages.
    In the left column we present the fractions assuming the
    original classifications used by L11, in the center
    column we present our updated fractions, and
    in the right column we highlight the most notable updates. \\ \\
$^{\alpha}$L11 included SNe Ic-BL and Ca-Rich transients with the Ibc-pec class.
        In our updated fractions we list the SNe Ic-BL seperately, and we do not group the Ca-Rich events with core-collapse SNe. \\
$^{\beta}$Including the II-L and II-P subclasses of L11. 
\end{deluxetable}
\end{table}

As with L11, our calculations for several of the rarer subtypes suffer from the effects of small-number statistics, but
Table~\ref{tab:fractions} and Figure~\ref{fig:pie} indicate an important update: the percentage of SNe Ic 
is reduced while the percentage of SNe Ib is increased.
Adopting our updated classifications, 83\% of our Monte Carlo trials indicate that normal SNe~Ib are more common than normal SNe~Ic,
while 99\% of the trials using the L11 classifications indicate the opposite.

L11 and \citet{2011MNRAS.412.1522S} found that SNe Ic are more than twice as common as SNe Ib
(grouping the SNe Ic-BL with the SNe Ic, which only affects these rates by a small amount).
L11 calculate a ratio of SNe~Ic/SNe~Ib = $54.2 \pm 9.8\% / 21.2^{+8.4}_{-7.7}\% = 2.6 \pm 1.1$, while
\citet[excluding SNe from highly inclined galaxies]{2011MNRAS.412.1522S}
calculated SNe Ic/Ib = $14.9^{+4.2}_{-3.8}\% / 7.1^{+3.1}_{-2.6}\% = 2.1 \pm 1.1$
(in all cases the errors listed are statistical only, and were derived from Monte Carlo simulations 
similar to those described above).

We now calculate a ratio of normal SNe Ic to normal SNe Ib of $0.6 \pm 0.3$ and, if
we include the SNe Ic-BL and other peculiar subtypes with the normal SNe Ib and Ic,
we find a (SN~Ic+Ic-BL+Ic-pec)/SN~Ib ratio of  $0.8 \pm 0.4$.

This update to the population fractions is driven by our reclassifications of seven stripped-envelope events.
First, we relabeled four events from a Ic subtype to a Ib or IIb subtype (SNe 2001M, 2001ci, 2004C, and 2005lr).
In each of these cases, the need for reclassification is easily understood:
three of these events had spectra severely reddened by host-galaxy dust and were originally classified by eye
without the aid of SNID, and one showed only weak \ion{He}{1} lines in the spectrum (SN 2001M). 
Second, we created the SN~Ib/Ic (unsure) category, which includes an additional two events
that show weak \ion{He}{1} lines with some uncertainty on their identification (SNe 2002jz and 2004cc)
and one event with only sparse and noisy observations (SN~2006eg).
If we assume that all of the SNe in the latter category deserve the Ic label, our Monte Carlo trials indicate that
normal SNe Ib and SNe Ic (excluding peculiar subtypes) occur at similar rates: SNe Ic/Ib = $0.9 \pm 0.5$.
If we rather assume that they are all SNe Ib, we get a ratio of normal SNe Ic/Ib = $0.5 \pm 0.3$.

These results have implications for our understanding of the progenitors of stripped-envelope SNe, as we discuss below, and
may affect other works that use the LOSS rates as input \citep[e.g.,][]{2013ApJ...778..167F}.

\section{Progenitor Constraints on Stripped-Envelope SNe}
\label{sec:progenitor}

Wolf-Rayet (WR) stars have long been discussed as Galactic analogues of SN~Ib/c progenitors
\citep[e.g.,][]{2003AA...404..975M,2007ARAA..45..177C}, though many authors have argued
that binary stars which undergo mass loss via Roche-lobe overflow before core collapse
are likely the most common SN Ib/c progenitor \citep[e.g.,][]{1992ApJ...391..246P,2009ARAA..47...63S,2011MNRAS.412.1522S,2013MNRAS.436..774E}.
Regardless, stellar modeling efforts have found it difficult to match the SN Ic/Ib fractions presented
by L11 and \citet{2011MNRAS.412.1522S}, which demand more SN Ic progenitors (stars that lose
both their hydrogen envelope and a large fraction of their helium envelopes) than Ib progenitors
 \citep[stars that lose just the hydrogen; e.g.,][]{2009AA...502..611G,2010ApJ...725..940Y,2015PASA...32...15Y},
though some success has been achieved by invoking rapid rotation of the progenitors \citep[e.g.,][]{2013ApJ...775L...7C,2013AA...558A.131G}.

To address this putative issue, some authors have proposed that some amount of helium in SNe Ic may be ``hidden'' and remain neutral
if the $^{56}$Ni (which provides nonthermal excitations via radioactive decay) is insufficiently mixed with the helium-rich ejecta
\citep[e.g.,][]{2011MNRAS.414.2985D,2012MNRAS.424.2139D}.
Comparisons to observation do not find evidence for large amounts of hidden helium in SNe Ic, however,
and it is unclear from the models how much helium could truly be hidden in this way
\citep[e.g.,][]{2012MNRAS.422...70H,2015AA...574A..60T,2016ApJ...827...90L}.
Our updated stripped-envelope fractions argue that this problem is less egregious than previously indicated.

Other discrepancies have arisen within the single WR-like progenitor scenario.
The observed ejecta masses of
normal SNe~Ib/c \citep[$M_{\rm ej} \approx 2.0$--4\,M$_{\odot}$;][]{2011ApJ...741...97D,2013MNRAS.434.1098C,2016MNRAS.457..328L}
are not in good agreement with the estimated masses of WR stars at the time of core collapse
\citep[$M \gtrsim 10$\,M$_{\odot}$;][]{2003AA...404..975M,2015PASA...32...15Y}, assuming that
most SNe~Ib/c produce neutron star remnants rather than black hole remnants.
Note that SNe~Ic-BL may have larger ejecta masses, and so this reasoning holds for normal SNe Ib/c only \citep{2013MNRAS.434.1098C}.
In addition, the rates of SNe~Ib/c compared to those of Type II SNe are inconsistent with WR star progenitors
\citep[the incidence rate of WR stars is too low to explain the high fraction of SNe~Ib/c; e.g.,][]{2011MNRAS.412.1522S}, and the search for 
SN~Ib/c progenitors in pre-explosion images has, in several instances, ruled out normal WR stars
(\citealp{2013MNRAS.436..774E}, though see also \citealp{2013AA...558A.131G}).

The binary progenitor system scenario for normal SNe Ib/Ic does not suffer from the same problems.
The modeled masses at the time of core collapse for post-mass-transfer binary members are generally in agreement with the observed 
SN~Ib/c ejecta masses \citep{2013MNRAS.436..774E,2015PASA...32...15Y}, constraints from SN~Ib/c progenitor searches are largely compatible
with binary progenitors \citep{2013MNRAS.436..774E}, and the identified progenitors of some SNe~IIb have been
shown to be the products of binary evolution \citep[SNe 1993J and 2011dh; e.g.,][]{2004Natur.427..129M,2011ApJ...739L..37M,2011ApJ...741L..28V,2012ApJ...757...31B}.
%\citep[Note, however, that some exceptional events remain plausibly understood as the explosions of WR stars; e.g.,][]{2013ApJ...775L...7C,2016AA...592A..89T}.

\citet{2012Sci...337..444S} show that more than 70\% of O-type stars
(zero-age main sequence $M \gtrsim 15$\,M$_{\odot}$) are formed
in binary pairs that will undergo significant binary interaction (either mass gain or mass stripping)
before core collapse, and so the population of core-collapse SN progenitors must necessarily be dominated by post-binary-interaction stars.  
We calculate a stripped-envelope fraction of $30 \pm 5$\% amongst all core-collapse SNe, similar to fractions found by previous authors. As \citet{2011MNRAS.412.1522S} note,
this value is in remarkably good agreement with the $\sim$33\% of O-type stars in our Galaxy found to experience envelope stripping
via binary interaction before their deaths \citep{2012Sci...337..444S}.

\section{Conclusion}
\label{sec:conclusion}

We have re-examined every SN classification within the LOSS volume-limited sample published by L11,
have discussed the peculiar and rare events within the sample,
and have found that several of the stripped-envelope SNe originally labeled as SNe Ic show clear signatures
of helium and (in two cases) hydrogen.
After relabeling these SNe as Type Ib or IIb appropriately, and discussing
the intrinsically peculiar events and those for which we cannot assign a clear classification,
we recalculate the implied fractions of these subtypes.
We find that the relative fractions of Type Ia SNe, Type II SNe, and stripped-envelope SNe 
are unchanged, but the relative fractions between different stripped-envelope SN subtypes are.

Based on the prior spectral identifications, L11 and \citet{2011MNRAS.412.1522S}
found that SNe~Ic are roughly twice as common as SNe~Ib.
We show that this measurement was hampered by the above misclassifications and,
additionally, that the SN~Ib/SN~Ic ratio is strongly dependent on exactly where one draws the line
between these subclasses.  We find
that SNe Ib are at least as common as SNe Ic in the local universe and in fact are likely to be more common.
We present a best-estimate normal SN~Ic/SN~Ib ratio of  $0.6 \pm 0.3$ ---
i.e., spectroscopically normal SNe Ib occur in the local universe $1.7 \pm 0.9$ times more often than do normal SNe Ic.

Other efforts \citep[e.g.,][]{2009MNRAS.395.1409S,2013MNRAS.436..774E} found SN~Ic/SN~Ib ratios similar to those of L11, and we believe 
they may also have been plagued by systematically mislabeled stripped-envelope events.
The updated stripped-envelope SN fractions published here should prove important for constraining the elusive progenitors of the
various subtypes of stripped-envelope SNe,
and we hope the public release of these data will be useful when exploring this valuable sample going forward.

\acknowledgments

We thank the many observers who assisted in obtaining the data published here,
especially the UC Berkeley undergraduates who have worked to discover new 
SNe within the KAIT data.
We are grateful to J.~Mauerhan, M.~Graham, P.~Kelly,
P.~Challis, R.~McCray, and I.~Kleiser
for useful discussions, and to the researchers
who shared their archival data so as to make this project possible,
including M.~Phillips, M.~Hamuy, R.~Kirshner, C.~Li, and Y. L.~Qiu.
W.~Li is remembered for his friendship, his tireless 
and excellent work on LOSS, and his
many contributions to our understanding of SNe.

Some of the data presented herein were obtained at the W. M. Keck
Observatory, which is operated as a scientific partnership among the
California Institute of Technology, the University of California, and
NASA; the observatory was made possible by the generous financial
support of the W. M. Keck Foundation.  KAIT and its ongoing operation
were made possible by donations from Sun Microsystems, Inc., the
Hewlett-Packard Company, AutoScope Corporation, Lick Observatory, the
National Science Foundation (NSF), the University of California, the Sylvia and Jim Katzman
Foundation, and the TABASGO Foundation.  Research at Lick Observatory
is partially supported by a generous gift from Google.
Some of the data presented herein were collected with the Copernico telescope (Asiago, Italy) of the
INAF -- Osservatorio Astronomico di Padova and the Galileo telescope (Asiago, Italy) of the
Dipartimento di Fisica e Astronomia -- Padova University.
This research has made use of the NASA/IPAC Extragalactic Database (NED) which is operated by 
the Jet Propulsion Laboratory, California Institute of Technology, under contract with NASA.
IRAF is distributed by the National Optical Astronomy Observatory, which is operated by the
Association of Universities for Research in Astronomy (AURA) under a cooperative agreement
with the NSF.

A.V.F.'s SN group at UC Berkeley has received generous financial assistance from the Christopher R. Redlich Fund, the TABASGO Foundation, and NSF grant AST-1211916. 
M.~Modjaz and the SNYU group are supported in part by NSF CAREER award AST-1352405 and by NSF award AST-1413260.
Y.~Liu is supported in part by a NYU/CCPP James Arthur Graduate Fellowship.
J.~M.~Silverman is supported by an NSF Astronomy and Astrophysics Postdoctoral Fellowship under award AST-1302771.
R.J.F.\ gratefully acknowledges support from NSF grant AST-1518052, the Alfred P.\ Sloan Foundation, and the David and Lucile Packard Foundation.
O.G. is supported in part by NSF award AST-1413260 and by an
NSF Astronomy and Astrophysics Fellowship under award AST-1602595.
S.~Benetti and A.~Pastorello are partially supported by the
PRIN-INAF 2014 project Transient Universe: unveiling new types of stellar explosions with PESSTO.

\bibliographystyle{apj}
\bibliography{bib}

\appendix
\section{Journal of Data Presented Here}
\label{appendix}
Table~\ref{tab:allspecs} lists every spectrum published here for the first time: a total of 151 spectra of 71 SNe.
Table~\ref{tab:allphot} lists all light curves rereduced from images and published here, including data for 20 SNe.
See \S\ref{sec:data} for a description of the observing and data-acquisition efforts.  All data will be made public
via the Berkeley SNDB (\url{http://heracles.astro.berkeley.edu/sndb}), WiseREP (\url{http://wiserep.weizmann.ac.il}),
and the Open Supernova Catalog (\url{https://sne.space/}).

\begin{longtable}{ l l | l c c c | c }
\caption{Log of Spectra Published Herein\label{tab:allspecs}} \\
\hline
\multirow{ 2}{*}{SN Name} & \multirow{ 2}{*}{UT Date} & \multirow{ 2}{*}{File Name$^{\alpha}$} &
\multirow{ 2}{*}{Instrument$^{\beta}$} & WL Range & Resolution$^{\gamma}$ & \multirow{ 2}{*}{Source$^{\beta}$} \\
                          &                           &                                        &
                                       & ($\rm{\AA}$) & ($\rm{\AA}$)      &                  \\
\hline
\endfirsthead

\multicolumn{7}{c}{{\bfseries \tablename\ \thetable{} -- continued from previous page}} \\
\hline 
\multirow{ 2}{*}{SN Name} & \multirow{ 2}{*}{UT Date} & \multirow{ 2}{*}{File Name$^{\alpha}$} &
\multirow{ 2}{*}{Instrument$^{\beta}$} & WL Range & Resolution$^{\gamma}$ & \multirow{ 2}{*}{Source$^{\beta}$} \\
                          &                           &                                        &
                                       & ($\rm{\AA}$) & ($\rm{\AA}$)      &                  \\

\hline 
\endhead

\hline
\multicolumn{7}{r}{{Continued on next page}} \\
\hline
\endfoot

\hline \hline
\multicolumn{7}{l}{$^{\alpha}$Different groups utilize different naming conventions for their data; we preserve these differences and the original names of these files.} \\
\multicolumn{7}{l}{$^{\beta}$See \S\ref{sec:data} for a description of the instruments and observational efforts listed here.} \\
\multicolumn{7}{l}{$^{\gamma}$Listed resolutions are estimates of the average resolution for the instrument (if two resolutions are given, they refer to the blue side} \\
\multicolumn{7}{l}{\phantom{$\gamma$}and red side of the spectrograph separately).} \\
\multicolumn{7}{l}{$^{\delta}$Traced from image of plot; see \S\ref{sec:05mg}.} \\
\endlastfoot
SN 1999an & 1999-03-10 & sn1999an-19991003.flm & OMR & 3610--8580\phantom{,}\phantom{1} & 10 & NAOC \\ 
SN 1999br & 1999-04-24.0 & sn1999br-19990424-opt.flm & Kast & 4300--7000\phantom{,}\phantom{1} & 6/5 & UCB \\ 
SN 1999bu & 1999-04-18.21 & sn1999bu-19990418.flm & FAST & 3720--7540\phantom{,}\phantom{1} & 7 & CfA \\ 
SN 1999cd & 1999-05-15.32 & sn1999cd-19990515.flm & FAST & 3720--7540\phantom{,}\phantom{1} & 7 & CfA \\ 
SN 1999cd & 1999-05-16.27 & sn1999cd-19990516.flm & FAST & 3720--7540\phantom{,}\phantom{1} & 7 & CfA \\ 
SN 1999el & 1999-11-05.0 & sn1999el-19991105-ui.flm & Kast & 3380--10,460 & 6/11 & UCB \\ 
SN 1999gi & 1999-12-10.0 & sn1999gi-19991210.flm & Kast & 3720--7540\phantom{,}\phantom{1} & 6/5 & UCB \\ 
SN 1999gi & 1999-12-12.0 & sn1999gi-19991212.flm & Kast & 3720--7540\phantom{,}\phantom{1} & 6/5 & UCB \\ 
SN 1999gi & 1999-12-13.0 & sn1999gi-19991213.flm & Kast & 3720--7540\phantom{,}\phantom{1} & 6/5 & UCB \\ 
SN 1999gi & 2000-01-05.0 & sn1999gi-20000105.flm & Kast & 3720--7540\phantom{,}\phantom{1} & 6/5 & UCB \\ 
SN 1999gi & 2000-01-10.0 & sn1999gi-20000110.flm & Kast & 3720--7540\phantom{,}\phantom{1} & 6/5 & UCB \\ 
SN 1999gi & 2000-01-13.0 & sn1999gi-20000113.flm & Kast & 3720--7540\phantom{,}\phantom{1} & 6/5 & UCB \\ 
SN 1999gi & 2000-03-04.0 & sn1999gi-20000304.flm & Kast & 3720--7540\phantom{,}\phantom{1} & 6/5 & UCB \\ 
SN 1999gi & 2000-03-08.0 & sn1999gi-20000308.flm & Kast & 3720--7540\phantom{,}\phantom{1} & 6/5 & UCB \\ 
SN 1999gi & 2000-03-15.0 & sn1999gi-20000315-ui.flm & Kast & 3300--10,500 & 6/11 & UCB \\ 
SN 1999gi & 2000-03-25.0 & sn1999gi-20000325-opts.flm & LRIS & 4380--6840\phantom{,}\phantom{1} & 7 & UCB \\ 
SN 1999gi & 2000-03-29.0 & sn1999gi-20000329-ui.flm & Kast & 3300--10,500 & 6/11 & UCB \\ 
SN 1999gi & 2000-04-25.0 & sn1999gi-20000425.flm & Kast & 3720--7540\phantom{,}\phantom{1} & 6/5 & UCB \\ 
SN 1999gi & 2000-04-27.0 & sn1999gi-20000427-ui.flm & Kast & 3300--10,350 & 6/11 & UCB \\ 
SN 1999gi & 2000-05-10.0 & sn1999gi-20000510.flm & Kast & 3720--7540\phantom{,}\phantom{1} & 6/5 & UCB \\ 
SN 1999gi & 2000-05-26.0 & sn1999gi-20000526.flm & Kast & 3720--7540\phantom{,}\phantom{1} & 6/5 & UCB \\ 
SN 1999go & 1999-12-28 & 1999go\_19991228.flm & DFOSC & 3330--9040\phantom{,}\phantom{1} & 14 & Asiago \\ 
SN 1999go & 1999-12-29 & 1999go\_19991229.flm & DFOSC & 3330--9040\phantom{,}\phantom{1} & 14 & Asiago \\ 
SN 1999go & 2000-01-02.23 & sn1999go-20000102.flm & FAST & 3720--7540\phantom{,}\phantom{1} & 7 & CfA \\ 
SN 2000C & 2000-01-25 & 2000C\_20000125.flm & B\&C$_{1.2}$ & 4490--9090\phantom{,}\phantom{1} & 22 & Asiago \\ 
SN 2000C & 2000-01-27 & 2000C\_20000127.flm & AFOSC & 3520--7560\phantom{,}\phantom{1} & 18 & Asiago \\ 
SN 2000C & 2000-01-28.14 & sn2000c-20000128.flm & FAST & 3720--7540\phantom{,}\phantom{1} & 7 & CfA \\ 
SN 2000C & 2000-01-29 & 2000C\_20000129.flm & AFOSC & 3520--7490\phantom{,}\phantom{1} & 18 & Asiago \\ 
SN 2000C & 2000-01-29.20 & sn2000c-20000129.flm & FAST & 3720--7540\phantom{,}\phantom{1} & 7 & CfA \\ 
SN 2000C & 2000-02-01 & 2000C\_20000201.flm & AFOSC & 3550--7490\phantom{,}\phantom{1} & 18 & Asiago \\ 
SN 2000C & 2000-02-05.25 & sn2000c-20000205.flm & FAST & 3720--7540\phantom{,}\phantom{1} & 7 & CfA \\ 
SN 2000C & 2000-02-11 & 2000C\_20000211.flm & AFOSC & 3620--7590\phantom{,}\phantom{1} & 18 & Asiago \\ 
SN 2000L & 2000-03-01.29 & sn2000l-20000301.flm & FAST & 3720--7540\phantom{,}\phantom{1} & 7 & CfA \\ 
SN 2000L & 2000-03-15.0 & sn2000l-20000315-ui.flm & Kast & 3300--10,500 & 6/11 & UCB \\ 
SN 2000N & 2000-03-08.43 & sn2000n-20000308.flm & FAST & 3720--7540\phantom{,}\phantom{1} & 7 & CfA \\ 
SN 2000N & 2000-03-09 & 2000N\_20000309.flm & DFOSC & 3850--6840\phantom{,}\phantom{1} & 6 & Asiago \\ 
SN 2000N & 2000-03-13 & 2000N\_20000313.flm & DFOSC & 3550--9040\phantom{,}\phantom{1} & 11 & Asiago \\ 
SN 2000el & 2000-11-29.0 & sn2000el-20001129-ui.flm & Kast & 3300--10,450 & 6/11 & UCB \\ 
SN 2000eo & 2000-12-21.0 & sn2000eo-20001221-ur.flm & Kast & 3250--7810\phantom{,}\phantom{1} & 6/5 & UCB \\ 
SN 2000ex & 2000-11-28.29 & sn2000ex-20001128.flm & FAST & 3760--7540\phantom{,}\phantom{1} & 7 & CfA \\ 
SN 2000ex & 2000-11-29.0 & sn2000ex-20001129-ur.flm & Kast & 3300--7800\phantom{,}\phantom{1} & 6/5 & UCB \\ 
SN 2001J & 2001-01-19.45 & sn2001j-20010119.flm & FAST & 3720--7540\phantom{,}\phantom{1} & 7 & CfA \\ 
SN 2001K & 2001-03-30.0 & sn2001k-20010330-ur.flm & Kast & 3300--7800\phantom{,}\phantom{1} & 6/5 & UCB \\ 
SN 2001M & 2001-02-01.0 & sn2001m-20010201-ui.flm & Kast & 3260--10,600 & 6/11 & UCB \\ 
SN 2001Q & 2001-02-01.0 & sn2001q-20010201-ui.flm & Kast & 3260--10,600 & 6/11 & UCB \\ 
SN 2001ac & 2001-03-14.45 & sn2001ac-20010314.flm & FAST & 3720--7540\phantom{,}\phantom{1} & 7 & CfA \\ 
SN 2001ac & 2001-03-25.36 & sn2001ac-20010325-blue-mmt.flm & MMT-Blue & 3250--8850\phantom{,}\phantom{1} & 8 & CfA \\ 
SN 2001ci & 2001-05-30.285 & sn2001ci-com-20010530.285-joined.flm & ESI & 3920--10,190 & 0.5 & UCB \\ 
SN 2001fz & 2001-11-19.50 & sn2001fz-20011119.flm & FAST & 3720--7540\phantom{,}\phantom{1} & 7 & CfA \\ 
SN 2001hf & 2002-01-14.0 & sn2001hf-20020114-ur.flm & Kast & 3260--7940\phantom{,}\phantom{1} & 6/5 & UCB \\ 
SN 2001is & 2002-01-06.29 & sn2001is-20020106.flm & FAST & 3720--7540\phantom{,}\phantom{1} & 7 & CfA \\ 
SN 2001is & 2002-01-07.31 & sn2001is-20020107.flm & FAST & 3720--7540\phantom{,}\phantom{1} & 7 & CfA \\ 
SN 2001is & 2002-01-08 & 2001is\_20020108.flm & AFOSC & 3360--7720\phantom{,}\phantom{1} & 25 & Asiago \\ 
SN 2001is & 2002-01-09 & 2001is\_20020109.flm & AFOSC & 3360--7720\phantom{,}\phantom{1} & 25 & Asiago \\ 
SN 2001is & 2002-01-14.0 & sn2001is-20020114-ui.flm & Kast & 3260--10,570 & 6/11 & UCB \\ 
SN 2002J & 2002-02-11.0 & sn2002j-20020211-ui.flm & Kast & 3300--10,400 & 6/11 & UCB \\ 
SN 2002J & 2002-02-14.0 & sn2002j-20020214-os.flm & LRIS & 3950--8830\phantom{,}\phantom{1} & 7 & UCB \\ 
SN 2002ce & 2002-04-12.18 & sn2002ce-20020412.flm & FAST & 3720--7520\phantom{,}\phantom{1} & 7 & CfA \\ 
SN 2002ce & 2002-05-07.0 & sn2002ce-20020507-ui.flm & Kast & 3300--10,300 & 6/11 & UCB \\ 
SN 2002dq & 2002-08-09.0 & sn2002dq-20020809-ui.flm & Kast & 3150--10,400 & 6/11 & UCB \\ 
SN 2002gw & 2002-10-29 & sn02gw.blue.29oct02.flm & LDSS-2 & 3600--9000\phantom{,}\phantom{1} & 13.5 & Asiago \\ 
SN 2002jj & 2002-12-05.35 & sn2002jj-20021205.flm & FAST & 3720--7520\phantom{,}\phantom{1} & 7 & CfA \\ 
SN 2002jj & 2002-12-12.0 & sn2002jj-20021212-ui.flm & Kast & 3170--10,400 & 6/11 & UCB \\ 
SN 2002jj & 2003-01-28.0 & sn2002jj-20030128-ui.flm & Kast & 3290--10,400 & 6/11 & UCB \\ 
SN 2002jz & 2003-01-07.0 & sn2002jz-20030107-br.flm & LRIS & 3080--9430\phantom{,}\phantom{1} & 7 & UCB \\ 
SN 2002jz & 2003-01-28.0 & sn2002jz-20030128-ui.flm & Kast & 3310--10,400 & 6/11 & UCB \\ 
SN 2002jz & 2003-02-28.0 & sn2002jz-20030228-br.flm & LRIS & 3120--9420\phantom{,}\phantom{1} & 7 & UCB \\ 
SN 2003E & 2003-01-28.0 & sn2003E-20030128-ui.flm & Kast & 3270--10,400 & 6/11 & UCB \\ 
SN 2003G & 2003-01-28.0 & sn2003G-20030128-ui.flm & Kast & 3100--10,400 & 6/11 & UCB \\ 
SN 2003G & 2003-02-04.0 & sn2003G-20030204-ui.flm & Kast & 3100--10,400 & 6/11 & UCB \\ 
SN 2003G & 2003-02-28.0 & sn2003g-20030228-br.flm & LRIS & 3040--9420\phantom{,}\phantom{1} & 7 & UCB \\ 
SN 2003H & 2003-02-04.0 & sn2003H-20030204-ui.flm & Kast & 3210--10,400 & 6/11 & UCB \\ 
SN 2003H & 2003-02-28.0 & sn2003h-20030228-br.flm & LRIS & 3170--9420\phantom{,}\phantom{1} & 7 & UCB \\ 
SN 2003aa & 2003-02-04.0 & sn2003aa-20030204-ui.flm & Kast & 3120--10,400 & 6/11 & UCB \\ 
SN 2003aa & 2003-02-28.0 & sn2003aa-20030228-br.flm & LRIS & 3050--9420\phantom{,}\phantom{1} & 7 & UCB \\ 
SN 2003aa & 2003-04-08.0 & sn2003aa-20030408-ui.flm & Kast & 3200--10,400 & 6/11 & UCB \\ 
SN 2003aa & 2003-05-25.0 & sn2003aa-20030525-brs.flm & LRIS & 3220--9420\phantom{,}\phantom{1} & 7 & UCB \\ 
SN 2003ao & 2003-02-22.31 & sn2003ao-20030222.flm & FAST & 3720--7540\phantom{,}\phantom{1} & 7 & CfA \\ 
SN 2003bk & 2003-03-04 & 2003bk\_20030304\_E3p6\_EFOSC2.flm & EFOSC & 3380--10,030 & 14 & Asiago \\ 
SN 2003br & 2003-03-11.48 & sn2003br-20030311.flm & FAST & 3720--7540\phantom{,}\phantom{1} & 7 & CfA \\ 
SN 2003dr & 2003-06-29.0 & sn2003dr-20030629-brs.flm & LRIS & 3250--9200\phantom{,}\phantom{1} & 7 & UCB \\ 
SN 2003dv & 2003-07-06.366 & sn2003dv-20030706.366-ui.flm & Kast & 3300--10,500 & 6/11 & UCB \\ 
SN 2003ed & 2003-05-25.0 & sn2003ed-20030525-brs.flm & LRIS & 3220--9420\phantom{,}\phantom{1} & 7 & UCB \\ 
SN 2003ed & 2003-05-30.0 & sn2003ed-20030530-ui.flm & Kast & 3230--10,400 & 6/11 & UCB \\ 
SN 2003ed & 2003-06-07.378 & sn2003ed-20030607.378-ui.flm & Kast & 3600--10,300 & 6/11 & UCB \\ 
SN 2003ed & 2003-07-06.329 & sn2003ed-20030706.329-ui.flm & Kast & 3350--10,310 & 6/11 & UCB \\ 
SN 2003ed & 2003-09-03.0 & sn2003ed-20030903-ui.flm & Kast & 3250--10,400 & 6/11 & UCB \\ 
SN 2003ef & 2003-05-21.19 & sn2003ef-20030521.flm & FAST & 3720--7540\phantom{,}\phantom{1} & 7 & CfA \\ 
SN 2003id & 2003-09-19.41 & sn2003id-20030919.flm & FAST & 3720--7540\phantom{,}\phantom{1} & 7 & CfA \\ 
SN 2003id & 2003-10-23.432 & sn2003id-20031023.432-ui.flm & Kast & 3830--10,500 & 6/11 & UCB \\ 
SN 2003ld & 2004-01-13.16 & sn2003ld-20040113.flm & FAST & 3720--7540\phantom{,}\phantom{1} & 7 & CfA \\ 
SN 2004C & 2004-01-15.50 & sn2004c-20040115.flm & FAST & 3720--7540\phantom{,}\phantom{1} & 7 & CfA \\ 
SN 2004C & 2004-01-17.0 & sn2004c-20040117-ui.flm & Kast & 3320--10,400 & 6/11 & UCB \\ 
SN 2004C & 2004-01-18.49 & sn2004c-20040118-mmt.flm & MMT-Blue & 3270--8460\phantom{,}\phantom{1} & 8 & CfA \\ 
SN 2004C & 2004-01-18.49 & sn2004c-20040118.flm & MMT-Blue & 3720--7540\phantom{,}\phantom{1} & 8 & CfA \\ 
SN 2004C & 2004-01-19.44 & sn2004c-20040119.flm & FAST & 3720--7520\phantom{,}\phantom{1} & 7 & CfA \\ 
SN 2004C & 2004-03-16.599 & sn2004c-20040316.599-br.flm & LRIS & 3350--9230\phantom{,}\phantom{1} & 7 & UCB \\ 
SN 2004C & 2004-03-17.40 & sn2004c-20040317.flm & FAST & 3720--7540\phantom{,}\phantom{1} & 7 & CfA \\ 
SN 2004C & 2004-04-26.336 & sn2004c-20040426.336-br.flm & LRIS & 3150--9410\phantom{,}\phantom{1} & 7 & UCB \\ 
SN 2004C & 2004-11-14.0 & sn2004c-20041114-br.flm & LRIS & 3070--9400\phantom{,}\phantom{1} & 7 & UCB \\ 
SN 2004al & 2004-03-13.26 & sn2004al-20040313.flm & FAST & 3720--7540\phantom{,}\phantom{1} & 7 & CfA \\ 
SN 2004aq & 2004-03-16.618 & sn2004aq-20040316.618-br.flm & LRIS & 3420--9230\phantom{,}\phantom{1} & 7 & UCB \\ 
SN 2004be & 2004-04-13.22 & sn2004be-20040413.flm & FAST & 3720--7540\phantom{,}\phantom{1} & 7 & CfA \\ 
SN 2004bm & 2004-05-12.226 & sn2004bm-20040512.226-ui.flm & Kast & 3390--10,580 & 6/11 & UCB \\ 
SN 2004cc & 2004-06-13.241 & sn2004cc-20040613.241-ui.flm & Kast & 3320--10,400 & 6/11 & UCB \\ 
SN 2004cc & 2004-06-20.0 & sn2004cc-20040620-ui.flm & Kast & 3300--10,400 & 6/11 & UCB \\ 
SN 2004ci & 2004-06-20.0 & sn2004ci-20040620-ui.flm & Kast & 3300--10,400 & 6/11 & UCB \\ 
SN 2004dd & 2004-07-18.0 & sn2004dd-20040718-ui.flm & Kast & 3300--10,400 & 6/11 & UCB \\ 
SN 2004dk & 2004-08-08.0 & sn2004dk-20040808-ui.flm & Kast & 3310--10,400 & 6/11 & UCB \\ 
SN 2004dk & 2004-08-16.174 & sn2004dk-20040816.174-ui.flm & Kast & 3320--10,400 & 6/11 & UCB \\ 
SN 2004dk & 2004-09-10.154 & sn2004dk-20040910.154-ui.flm & Kast & 3310--10,500 & 6/11 & UCB \\ 
SN 2004dk & 2004-09-24.15 & sn2004dk-20040924.150-ui.flm & Kast & 3320--10,500 & 6/11 & UCB \\ 
SN 2004dk & 2005-05-11.568 & sn2004dk-20050511.568-br.flm & LRIS & 3100--9350\phantom{,}\phantom{1} & 7 & UCB \\ 
SN 2004er & 2004-10-24 & SN04er\_b01\_CLA\_LD\_24oct04.flm & LDSS-2 & 3600--8990\phantom{,}\phantom{1} & 13.5 & CSP \\ 
SN 2004fc & 2004-12-17.168 & sn2004fc-20041217.168-ui.flm & Kast & 3320--10,600 & 6/11 & UCB \\ 
SN 2004fx & 2004-11-14.0 & sn2004fx-20041114-br.flm & LRIS & 3080--9400\phantom{,}\phantom{1} & 7 & UCB \\ 
SN 2004fx & 2004-12-17.342 & sn2004fx-20041217.342-ui.flm & Kast & 3340--10,550 & 6/11 & UCB \\ 
SN 2004gq & 2004-12-12.482 & sn2004gq-20041212.482-br.flm & LRIS & 3200--9320\phantom{,}\phantom{1} & 7 & UCB \\ 
SN 2004gq & 2004-12-13 & SN04gq\_b01\_DUP\_WF\_13dec04.flm & WFCCD & 3800--9230\phantom{,}\phantom{1} & 6 & CSP \\ 
SN 2004gq & 2004-12-17.317 & sn2004gq-20041217.317-ui.flm & Kast & 3370--10,300 & 6/11 & UCB \\ 
SN 2004gq & 2005-01-16.287 & sn2004gq-20050116.287-ui.flm & Kast & 3300--10,500 & 6/11 & UCB \\ 
SN 2004gq & 2005-02-12.395 & sn2004gq-20050212.395-br.flm & LRIS & 3780--9250\phantom{,}\phantom{1} & 7 & UCB \\ 
SN 2004gq & 2005-03-11.29 & sn2004gq-20050311.290-br.flm & LRIS & 3400--9260\phantom{,}\phantom{1} & 7 & UCB \\ 
SN 2005E & 2005-01-15.353 & sn2005e-20050115.353-br.flm & LRIS & 3380--9250\phantom{,}\phantom{1} & 7 & UCB \\ 
SN 2005E & 2005-02-01.125 & sn2005e-20050201.125-ui.flm & Kast & 3310--10,500 & 6/11 & UCB \\ 
SN 2005E & 2005-03-11.257 & sn2005e-20050311.257-br.flm & LRIS & 3400--9260\phantom{,}\phantom{1} & 7 & UCB \\ 
SN 2005H & 2005-01-17.113 & sn2005h-20050117.113-ui.flm & Kast & 3300--10,500 & 6/11 & UCB \\ 
SN 2005H & 2005-03-11.238 & sn2005h-20050311.238-br.flm & LRIS & 3390--9260\phantom{,}\phantom{1} & 7 & UCB \\ 
SN 2005J & 2005-02-04 & SN05J\_b01\_DUP\_WF\_04feb05.flm & WFCCD & 3800--9230\phantom{,}\phantom{1} & 6 & CSP \\ 
SN 2005ad & 2005-03-11.222 & sn2005ad-20050311.222-br.flm & LRIS & 3400--9260\phantom{,}\phantom{1} & 7 & UCB \\ 
SN 2005an & 2005-03-15 & SN05an\_b01\_DUP\_WF\_15mar05.flm & WFCCD & 3800--9230\phantom{,}\phantom{1} & 6 & CSP \\ 
SN 2005aq & 2005-03-11.284 & sn2005aq-20050311.284-br.flm & LRIS & 3400--9260\phantom{,}\phantom{1} & 7 & UCB \\ 
SN 2005bb & 2005-04-07 & SN05bb\_b01\_DUP\_WF\_07apr05.flm & WFCCD & 3800--9230\phantom{,}\phantom{1} & 6 & CSP \\ 
SN 2005io & 2005-11-05.636 & sn2005io-20051105.636-br.flm & LRIS & 3200--9240\phantom{,}\phantom{1} & 7 & UCB \\ 
SN 2005lr & 2005-12-18 & SN05lr\_g01\_NTT\_EM\_18dec05.flm & EMMI & 4000--10,200 & 9 & CSP \\ 
SN 2005lr & 2005-12-20 & SN05lr\_b01\_DUP\_WF\_20dec05.flm & WFCCD & 3800--9230\phantom{,}\phantom{1} & 6 & CSP \\ 
SN 2005mg & 2005-12-28 & sn2005mg-20051228.flm.gif.traced.flm$^{\delta}$ & FAST & 3490--7400\phantom{,}\phantom{1} & 7 & CfA \\ 
SN 2005mg & 2005-12-29.12 & sn2005mg-20051229.flm.gif.traced.flm$^{\delta}$ & FAST & 3490--7420\phantom{,}\phantom{1} & 7 & CfA \\ 
SN 2005mg & 2006-01-05.145 & sn2005mg-20060105.145-ui.flm & Kast & 3310--10,500 & 6/11 & UCB \\ 
SN 2006F & 2006-01-16 & SN06F\_g01\_NTT\_EM\_16jan06.flm & EMMI & 4000--10,200 & 9 & CSP \\ 
SN 2006T & 2006-02-13 & SN06T\_g01\_NTT\_EM\_13feb06.flm & EMMI & 4000--10,200 & 9 & CSP \\ 
SN 2006T & 2006-02-22.32 & sn2006t-20060222.320-ui.flm & Kast & 3320--10,300 & 6/11 & UCB \\ 
SN 2006be & 2006-03-30 & SN06be\_b01\_DUP\_WF\_30mar06.flm & WFCCD & 3800--9230\phantom{,}\phantom{1} & 6 & CSP \\ 
SN 2006be & 2006-05-05.419 & sn2006be-20060505.419-ui.flm & Kast & 3350--10,400 & 6/11 & UCB \\ 
SN 2006bp & 2006-04-28.26 & sn2006bp-20060428.260-ui.flm & Kast & 3320--10,500 & 6/11 & UCB \\ 
SN 2006ca & 2006-05-05.399 & sn2006ca-20060505.399-ui.flm & Kast & 3350--10,400 & 6/11 & UCB \\ 
SN 2006ca & 2006-06-20.367 & sn2006ca-20060620.367-ui.flm & Kast & 3330--10,580 & 6/11 & UCB \\ 
SN 2006ca & 2006-07-04.363 & sn2006ca-20060704.363-ui.flm & Kast & 3310--10,700 & 6/11 & UCB \\ 
SN 2006eg & 2006-08-24.399 & sn2006eg-20060824.399-ui.flm & Kast & 3320--10,400 & 6/11 & UCB \\ 
SN 2006qr & 2006-12-01.49 & sn2006qr-20061201.490-ui.flm & Kast & 3310--10,500 & 6/11 & UCB \\ 
SN 2006qr & 2006-12-13 & SN06qr\_b01\_DUP\_BC\_13dec06.flm & B\&C$_{2.5}$ & 3620--9820\phantom{,}\phantom{1} & 8 & CSP \\ 

\end{longtable}

\clearpage

\begin{table}
\caption{Log of Light Curves Published Herein\label{tab:allphot}}
\centering
\begin{tabular}{ l | c c | c c }
\hline
SN Name & Telescope & Filters & N Detections & Date Range \\
\hline
SN 2000N & KAIT & clear & 9 & 2000-03-04 -- 2000-06-03 \\ 
SN 2001J & KAIT & clear & 2 & 2001-01-15 -- 2001-01-16 \\ 
SN 2001M & KAIT & clear & 4 & 2001-01-21 -- 2001-02-03 \\ 
SN 2001ci & KAIT & clear & 8 & 2001-04-25 -- 2001-05-12 \\ 
SN 2002ds & KAIT & clear & 6 & 2002-06-25 -- 2002-07-25 \\ 
SN 2002jj & KAIT & clear & 6 & 2002-10-24 -- 2003-01-17 \\ 
SN 2002jz & KAIT & clear & 6 & 2002-12-24 -- 2003-01-31 \\ 
SN 2003bk & KAIT & clear & 6 & 2003-02-28 -- 2003-05-31 \\ 
SN 2003br & KAIT & clear & 8 & 2003-03-07 -- 2003-06-03 \\ 
SN 2003bw & KAIT & clear & 4 & 2003-03-03 -- 2003-03-28 \\ 
SN 2003id & KAIT,Nickel & B,V,R,I,clear & 8 & 2003-09-16 -- 2003-10-28 \\ 
SN 2004C & KAIT & clear & 9 & 2004-01-21 -- 2004-05-22 \\ 
SN 2004al & KAIT & clear & 5 & 2004-03-03 -- 2004-04-22 \\ 
SN 2004bm & KAIT & clear & 4 & 2004-04-13 -- 2004-05-08 \\ 
SN 2004er & KAIT & clear & 15 & 2004-09-25 -- 2005-02-07 \\ 
SN 2005ci & KAIT & clear & 20 & 2005-06-10 -- 2006-05-02 \\ 
SN 2005io & KAIT & clear & 13 & 2005-11-03 -- 2006-02-24 \\ 
SN 2005lr & KAIT & clear & 5 & 2005-12-04 -- 2006-01-06 \\ 
SN 2005mg & KAIT & clear & 2 & 2005-12-29 -- 2006-01-10 \\ 
SN 2006eg & KAIT & clear & 7 & 2006-07-31 -- 2006-10-15 \\ 

\hline
\end{tabular}
\end{table}

\end{document}